\documentclass[12pt]{article}
\usepackage{amsfonts,amssymb,epsfig,amsmath}
\usepackage{color}

\usepackage[vcentermath]{youngtab}

\renewcommand{\baselinestretch}{1.2}

\begin{document}

\makeatletter \@addtoreset{equation}{section} \makeatother
\renewcommand{\theequation}{\thesection.\arabic{equation}}
\renewcommand{\thefootnote}{\alph{footnote}}

\begin{titlepage}

\begin{center}
\hfill {\tt SNUTP11-007}\\
\hfill {\tt KIAS-P11047}\\
\hfill {\tt DAMTP-2011-57}

\vspace{1.5cm}

{\Large\bf On instantons as Kaluza-Klein modes of M5-branes}

\vspace{1.5cm}

\renewcommand{\thefootnote}{\alph{footnote}}

{
Hee-Cheol Kim$^{1,2}$, Seok Kim$^1$, Eunkyung Koh$^2$, Kimyeong Lee$^2$
and Sungjay Lee$^3$}

\vspace{1cm}

\textit{$^1$Department of Physics and Astronomy \& Center for
Theoretical Physics,\\
Seoul National University, Seoul 151-747, Korea.}

\textit{$^2$School of Physics, Korea Institute for Advanced Study,
Seoul 130-722, Korea.}

\textit{$^3$DAMTP, Centre for Mathematical Sciences, Cambridge University,\\
Wilberforce Road, Cambridge CB3 0WA, United Kingdom.}\\

\vspace{0.7cm}

E-mails: {\tt heecheol1@gmail.com, skim@phya.snu.ac.kr, ekoh@kias.re.kr,\\
klee@kias.re.kr, s.lee@damtp.cam.ac.uk}

\end{center}

\vspace{0.8cm}

\begin{abstract}

Instantons and W-bosons in 5d maximally supersymmetric Yang-Mills theory arise from
a circle compactification of the 6d (2,0) theory as Kaluza-Klein modes and winding
self-dual strings, respectively. We study an index which counts BPS instantons
with electric charges in Coulomb and symmetric phases. We first prove the existence of
unique threshold bound state  of (noncommutative) $U(1)$ instantons for any instanton number, and also show that charged instantons in the Coulomb phase correctly give the degeneracy of $SU(2)$ self-dual strings. By studying $SU(N)$ self-dual strings in the Coulomb phase, we
find novel momentum-carrying degrees on the worldsheet. The total number of these degrees
equals the anomaly coefficient of $SU(N)$ (2,0) theory. We finally show that our index can
be used to study the symmetric phase of this theory, and provide an interpretation as the
superconformal index of the sigma model on instanton moduli space.

\end{abstract}

\end{titlepage}

\renewcommand{\thefootnote}{\arabic{footnote}}

\setcounter{footnote}{0}

\renewcommand{\baselinestretch}{1}

\tableofcontents

\renewcommand{\baselinestretch}{1.2}

\section{Introduction}

Among many of the mysteries of M-theory \cite{Hull:1994ys}, M5-branes probably remain
to be the least understood object to date. The low energy dynamics of multiple M5-branes
is described by the 6d (2,0) superconformal field theory, whose details are mostly unknown
to date. The presence of $N^3$ degrees of freedom on $N$ coincident M5-branes
\cite{Klebanov:1996un} is at the center of the puzzle.

Reducing the (2,0) theory on a circle, one obtains the 5d maximal
super Yang-Mills theory which is the low energy description of D4-branes.
One would have thought that the 5d theory is the (2,0) theory
without all Kaluza-Klein (KK) momentum modes. However, instanton solitons,
being the threshold bound states of the D0-D4 branes, turn out to carry
all KK momenta along the circle \cite{Aharony:1997th,Aharony:1997an}. Even though
the 5d gauge theory appears to be non-renormalizable, it has been suggested
to have a UV fixed point which is given by the 6d theory. The question whether the
5d theory is UV complete by its own, by perhaps including instantons and
tensionless monopole strings,  is currently a major challenge
\cite{Lambert:2010wm,Douglas:2010iu,Lambert:2010iw}. See also \cite{Lambert:2011gb}
for a recent study.

In supersymmetric theories, there are nontrivial observables which are not very much
sensitive to the details of UV completion. Quantum effects which nontrivially contribute to
such BPS observables are often highly constrained. In this paper, rather than trying to address
the issues of UV completeness in full generality, we study   BPS observables of the circle
compactified $(2,0)$ theory which can be calculated in the 5 dimensional theory without
any ambiguity.

\begin{figure}[t]
  \begin{center}
    \includegraphics[width=13cm]{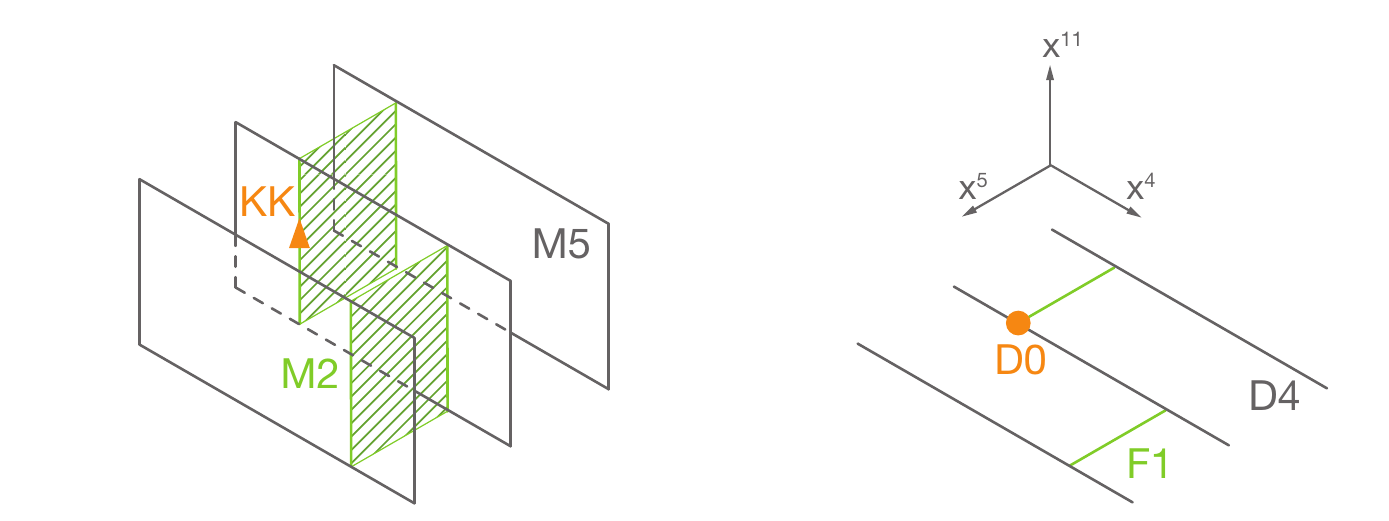}
    \caption{Instantons and elementary particles uplift to momenta and M2 self-dual
    strings.}\label{m2-m5}
  \end{center}
\end{figure}
More concretely, we study the problem of counting the BPS bound states of instantons
with other charged particles of the 5d maximally supersymmetric Yang-Mills theory in its
Coulomb or symmetric phase. These bound states are all at threshold, having zero binding
energies. From the 6d point of view, we are counting the BPS states of KK momentum
modes and winding self-dual strings \cite{Strominger:1995ac}, which come from intersecting
M2-M5 brane systems. See Fig \ref{m2-m5}. \cite{Arvidsson:2002tr} discusses some worldvolume
descriptions of these strings, and \cite{Berman:2004ew} calculates the anomalies oo self-dual
strings in the Coulomb phase. A more complete list of references can be found in
\cite{Berman:2007bv}. Compactifying the 5d theory further along another circle,
one can also view the latter system as coming from the magnetic monopole strings of D2-D4
brane systems by changing the role of the M-theory circle. One can count the BPS states on
these monopole strings with momenta. We find exact matches of these two calculations in some
simple cases, which supports that the 5d theory compacfitied on a circle is S-duality
invariant \cite{Douglas:2010iu,Lambert:2010iw,Tachikawa:2011ch}. We also find interesting
predictions on the quantum bound states of multiple monopole (or self-dual) strings at threshold.

At this point, we should mention that BPS bound states in 5 dimensional gauge theories
with $8$ supercharges have been studied quite extensively, as Nekrasov's instanton partition
function of these 5d theories on a circle can be interpreted as an index which counts such
bound states \cite{Nekrasov:2002qd,Nekrasov:2003rj,Gopakumar:1998ii}. Similar
studies for the maximally supersymmetric theory are relatively rare. See \cite{Sethi:1996kj}
for earlier works on these D0-D4 bound states. The bound states of
instantons with charged particles in the Coulomb phase are sometimes called dyonic
instantons, whose classical soliton solutions were first studied in \cite{Lambert:1999ua}.

We find that our index is closely related to the Nekrasov's partition function
for the 5 dimensional $\mathcal{N}\!=\!2^*$ theory compactified on $S^1$. Recall that
the $\mathcal{N}\!=\!2^*$ theory is obtained from maximally supersymmetric theory by
turning on a hypermultiplet mass. Among others, this relation was recently used by
Okuda and Pestun \cite{Okuda:2010ke}, by relating the chemical potentials of the index of
maximally supersymmetric theory to the parameters appearing in Nekrasov's partition function.
See our eqns.(\ref{parameter-map}), (\ref{mass}). The first part of this paper explicitly
verifies this proposal by a detailed calculation, in which we brutally compute the index up
to 3 instanton orders and show the agreement.

In the remaining part of this paper, using these results, we address some interesting
issues on the 5d Yang-Mills theory as the $(2,0)$ theory on a circle.

Firstly, although our index generally counts $\frac{1}{4}$-BPS particles with
electric/instanton charges, it also captures neutral $\frac{1}{2}$-BPS states with instanton
charges only. In particular, for the $U(1)$ SYM, there are no charged states as all the
fields are in adjoint representations. In this case, our index can be used to provide an
evidence of the conjecture on M-theory that these instantons form unique bound states at
all instanton numbers.\footnote{Although $U(1)$ instantons are `small' or singular in
field theory, we can treat them with a non-commutative deformation \cite{Nekrasov:1998ss}.
As we are computing an index, this continuous parameter does not affect the index, while
providing a mild UV completion for small instantons.}
Recall that, as D0-branes on a D4 are supposed to provide the KK states of the free
6d tensor multiplet along a circle, we expect there to be exactly one supermultiplet of
bound states at each KK momentum (or instanton number) \cite{Hull:1994ys,Aharony:1997th}.
The single particle index obtained from our $U(1)$ index exactly shows this desired property,
which we think provides the most nontrivial and concrete microscopic evidence for this
long-standing conjecture.

Secondly, we study various charged bound states in the Coulomb phase and
relate them to the BPS spectra of self-dual strings (M2-M5) or the magnetic monopole strings
(D2-D4) via S-duality. In particular, we show that the spectrum of a single W-boson in the
$SU(2)$ theory bound to many instantons exactly matches that of the magnetic monopole
string with many units of momentum on its worldsheet. This provides another evidence that the
5d theory is sufficient to reproduce the required KK spectrum. This example also
supports the fact that the S-duality of the $(2,0)$ theory on a 2-torus is visible from 5d
SYM, as the F1-D0 bounds are S-dual to the D2-momentum bounds.

We further study our index for more nontrivial charged bound states. We first study
an index counting BPS states of self-dual strings connecting $\frac{N(N-1)}{2}$ possible
pairs of M5-branes. From the perspective of the monopole strings, note that these monopoles
without KK momentum are visible as threshold bound states of $N-1$ distinct fundamental
monopoles \cite{Lee:1996if,Yi:1996bb}. Under a torus compactification, they are S-dual to
the $\frac{N(N-1)}{2}$ W-bosons. We observe that instantons provide some novel `partonic'
excitations on the worldsheets of these strings with nonzero momentum. These
degrees might be the basic constituents for all BPS states in the Coulomb phase, although we
only have studied a small subset of them. The number of these degrees scales faster
than $N^2$. Curiously, the total number of these degrees on 2d worldsheet turns out to
be $N(N^2-1)$, coinciding with the anomaly coefficient of the $A_{N\!-\!1}$ type $(2,0)$
theory \cite{Henningson:1998gx,Harvey:1998bx}.\footnote{The approach here is somewhat
different from the study of $\frac{1}{4}$-BPS junctions \cite{Bolognesi:2011rq}. As all
BPS monopole strings are parallel here, one might be able view them as degenerated
$\frac{1}{2}$-BPS junctions.}

We also find that novel bound states of identical multi-monopole strings are allowed
when (and only when) we turn on nonzero momentum. See section 4.1 for some examples.

Finally, we show that our index is meaningful and calculable in the symmetric
phase, in which the scalar VEV is zero so that the $SU(N)$ remains unbroken. The
chemical potentials that we introduce still makes the index calculable.
A complete physical interpretation of this index is not obvious to us at the moment,
for reasons summarized in section 5. However, we show that our symmetric phase index can
be intepreted as a `superconformal index,' counting BPS operators of the superconformal
quantum mechanics of the low energy sigma model whose target space is given by the
instanton moduli space.

Perhaps we should emphasize that the study of Witten index for threshold bound states
is very subtle, as there is a continuum of spectrum above the threshold without a mass
gap. In this situation, Witten index generally loses its topological robustness against
the change of various continuous parameters. In fact, the threshold D0-brane bound states
in type IIA string theory were studied for two D0-branes
\cite{Yi:1997eg,Sethi:1997pa} and then for general number of D0-branes \cite{Moore:1998et},
which actually show such subtleties. Fortunately, we have a way to circumvent this problem
in our D0-D4 system. Firstly, the position
zero modes of the instantons on D4-branes are lifted by introducing the chemical potentials
for the $SO(4)$ angular momentum, which is equivalent to the Omega deformation
\cite{Nekrasov:2002qd}. This only works for even dimensions and fails to
completely localize odd dimensional zero modes, say in 9 spatial dimensions for D0-branes
\cite{Moore:1998et}. Secondly, the D0-branes' position zero modes away from the D4-branes
are lifted by introducing non-commutativity \cite{Nekrasov:1998ss}. Thirdly, instantons also
have internal noncompact directions
from their size moduli. They are lifted by introducing the chemical potentials for the
$U(N)$ electric charges. Especially, with the $SO(4)$ chemical potentials, our index
counts both single- and multi-particle states, either bound or unbound.
For each particle, the $SO(4)$ chemical potentials provide a factor of index coming from
its center-of-mass supermultiplet, which we call $I_{com}$. By counting how many factors of
$I_{com}$ appear in a term, we can see the particle number of that contribution.

The remaining part of this paper is organized as follows. In section 2, we explain
the D0-D4 quantum mechanics. We also calculate the index and relate it to the instanton
partition function of the 5d $\mathcal{N}=2^\ast$ theory. In section 3, we study the
threshold bound states of $U(1)$ instantons, or the bound states of one D4-brane
with many D0's, and prove that there exist unique bound states at all instanton number.
In section 4, we study various charged bound states in the Coulomb phase. Especially, we
show that the bound states of multi-instantons with a W-boson in the $SU(2)$ theory
completely reproduce the degeneracy of an $SU(2)$ monopole (or self-dual) string with momenta.
We also study the threshold bound states of $SU(N)$ strings with momenta and find novel
`partonic' degrees of freedom. Section 5 explains various interpretations of the instanton
index in the symmetric phase, focusing on the superconformal index interpretation. Section 6
concludes with discussions. Two appendices explain the technical details of the saddle
points and the determinants in the index calculation.

\section{The instanton index of 5d maximal SYM}

5d maximal SYM for $N$ D4-branes has a dimensionful coupling constant $g_{YM}^2$.
This theory has `instanton' particles, classically satisfying
$F_{\mu\nu}=\pm  {\star_4}  F_{\mu\nu}$ in the spatial part. They are D0-branes bound to
the D4-branes which can be uplifted to  the KK momenta on M5-branes along the M-theory circle. The
mass of an instanton is thus identified with the radius of the M-theory circle as
\begin{equation}
  \frac{8\pi^2}{g_{YM}^2}=\frac{1}{R_{11}}\ .
\end{equation}
Elementary excitations or W-bosons are uplifted to self-dual strings on M5-brane.
See Fig \ref{m2-m5}. As $k$ D0-branes bound to $N$ D4-branes can be described
by a matrix quantum mechanics, we calculate the index from this mechanical system. We first
explain this system in subsection 2.1. In subsection 2.2, we evaluate  the index which counts
these BPS particles.

\subsection{The D0-D4 quantum mechanics}

The quantum mechanics for $k$ D0-branes on $N$ D4-branes has a $U(k)$ vector multiplet,
an adjoint hypermultiplet and $N$ fundamental hypermultiplets. The global symmetry
$SO(4)_1\sim SU(2)_{1L}\times SU(2)_{2R}$ rotates the $4$ spatial directions on D4-branes,
and $SO(4)_2 \sim SU(2)_{2L} \times SU(2)_{2R}$ is a subgroup of $SO(5)$ R-symmetry
unbroken by a nonzero scalar VEV. We mostly follow the notations of \cite{Dorey:2002ik}.
Before adding fundamental hypermultiplets, the Lagrangian is simply that for $k$ D0-branes,
a reduction of 10d SYM theory with $U(k)$ gauge group. This action is given by
\begin{eqnarray}\label{sym-action}
  L_{SYM}&=&{\rm tr}_k\left(\frac{1}{2}D_t\varphi_ID_t\varphi_I
  +\frac{1}{2}D_ta_mD_ta_m+\frac{1}{4}[\varphi_I,\varphi_J]^2
  +\frac{1}{2}[a_m,\varphi_I]^2+\frac{1}{4}[a_m,a_n]^2\right.\nonumber\\
  &&\left.+\frac{i}{2}(\bar\lambda^{i\dot\alpha})^\dag
  D_t\bar\lambda^{i\dot\alpha}
  +\frac{1}{2}(\bar\lambda^{i\dot\alpha})^\dag(\gamma^I)^i_{\ j}
  [\varphi_I,\bar\lambda^{j\dot\alpha}]
  +\frac{i}{2}(\lambda^i_\alpha)^\dag D_t\lambda^i_\alpha
  -\frac{1}{2}(\lambda^i_\alpha)^\dag(\gamma^I)^i_{\ j}
  [\varphi_I,\lambda^j_\alpha]\right.\nonumber\\
  &&\left.-\frac{i}{2}(\lambda^i_\alpha)^\dag(\sigma^m)_{\alpha\dot\beta}
  [a_m,\bar\lambda^{i\dot\beta}]+\frac{i}{2}(\bar\lambda^{i\dot\alpha})^\dag
  (\bar\sigma^m)^{\dot\alpha\beta}[a_m,\lambda^i_\beta]\right)\ .
\end{eqnarray}
$I\!=\!1,2,3,4,5$ and $i\!=\!1,2,3,4$ are vector and spinor indices,
respectively, for the $SO(5)$ R-symmetry of 5d SYM.
$m\!=\!1,2,3,4$ are for $SO(4)_1$ vectors along the spatial directions of D4-branes,
$\alpha\!=\!1,2$ and $\dot\alpha\!=\!1,2$ are for $SU(2)_{1L}\times SU(2)_{2R}$ indices,
$\sigma^m=(i\vec\tau,1)$,  $\bar\sigma^m=(-i\vec\tau,1)$ with
the Pauli matrices $\vec\tau$, and finally $D_t=\partial_t-i[A_t,\ ]$.
We take the gamma matrices for $Sp(4)\sim SO(5)$ in the following representation,
\begin{equation}
  \gamma^I:\ \gamma^5=\left(\begin{array}{cc}\delta_a^{\ b}&0\\
  0&-\delta^{\dot{a}}_{\ \dot{b}}\end{array}\right)\ ,\ \
  \gamma^m=\left(\begin{array}{cc}0&(\sigma^m)_{a\dot{b}}\\
  (\bar\sigma^m)^{\dot{a}b}&0\end{array}\right)\ ,\ \
  \gamma^{12345}=-1\ ,
\end{equation}
where $a,\dot{a}=1,2$ etc. denote indices for $SU(2)_{2L}\times SU(2)_{2R}$ subgroup
of $Sp(4)$. We deliberately chose the first four components of the
internal $SO(5)$ vectors to be labeled by the same index $m$ as the spatial $SO(4)_1$,
for a minor technical reason to be explained below. The $Sp(4)$ invariant tensor
$\omega$ takes the following from
\begin{equation}
  \omega\equiv-\gamma^1\gamma^3=\left(\begin{array}{cc}\epsilon&0\\
  0&\epsilon\end{array}\right)\ ,\ \
  \omega^T=-\omega\ ,\ \ \omega(\gamma^I)^T\omega^{-1}=+\gamma^I\ ,
\end{equation}
where $\epsilon\equiv i\tau^2$. We also define the anti-symmetric tensors
$\epsilon^{\alpha\beta}$, $\epsilon_{\alpha\beta}$, $\epsilon^{\dot\alpha\dot\beta}$,
$\epsilon_{\dot\alpha\dot\beta}$ by $\epsilon^{12}=-\epsilon_{12}=1$ and so on.
Fermions satisfy the symplectic-Majorana reality condition using $SU(2)_{1L}\times SO(5)$
or $SU(2)_{1R}\times SO(5)$ (overbars on spinors are used for $SU(2)_{1R}$, not for conjugates):
\begin{equation}\label{reality}
  \lambda^i_\alpha=\epsilon_{\alpha\beta}\omega^{ij}(\lambda^j_\beta)^\dag\
  ,\ \ \bar\lambda^{i\dot\alpha}=\epsilon^{\dot\alpha\dot\beta}\omega^{ij}
  (\bar\lambda^{j\dot\beta})^\dag\ .
\end{equation}
The supercharges to be explained below also satisfy these reality conditions.
The terms on the first line of (\ref{sym-action})
including $a_m$ may be written as
\begin{equation}
  \frac{1}{2}D_ta_{\alpha\dot\alpha}D_ta^{\dot\alpha\alpha}
  +\frac{1}{2}[\varphi_I,a_{\alpha\dot\alpha}][\varphi_I,a^{\dot\alpha\alpha}]
  -\hat{D}^{\dot\alpha}_{\ \dot\beta}\hat{D}^{\dot\beta}_{\ \dot\alpha}
\end{equation}
with
\begin{equation}
  \hat{D}^{\dot\alpha}_{\ \dot\beta}=\frac{1}{2}\left([a^{\dot\alpha\alpha},a_{\alpha\dot\beta}]-\frac{1}{2}
  \delta^{\dot\alpha}_{\ \dot\beta}[a^{\dot\gamma\alpha},
  a_{\alpha\dot\gamma}]\right)\ ,
\end{equation}
while the last line may be written as
\begin{equation}
  -\frac{i}{\sqrt{2}}(\lambda^i_\alpha)^\dag[a_{\alpha\dot\beta},
  \bar\lambda^{i\dot\beta}]+\frac{i}{2}(\bar\lambda^{i\dot\alpha})^\dag
  [a^{\dot\alpha\beta},\lambda^i_\beta]\ ,
\end{equation}
where $a_{\alpha\dot\alpha}=\frac{1}{\sqrt{2}}(\sigma^m)_{\alpha\dot\alpha}a_m$,
$a^{\dot\alpha\alpha}=\frac{1}{\sqrt{2}}(\bar\sigma^m)^{\dot\alpha\alpha}a_m$,
$a^{\dot\alpha\alpha}=\epsilon^{\alpha\beta}\epsilon^{\dot\alpha\dot\beta}a_{\beta\dot\beta}
=(a_{\alpha\dot\alpha})^\dag$.

Adding $N$ fundamental hypermultiplets for the D0-D4 open strings, which we call
$(q_{\dot\alpha},\psi^i)$, the total action takes the form of $L=L_{SYM}+L_f$ with
\begin{eqnarray}
  L_f&=&D_tq_{\dot\alpha}D_t\bar{q}^{\dot\alpha}-
  (\varphi_I\bar{q}^{\dot\alpha}\!-\!\bar{q}^{\dot\alpha}v_I)
  (q_{\dot\alpha}\varphi_I\!-\!v_Iq_{\dot\alpha})
  +i(\psi^i)^\dag D_t\psi^i+(\psi^i)^\dag(\gamma^I)^i_{\ j}
  \left(\psi^j\varphi_I-v_I\psi^j\right)\nonumber\\
  &&+\sqrt{2}i\left((\bar\lambda^{i\dot\alpha})^\dag\bar{q}^{\dot\alpha}
  \psi^i-(\psi^i)^\dag q_{\dot\alpha}\bar\lambda^{i\dot\alpha}\right)\ ,
\end{eqnarray}
and then replacing $\hat{D}$ above for the adjoint hypermultiplet potential
by $D$ given as follows:
\begin{equation}
  D^{\dot\alpha}_{\ \dot\beta}= \left(\bar{q}^{\dot\alpha}
  q_{\dot\beta}-\frac{1}{2}\delta^{\dot\alpha}_{\ \dot\beta}
  (\bar{q}^{\dot\gamma}q_\gamma)-\frac{1}{2}\zeta^A(\tau^A)^{\dot\alpha}_{\ \dot\beta}\right)
  +\frac{1}{2}\left([a^{\dot\alpha\alpha},a_{\alpha\dot\beta}]-\frac{1}{2}
  \delta^{\dot\alpha}_{\ \dot\beta}[a^{\dot\gamma\alpha},
  a_{\alpha\dot\gamma}]\right)\ .
\end{equation}
The covariant derivatives are defined as $D_tq_{\dot\alpha}=\partial_tq_{\dot\alpha}
+iq_{\dot\alpha}A_t$, etc. The $N\times N$ matrix parameters $v_I$ represent the VEV of
the five real scalar fields in the 5 dimensional theory. As the five matrices should commute
to represent the vacuum, we take all of them to be diagonalized. This breaks the $U(N)$
symmetry to $U(1)^N$. As mentioned at the beginning, we shall consider the case where only
one scalar VEV $v_5$ could be nonzero. This amounts to separating D4-branes along the fifth
direction as in Fig \ref{m2-m5}. We also added a deformation of Fayet-Iliopoulos term
($\propto\zeta^A$) for non-commutative instantons. The $SU(2)_{1R}$ triplet $D$ may be
written as
\begin{equation}\label{ADHM-D}
  D^A\equiv(\tau^{A})^{\dot\beta}_{\ \dot\alpha}D^{\dot\alpha}_{\ \dot\beta}
  =\bar{q}^{\dot\alpha}q_{\dot\beta}(\tau^{A})^{\dot\beta}_{\ \dot\alpha}-\zeta^A
  +\frac{i}{2}\bar\eta^A_{mn}[a_m,a_n]\ ,
\end{equation}
where $\bar\sigma_{mn}=i\bar\eta^a_{mn}\tau^a$ with anti-self-dual 't Hooft symbol
$\bar\eta^a_{mn}$.

5d SYM preserves $16$ supersymmetries. We write them as $Q^i_\alpha$
and $\bar{Q}^{i\dot\alpha}$, which satisfy reality conditions like (\ref{reality}).
Combining these into a $SO(4,1)$ spinor $Q^i_M$ with $M=1,2,3,4$, the superalgebra is
given by
\begin{equation}
  \{Q^i_M,Q^j_N\}=P_\mu(\Gamma^\mu C)_{MN}\omega^{ij}+i\frac{8\pi^2k}{g_{YM}^2}
  C_{MN}\omega^{ij}-i{\rm tr}(qv_I)(\Gamma^I\omega)^{ij}C_{MN}
\end{equation}
where $k$ is the instanton charge and $q$ is the electric charge. Among these, only $8$
of them are realized in the mechanical model for the half-BPS instantons. The preserved
supercharge is taken to be $\bar{Q}_{\dot\alpha}^i$ for self-dual instantons.
The fields $(A_t,\bar\lambda^{i\dot\alpha}, \varphi_I)$ form a vector
multiplet, while $(a_{\alpha\dot\beta},\lambda^i_\alpha)$ and $(q_{\dot\alpha},\psi^i)$
form hypermultiplets in $U(k)$ adjoint and fundamental, respectively.
The $\bar{Q}^i_{\dot\alpha}$ transformations are given by
\begin{eqnarray}\label{SUSY-vector}
  \bar{Q}^{i\dot\alpha}A_t&=&i\bar\lambda^{i\dot\alpha}\ ,\ \
  \bar{Q}^{i\dot\alpha}\varphi^I=-i(\gamma^I)^i_{\ j}\bar\lambda^{j\dot\alpha}\\
  \bar{Q}^{i\dot\alpha}\bar\lambda^{j\dot\beta}&=&
  \epsilon^{\dot\alpha\dot\beta}(\gamma^I\omega)^{ij}D_0\varphi^I
  -\frac{i}{2}\epsilon^{\dot\alpha\dot\beta}(\gamma^{IJ}\omega)^{ij}[\varphi^I,\varphi^J]
  -2i\omega^{ij}D^{\dot\alpha}_{\ \ \dot\gamma}\epsilon^{\dot\gamma\dot\beta}\nonumber
\end{eqnarray}
for the vector multiplet,
\begin{eqnarray}\label{SUSY-adjoint}
  &&\bar{Q}^{i\dot\alpha}a_{\alpha\dot\beta}
  =\sqrt{2}\delta^{\dot\alpha}_{\dot\beta}\lambda^i_\alpha\ \ \
  (\ {\rm or}\ \ \bar{Q}^{i\dot\alpha}a^m=(\bar\sigma^m)^{\dot\alpha\beta}
  \lambda^i_\beta\ )\\
  &&\bar{Q}^{i\dot\alpha}\lambda^j_\beta=
  (\bar\sigma^m)^{\dot\alpha\gamma}\epsilon_{\gamma\beta}
  \left(i\omega^{ij}D_ta_m+(\gamma^I\omega)^{ij}[\varphi_I,a_m]\right)=
  \sqrt{2}\left(i\omega^{ij}D_ta_{\beta\dot\gamma}+(\gamma^I\omega)^{ij}
  [\varphi_I,a_{\beta\dot\gamma}]\right)\epsilon^{\dot\gamma\dot\alpha}
  \nonumber
\end{eqnarray}
for the adjoint hypermultiplet, and
\begin{equation}\label{SUSY-fundamental}
  \bar{Q}^{i\dot\alpha}q_{\dot\beta}=\sqrt{2}\delta^{\dot\alpha}_{\dot\beta}\psi^i\ ,\ \
  \bar{Q}^{i\dot\alpha}\psi^j=\sqrt{2}\left[i\omega^{ij}D_t q_{\dot\beta}-(\gamma^I\omega)^{ij}
  \left(q_{\dot\beta}\varphi_I-v_Iq_{\dot\beta}\right)\right]\epsilon^{\dot\beta\dot\alpha}
\end{equation}
for the fundamental hypermultiplet. The half-BPS $k$ instantons, of either single or
multi-particle types, are supersymmetric ground states of this mechanical model.

There also exist quarter-BPS states carrying non-zero electric charges of
$U(1)^N\subset U(N)$ unbroken by the VEV $v\equiv v_5\neq 0$. Depending on the sign of the
electric charge, the particle preserves different components of supercharges.
Without losing generality, we consider the particles preserving 4 real supercharges
$\bar{Q}^{\dot{a}\dot\alpha}$ with $\dot{a}=1,2$, $\dot\alpha=1,2$: recall that
the $Sp(4)$ R-symmetry index $i=1,2,3,4$ decomposes to $a=1,2$ (for $i=1,2$)
and $\dot{a}=1,2$ (for $i=3,4$). Decomposing the fermions in fundamental hypermultiplets as
$\psi^i=(\psi_a,\psi^{\dot{a}})$, one obtains the following supersymmetry transformation
\begin{equation}\label{BPS-dyonic}
  \bar{Q}^{\dot{a}\dot\alpha}\psi^{\dot{b}}=
  \sqrt{2}\left[iD_tq_{\dot\beta}+(q_{\dot\beta}\varphi_5-vq_{\dot\beta})
  \right]\epsilon^{\dot{a}\dot{b}}\epsilon^{\dot\beta\dot\alpha}\ .
\end{equation}
This yields a BPS equation on the right hand side which agrees with those
studied in \cite{Lambert:1999ua,Kim:2008kn}. The classical BPS configurations invariant
under the supersymmetry (\ref{BPS-dyonic}) has a solution
\begin{equation}
  A_t=\varphi^5\ ,\ \ q_{\dot\alpha}(t)=e^{-ivt}q_{\dot\alpha}(0)\ .
\end{equation}
For a reason which will be clear shortly, we want to redefine variables to make these
quarter-BPS configuration to be time independent. We define variables $x_{\dot\alpha}$ as
\begin{equation}
  q_{\dot\alpha}(t)=e^{-ivt}x_{\dot\alpha}(t)\ .
\end{equation}
In this variable, the Lagrangian including fundamental variables is given by
\begin{eqnarray}\label{fund-action}
  L_f&=&
  \left(D_t\bar{x}^{\dot\alpha}+i\bar{x}^{\dot\alpha}v\right) \left(D_tx_{\dot\alpha}-ivx_{\dot\alpha}\right) -
  (\varphi_I\bar{x}^{\dot\alpha}\!-\!\bar{x}^{\dot\alpha}v_I)
  (x_{\dot\alpha}\varphi_I\!-\!v_Ix_{\dot\alpha})\\
  &&+i(\xi^i)^\dag\left(D_t\xi^i-iv\xi^i\right)+
  (\xi^i)^\dag(\gamma^I)^i_{\ j}\left(\xi^j\varphi_I-v_I\xi^j\right)+
  \sqrt{2}i\left((\bar\lambda^{i\dot\alpha})^\dag\bar{x}^{\dot\alpha}\xi^i-
  (\xi^i)^\dag x_{\dot\alpha}\bar\lambda^{i\dot\alpha}\right)\nonumber
\end{eqnarray}
where we defined $\xi^i\equiv e^{ivt}\psi^i$. In the next subsection, we will be
interested in the Euclidean version of this theory relevant for computing an index where
the time direction is taken to be periodic with radius $\beta$. Had we been not
redefining variables to $x_{\dot\alpha}$, time dependent saddle points in the Euclidean
theory would be $q_{\dot\alpha}\sim e^{-v\tau}$ with Euclidean time $\tau=it$, spoiling the
periodicity. This is fine as we can naturally work with non-periodic or twisted boundary
conditions along the circle. Going to the variable $x_{\dot\alpha}$ to restore periodicity is
sometimes called `untwisting', which introduces an external gauge field as in
(\ref{fund-action}), making the Euclidean action complex.

To define and calculate the index for these $\frac{1}{4}$-BPS particles,
it is convenient to choose one supercharge among $\bar{Q}^{\dot{a}\dot\alpha}$. We take it as
\begin{equation}\label{special-susy}
  Q\equiv\frac{1}{\sqrt{2}}\epsilon_{\dot{a}\dot\alpha}\bar{Q}^{\dot{a}
  \dot\alpha}\ ,\ \ Q=-Q^\ast\ .
\end{equation}
This may be regarded as the scalar supercharge in a twisted theory which
identifies $SU(2)_{1R}$ and $SU(2)_{2R}$. We shall use a subset of supercharges of
$\bar{Q}^{\dot{a}\dot\alpha}$, including $Q$ above, to localize the quantum mechanical
path integral for our index in the next subsection.

It is sometimes helpful to rewrite the above theory in a cohomological formulation
using $Q$. This is a straightforward generalization of \cite{Moore:1998et} by including
fundamental hypermultiplets and extra potential terms from nonzero VEV $v$. We
consider a Euclidean theory obtained by taking $t=-i\tau$,
$A_t=iA_\tau$. Following \cite{Moore:1998et}, we use the `matrix model' like notation by
replacing covariant time derivatives $D_\tau$ in the Euclidean theory by $A_\tau$. Whenever
necessary, one can restore time derivatives simply by replacing $A_\tau$ by $D_\tau$.\footnote{In \cite{Moore:1998et}, multi-instanton bound states (without D4's) were considered, generalizing
earlier works \cite{Yi:1997eg,Sethi:1997pa}. In that case, the quantum mechanical path integral
reduced down to an ordinary matrix integral. This will not be true in our case, so the matrix
model like notation should always be understood with this replacement.} Defining
\begin{eqnarray}\label{coh-redefine}
  &&\phi\equiv-i(A_\tau+i\varphi_5)\ ,\ \ \bar\phi\equiv i(A_\tau-i\varphi_5)\ ,\ \
  \eta\equiv\sqrt{2}i\epsilon_{\dot{a}\dot\alpha}\bar\lambda^{\dot{a}\dot\alpha}
  \ ,\nonumber\\
  &&\Psi_m\equiv Qa_m=\frac{1}{\sqrt{2}}\epsilon_{\dot{a}\dot\alpha}
  \left(\bar\sigma_m\right)^{\dot\alpha\beta}\lambda^{\dot{a}}_\beta\
  ,\ \ \Psi_{m\!+\!4}\equiv Q\varphi_m=-\frac{i}{\sqrt{2}}
  \epsilon_{\dot{a}\dot\alpha}\left(\bar\sigma_m\right)^{\dot{a}b}
  \bar\lambda^{\dot\alpha}_b
\end{eqnarray}
in the $U(k)$ adjoint sector, part of the supersymmery transformation under
$Q$ is given by
\begin{eqnarray}
  &&Q\phi=0\ ,\ \ Q\bar\phi=\eta\ ,\ \ Q\eta=[\phi,\bar\phi]\nonumber\\
  &&Q\Psi_m=[\phi,a_m]\ ,\ \ Q\Psi_{m\!+\!4}=[\phi,\varphi_m]\ ,
\end{eqnarray}
which is same as that in \cite{Moore:1998et} if one defines the
`$SO(8)$ vectors' $(a_m,\varphi_m)$ and $(\Psi_m,\Psi_{m\!+\!4})$. Note that $Q^2$ acting
on these variables yields $[\phi,\ \ ]$, implying that $Q$ is nilpotent up to a complexified
gauge transformation generated by $\phi$. In case time derivative is kept, this complex gauge transformation is accompanied by a time translation. In the variable $x_{\dot\alpha}$, time translation generator is simply $H-v^i\Pi_i$ with the $U(1)^N$ electric charges $\Pi_i$,
since we moved to a rotating frame in the $U(1)^N$ angles.

We also consider 4 components of $\lambda_{a\alpha}$ and 3 components
$(\epsilon^{-1}\bar\sigma^{mn})_{\dot{a}\dot\alpha} \bar\lambda^{\dot{a}\dot\alpha}$
of $\bar\lambda^{\dot{a}\dot\alpha}$, apart from $\eta$ considered in (\ref{coh-redefine}).
We reorganize them into a seven component vector $\vec\chi$ given by
\begin{equation}
  \vec\chi=\left(\chi^A_R,\chi^A_L,\chi\right)=
  \left(-\frac{1}{\sqrt{2}}\left(\epsilon^{-1}\tau^A\right)_{\dot{a}\dot\alpha}
  \bar\lambda^{\dot{a}\dot\alpha},\  - \frac{i}{\sqrt{2}}\left(
  \tau^A\epsilon\right)^{a\alpha}\lambda_{a\alpha},\ \frac{1}{\sqrt{2}}
  \epsilon^{a\alpha}\lambda_{a\alpha}\right)\ .
\end{equation}
Defining seven components of quadratures as
\begin{align}
  \vec{\mathcal{E}}=&\left(\mathcal{E}^A_R,
  \mathcal{E}^A_L,\mathcal{E}\right)\\
  \equiv&\left(\frac{i}{2}\bar\eta^A_{mn}\left([\varphi_m,\varphi_n]-
  [a_m,a_n]\right)-\bar{x}^{\dot\alpha}x_{\dot\beta}(\tau^A)^{\dot\beta}_{\ \dot\alpha}
  +\zeta^A,\frac{i}{2}\eta^A_{mn}\left([\varphi_m,a_n]+
  [a_m,\varphi_n]\right),-i[\varphi_m,a_m]\right)\nonumber
\end{align}
generalizing \cite{Moore:1998et}, with $A=1,2,3$, $\sigma_{mn}=i\eta^A_{mn}\tau^A$ and
$\bar\sigma_{mn}=i\bar\eta^A_{mn}\tau^A$, one obtains
\begin{equation}
  Q\vec\chi=i\vec{\mathcal{E}}\ .
\end{equation}
One also finds
\begin{equation}\label{7-potential}
  \frac{1}{2}{\rm tr}\left(\vec{\mathcal{E}}\cdot\vec{\mathcal{E}}\right)
  =\frac12 {\rm tr}\left(-\frac{1}{2}[\varphi_m,\varphi_n][\varphi_m,\varphi_n]
  -[\varphi_m,a_n][\varphi_m,a_n]+D^AD^A
  -[\varphi_m,\varphi_n]\bar{q}^{\dot\alpha}q_{\dot\beta}
  \left(\bar\sigma^{mn}\right)^{\dot\beta}_{\ \dot\alpha}\right)\ ,
\end{equation}
where $D^A$ is defined by (\ref{ADHM-D}). The right hand side is the bosonic potential
energy apart from the last term (which will be taken care of shortly).
After some algebra, and using equations of motion for fermions,
one finds that
\begin{equation}
  Q^2\vec\chi=Q\left(i\vec{\mathcal{E}}\right)=[\phi,\vec\chi]\ ,
\end{equation}
so that $Q^2$ acting on $\vec\chi$ is again a complexified gauge transformation.
To make $Q$ nilpotent (up to a gauge transformation) off-shell, we introduce seven
auxiliary scalars $\vec{H}$ which satisfy
\begin{equation}
  Q\vec\chi=\vec{H}\ ,\ \ Q\vec{H}=[\phi,\vec\chi]
\end{equation}
with the bosonic action containing $\vec{H},\vec{\mathcal{E}}$ given by
\begin{equation}
  \frac{1}{2}\vec{H}\cdot\vec{H}-i\vec{H}\cdot\vec{\mathcal{E}}\ .
\end{equation}
Integrating out $\vec{H}$ gives the potential energy (\ref{7-potential})
and supersymmetry.

Finally, fundamental variables transform under $Q$ as
\begin{eqnarray}
  &&Qx_{\dot\alpha}=-\epsilon_{\dot\alpha\dot{a}}e^{ivt}\psi^{\dot{a}}
  \equiv-\epsilon_{\dot\alpha\dot{a}}\xi^{\dot{a}}\ ,\nonumber\\
  &&Q\xi^{\dot{a}}=\epsilon^{\dot{a}\dot\alpha}x_{\dot\alpha}\phi
  \ ,\ \ Q\xi_a=  - \left(\sigma^m\right)_{a\dot\alpha}
  \epsilon^{\dot\alpha\dot\beta}x_{\dot\beta}\varphi_m
  \equiv i\mathcal{F}_a\ .
\end{eqnarray}
$Q^2$ acing on $x_{\dot{a}}$ and $\xi^{\dot{a}}$ is again a gauge transformation, and
\begin{equation}
  Q^2\xi_a=-\chi_a\phi \ ,
\end{equation}
using the equation of motion for $\xi_a$.
It is again useful to define complex variables $h_a$ so that
\begin{equation}
  Q\xi_a=h_a\ ,\ \ Qh_a=-\xi_a\phi\ .
\end{equation}
The action involving $h_a$ can be written as
\begin{equation}
  h_a(h_a)^\dag-i\mathcal{F}_a(h_a)^\dag-ih_a(\mathcal{F}_a)^\dag\ .
\end{equation}
After integrating out $h_a$ by setting $h_a=i\mathcal{F}_a$, one obtains
\begin{equation}
  {\rm tr}\left(\mathcal{F}_a(\mathcal{F}_a)^\dag\right)={\rm tr}\left(
  (\varphi_m\bar{x}^{\dot\alpha})(x_{\dot\alpha}\varphi_m)+\frac{1}{2}[\varphi_m,\varphi_n]
  \bar{x}^{\dot\alpha}x_{\dot\beta}\left(\bar\sigma^{mn}\right)^{\dot\beta}_{\ \dot\alpha}
  \right)\ .
\end{equation}
Collecting all the results, one can rewrite the bosonic part of the Lagrangian as follows
\begin{align}\label{bosonic}
  L_{\rm bos} = & \frac12 {\rm tr}\Big(\frac{1}{4}[\phi,\bar\phi]^2 -
  [\phi,a_m][\bar\phi,a_m] - [\phi,\varphi_m][\phi,\varphi_m]\!+\!
  \vec{\mathcal{E}}\cdot\vec{\mathcal{E}} + \mathcal{F}_a(\mathcal{F}_a)^\dag
  +\{\phi,\bar\phi\}\bar{x}^{\dot\alpha}x_{\dot\alpha}
  -4 \phi\bar{x}^{\dot\alpha}vx_{\dot\alpha}\Big) \nonumber\\
  \rightarrow & \ \ {\rm tr}\left(\frac{1}{8}[\phi,\bar\phi]^2-\frac{1}{2}[\phi,a_m][\bar\phi,a_m]
  -\frac{1}{2}[\phi,\varphi_m][\phi,\varphi_m]+\frac{1}{2}|\vec{H}|^2-i\vec{H}\cdot
  \vec{\mathcal{E}}\right.\nonumber\\
  & \hspace{2.3cm}\left.+ \ h_a(h_a)^\dag-i\mathcal{F}_a(h_a)^\dag-ih_a(\mathcal{F}_a)^\dag
  +\frac{1}{2}\{\phi,\bar\phi\}\bar{x}^{\dot\alpha}x_{\dot\alpha}
  -2\phi\bar{x}^{\dot\alpha}vx_{\dot\alpha}\right)\ ,
\end{align}
where the last step involves introducing auxiliary fields $\vec{H}$, $h_a$, $h_a^\dag$.
One should remember that in all supersymmetry transformations and the action, replacing
$A_\tau$, $\phi,\bar\phi$ appropriately by $D_\tau$ yields our mechanics action.


\subsection{The index}

We define and calculate a Witten index counting $\frac{1}{4}$-BPS states preserving
$\bar{Q}^{\dot{a}\dot\alpha}$. We first explain what kind of chemical potentials
we can introduce to weight these states.

Among the $SO(4)_1\times SO(4)_2\subset SO(4,1)\times SO(5)$ symmetry
unbroken by massive particles and the VEV $v_5$, the two $SU(2)_{1L}\times SU(2)_{2L}$
subgroups which come with undotted indices like $\alpha,a$ commute with all four supercharges $\bar{Q}^{\dot{a}\dot\alpha}$. So we can include the chemical potentials for their
Cartans. We denote by $\gamma_1$, $\gamma_2$ the chemical potentials for the
Cartans of $SU(2)_{1L} \times SU(2)_{2L}$, respectively. Also, since we have in mind
using a subset of the $4$ supercharges including $Q$ of (\ref{special-susy}) to localize
the path integral, there exists a \textit{calculable index} which also includes another
chemical potential for the diagonal subgroup of $SU(2)_{1R}\times SU(2)_{2R}$ under which
$Q$ is neutral. This diagonal $SU(2)_R$ rotates $\dot{a}$ and $\dot\alpha$
type indices simultaneously. We denote by $\gamma_R$ the chemical potential for its Cartan.
Note that the introduction of nonzero FI term $\sim\zeta^A$ in (\ref{ADHM-D}) breaks
$SU(2)_{1R}$ to $U(1)$. Even in this case, we can still introduce $\gamma_R$ for the
unbroken $U(1)$. There are two real supercharges $\bar{Q}^{\dot{1}\dot{2}}$,
$\bar{Q}^{\dot{2}\dot{1}}$ which commute with this $SU(2)_R$ Cartan. One combination
(\ref{special-susy}) is the scalar supercharge $Q$. We denote another combination by
$\tilde{Q}$. $Q,\tilde{Q}$ satisfy $\{Q,\tilde{Q}\}=0$ and $Q^2=\tilde{Q}^2=H-v^i\Pi_i$.
We consider the Witten index associated with $Q,\tilde{Q}$, given by
\begin{equation}\label{index}
  I_k(\mu^i,\gamma_1,\gamma_2,\gamma_R)={\rm Tr}_k\left[(-1)^Fe^{-\beta Q^2}e^{-\mu^i\Pi_i}e^{-i\gamma_1(2J_{1L})-i\gamma_2(2J_{2L})-i\gamma_R(2J_R)}\right]\ ,
\end{equation}
where
\begin{equation}\label{electric}
  \Pi_i=\left[-ix_{\dot\alpha}p^{\dot\alpha}+i\bar{p}_{\dot\alpha}\bar{x}^{\dot\alpha}
  \right]_{ii}+{\rm (fermionic)}
\end{equation}
for $i=1,2,\cdots,N$ are the Noether electric charges for the $U(1)^N\subset U(N)$
symmetry.\footnote{We hope the indices $i,j,\cdots$ for $U(N)$ are not confused with
similar indices used for $SO(5)$ spinors in the previous subsection. Similarly, we shall
later use $I,J,\cdots$ indices for $U(k)$ indices, which clashes with the $SO(5)$ vector
indices in the previous subsections. These indices from now on will not be used for $SO(5)$.} $p^{\dot\alpha}$ is the momentum
conjugate to $x_{\dot\alpha}$. $J_{1L}$, $J_{2L}$, $J_R=J_{1R}+J_{2R}$ are Cartans
for $SU(2)_{1L}$, $SU(2)_{2L}$ and the diagonal $SU(2)_R$, respectively. As all
the charges appearing in the trace (including $Q^2$) commute with $Q$, $\tilde{Q}$,
pairs of bosonic and fermionic states which are not annihilated by $Q$, $\tilde{Q}$
do not contribute to this index. Thus, $I_k$ does not depend on the parameter $\beta$ in
(\ref{index}). Also, the continuous parameters $v,\zeta^A$ appearing in the theory are
also expected not to affect the index. We can take these parameters to whatever
convenient values for calculation. It is also useful to consider
\begin{equation}
  I(q,\mu^i,\gamma_1,\gamma_2,\gamma_R)=\sum_{k=0}^\infty q^kI_k\ ,
\end{equation}
where $q$ is the fugacity of the instanton number charge $k$, and $I_0\equiv 1$.

We emphasize the condition we put on $\mu^i$. The separation of D4 branes is parametrized
by $v$. The order of these branes can be chosen to be $v^1>v^2>\cdots >v^N$. The positivity
of the electric charge contribution to the mass, $v^i\Pi_i>0$, puts a constraint on electric
charges $\Pi_i$. For instance, $\Pi_1=1,\Pi_2=-1$ corresponding to a single stretched string
is allowed since $v^1>v^2$, but $\Pi_1=-1,\Pi_2=1$ corresponding to a string with opposite
orientation is anti-BPS and does not appear in this BPS sector. The requirement that we only
admit these allowed charges in our index is implemented by setting $\mu^1>\mu^2>\cdots>\mu^N$,
which is the same order as $v^i$. Thus, only the topological information
of $v^i$ is encoded in the chemical potential $\mu^i$.

The above index admits a path integral representation over a periodic time direction
with radius $\beta$. Keeping $x_{\dot\alpha}$ and its conjugate momenta, the path integral
takes the following form,
\begin{eqnarray}
  I_k&=&\int_{\tau\sim\tau+\beta}\left[dx_{\dot\alpha} d\bar{x}^{\dot\alpha}
  dp^{\dot\alpha}d\bar{p}_{\dot\alpha}dA_\tau(\cdots)\right]
  e^{i\int d\tau(p^{\dot\alpha}D_\tau{x}_{\dot\alpha}+\bar{p}_{\dot\alpha}
  D_\tau{\bar{x}}^{\dot\alpha}+\cdots)}e^{-\int d\tau(H-v^i\Pi_i)}\nonumber\\
  &&\hspace{2cm}\times e^{-\mu^i\Pi_i-i\gamma_1(2 J_{1L})-i\gamma_2(2J_{2L})-i\gamma_R(2J_R)}
\end{eqnarray}
where $(\cdots)$ denotes appearances of other phase space variables in the theory.
One can integrate out the momentum variables to obtain a configuration space path integral.
For simplicity, we first illustrate this for the variables $x_{\dot\alpha},p^{\dot\alpha}$
in detail, as this part is most nontrivial. The extension to the full momentum variable
integral will be obvious. Since $\Pi_i$ is conserved along time evolution, one may replace
$x_{\dot\alpha}p^{\dot\alpha}$, etc. in $\Pi_i$ of (\ref{electric}) by
$\frac{1}{\beta}\int d\tau x_{\dot\alpha}p^{\dot\alpha}$, etc. $H$ is simply quadratic
in momenta, and especially contains $p^{\dot\alpha}\bar{p}_{\dot\alpha}$ conjugate to the
$x$, $\bar{x}$ variables. These momenta can be integrated out, after which one obtains a
measure given by the Euclidean action. Insertion of $v^i\Pi_i$ and $-\mu^i\Pi_i$ results
in shifts of the on-shell values of $p^{\dot\alpha}$, $\bar{p}_{\dot\alpha}$ as
\begin{equation}\label{chemical-shift}
  \bar{p}_{\dot\alpha}=iD_\tau x_{\dot\alpha}-i\left(v-\frac{\mu}{\beta}\right)x_{\dot\alpha}
  \ ,\ \ p^{\dot\alpha}=
  iD_\tau\bar{x}^{\dot\alpha}+i\bar{x}^{\dot\alpha}\left(v-\frac{\mu}{\beta}\right)
\end{equation}
where $v$, $\mu$ are regarded as diagonal $N\times N$ matrices. Thus, the Euclidean action and supersymmetry are twisted by covariantizing time derivatives with external gauge field given
by chemical potentials. The shift proportional to $v$ above, coming from the insertion
$v^i\Pi_i$ in the exponent, actually yields the canonical momentum obtained from
(\ref{fund-action}).
This is compatible with our early observation that $H-v^i\Pi_i$ is the Hamiltonian in these
variables. Now generalizing the above by including all other variables and chemical
potentials, the derivative is shifted as
\begin{equation}\label{twist}
  D_\tau\rightarrow D_\tau-\frac{\mu^i}{\beta}\Pi_i-i\frac{\gamma_1}{\beta}(2J_{1L})
  -i\frac{\gamma_2}{\beta}(2J_{2L})-i\frac{\gamma_R}{\beta}(2J_R)
\end{equation}
where $\Pi_i$, $J_{1L}$, $J_{2L}$, $J_R$ denote the charges of the variable on which
$D_\tau$  acts. For instance, the $i$'th element of $x_{\dot\pm}$ in $U(N)$ has
$\Pi_i=-1$ (others being zero), $J_{1L}\!=\!J_{2L}\!=\!0$ and $J_R=\pm\frac{1}{2}$.
Due to the appearance of the twist by $\mu^i$, $\gamma\equiv(\gamma_1,\gamma_2,\gamma_R)$,
we now have a deformed Lagrangian $L_{\mu,\gamma}$ in the path integral measure which
is invariant under the deformed supercharge $Q_{\mu,\gamma}$, covariantizing all time
derivatives as (\ref{twist}).

Now we consider the continuous parameters in the theory.
Without losing generality, we first take the FI parameter to be aligned along $A=3$,
and write $\zeta=\zeta^3$. We would like to compute the path integral after taking
$\beta\rightarrow 0^+$, $\zeta\rightarrow\infty$ in appropriate rate, to be specified
below during the calculation. One could also have taken $v_i\rightarrow\infty$, but the
last limit is not essential. The limit will localize the path integral to Gaussian
fluctuations around supersymmetric saddle points.

The saddle point configurations which preserve $Q$ can be classified by the $N$-colored
Young diagrams. Although this is well-known \cite{Nekrasov:2002qd,Nekrasov:2003rj},
we review it in our context in appendix A. Saddle points are first classified by how one
can distribute identical $k$ instantons to $N$ D4-branes. They are labeled by partitions
of $k$ into $N$ non-negative integers $k_i$ ($i=1,2,\cdots,N$) satisfying
\begin{equation}
  k_1+k_2+\cdots+k_N=k\ .
\end{equation}
Then, for the set of $k_i$ instantons on $i$'th D4-brane, possible saddle point
solutions in this part are labeled by Young diagrams $Y_{i}(k_i)$ with $k_i$ boxes.
The whole saddle point solutions are labeled by the collection of $N$ Young diagrams,
\begin{equation}
  \left(Y_1(k_1),Y_2(k_2),\cdots,Y_N(k_N)\right)\ \ ,\ \ \ \sum_{i=1}^Nk_i=k\ ,
\end{equation}
which is called $N$-colored Young diagram. The general form of the solution
as well as concrete examples for $k=1,2,3$ are explained in appendix A.

We start by studying the single instanton sector in some detail. There are $N$ saddle points,
\begin{equation}
  a_m=0\ ,\ \ x_{+}=\sqrt{\zeta}e^{i\theta}\ {\bf e}^i\ ,\ \ x_-=0\ ,\ \
  \phi=\frac{\mu^i-i\gamma_R}{\beta}\ ,\ \ \bar\phi=2v^i-\frac{\mu^i-i\gamma_R}{\beta}
\end{equation}
with $i=1,2,\cdots,N$, where $a_m,\phi,\bar\phi$ are just numbers, $x_{\dot\pm}$ are
complex $N\times 1$ matrices (row vectors), $\theta$ is a phase which corresponds
to the $U(1)$ gauge orbit, and ${\bf e}^i$ is an $N$ dimensional unit row vector with
nonzero $i$'th component. See appendix A.1 for its derivation. The path integral for
large $\zeta$ and small $\beta$ is calculated by Gaussian approximation. The result is
\begin{equation}\label{single-index}
  I_{k=1}=\left(\frac{\sin\frac{\gamma_1+\gamma_2}{2}\sin\frac{\gamma_1-\gamma_2}{2}}
  {\sin\frac{\gamma_1+\gamma_R}{2}\sin\frac{\gamma_1-\gamma_R}{2}}\right)\sum_{i=1}^N
  \prod_{j(\neq i)}\frac{\sinh\frac{\mu_{ij}+i\gamma_2-i\gamma_R}{2}
  \sinh\frac{\mu_{ij}-i\gamma_2-i\gamma_R}{2}}
  {\sinh\frac{\mu_{ij}}{2}\sinh\frac{\mu_{ij}-2i\gamma_R}{2}}\equiv
  I_{com}\sum_{i=1}^N\prod_{j(\neq i)}I(\mu_{ij})\ ,
\end{equation}
where $\mu_{ij}=\mu_i-\mu_j$,
\begin{equation}\label{com-index}
  I_{com}(\gamma_1,\gamma_2,\gamma_R)=
  \frac{\sin\frac{\gamma_1+\gamma_2}{2}\sin\frac{\gamma_1-\gamma_2}{2}}
  {\sin\frac{\gamma_1+\gamma_R}{2}\sin\frac{\gamma_1-\gamma_R}{2}}
\end{equation}
and
\begin{equation}\label{internal-index}
  I(\mu_{ij})\equiv I_{com}(\gamma_R\!+\!i\mu_{ij},\gamma_2,\gamma_R)
  =\frac{\sinh\frac{\mu_{ij}+i\gamma_2-i\gamma_R}{2}\sinh\frac{\mu_{ij}
  -i\gamma_2-i\gamma_R}{2}}{\sinh\frac{\mu_{ij}}{2}\sinh\frac{\mu_{ij}-2i\gamma_R}{2}}\ .
\end{equation}
The summation over $i=1,2,\cdots,N$ comes from contributions from $N$ different
saddle points. The factor $I_{com}$
comes from the center-of-mass supermultiplet, as we shall explain shortly.
This result is derived in appendix B.

One can interpret the factor $I_{com}$ as contributions to the index from the
center-of-mass supermultiplet for the half-BPS instantons. This multiplet
is a tensor super-multiplet which consists of two-form field $B_2$,
five scalar fields $\phi_I$ and their superpartners $\lambda$.
This multiplet can be generated by $8$ real supercharges $Q_{a\alpha}$, $Q^{\dot{a}}_\alpha$
of SYM broken by the half-BPS instantons, together with the center-of-mass position zero
modes.\footnote{Even with $\frac{1}{4}$-BPS states, this center-of-mass index appears in
the same form in the full index. However, as we shall explain in section 4, it is sometimes
more natural to change the viewpoint and explain the index in some sectors with the
center-of-mass index for $\frac{1}{2}$-BPS W-bosons.}
Their representations under various symmetries are given as follows:
\begin{center}
\begin{table}[h]
\vskip -0.3cm
$$
\begin{array}{c|cccc}
  \hline
  &SU(2)_{1L}&SU(2)_{1R}&SU(2)_{2L}&SU(2)_{2R}\\
  \hline
  B_{2} & {\bf 3} & {\bf 1} & {\bf 1} & {\bf 1} \\
  \hline\phi_I & {\bf 1} & {\bf 1} & {\bf 2} & {\bf 2} \\
  &{\bf 1} & {\bf 1} & {\bf 1} & {\bf 1} \\
  \hline\lambda &  {\bf 2} & {\bf 1} & {\bf 2} & {\bf 1} \\
  & {\bf 2} & {\bf 1} & {\bf 1} & {\bf 2}\\
  \hline
\end{array}
$$
\end{table}
\end{center}
%
\vskip -1.5cm
The index over these fields is generated by four fermionic
oscillators coming from $Q_{a\alpha}$, $Q^{\dot{a}}_\alpha$ and is given by
\begin{eqnarray}\label{fermion-zero}
  &&\left(e^{i\frac{\gamma_1\!+\!\gamma_2}{2}}-e^{-i\frac{\gamma_1\!+\!\gamma_2}{2}}\right)
  \left(e^{i\frac{\gamma_1\!-\!\gamma_2}{2}}-e^{-i\frac{\gamma_1\!-\!\gamma_2}{2}}\right)
  \left(e^{i\frac{\gamma_1\!+\!\gamma_R}{2}}-e^{-i\frac{\gamma_1\!+\!\gamma_R}{2}}\right)
  \left(e^{i\frac{\gamma_1\!-\!\gamma_R}{2}}-e^{-i\frac{\gamma_1\!-\!\gamma_R}{2}}\right)
  \nonumber\\
  &&=(2i)^4\sin\frac{\gamma_1+\gamma_2}{2}\sin\frac{\gamma_1-\gamma_2}{2}
  \sin\frac{\gamma_1+\gamma_R}{2}\sin\frac{\gamma_1-\gamma_R}{2}\ ,
\end{eqnarray}
where we have assumed a convention for the bosonic/fermionic nature of the Clifford
vacuum. This is proportional to the determinant contribution from the fermion zero modes
$\lambda_{a\alpha}$, $\lambda^{\dot{a}}_\alpha$ to the index that we obtained in
appendix B. We also have contributions from $4$ bosonic translational zero modes $a_m$.
These zero modes appear in the wave-function on $\mathbb{R}^4$. As we should weight all
these wave-functions with $U(1)^2\subset SO(4)$ chemical potentials, let us consider the
factorized bases in two orthogonal $\mathbb{R}^2$'s separately. In one $\mathbb{R}^2$,
say spanned by $x_1,x_2$, one can take the basis for the wave-function to have the form $f(x_1,x_2)e^{-(x_1^2+x_2^2)}$, where $f(x_1,x_2)$ is all possible polynomials
of $x_1,x_2$. As we want them to be $U(1)^2$ angular momentum eigenstates, we construct
the polynomial in terms of $x_{\mp\dot\pm}\equiv x_1\pm ix_2$, where the subscripts denote
the sign of charges for $SU(2)_L$ and $SU(2)_R$ Cartans. One weights a monomial $(x_{-\dot{+}})^m(x_{+\dot{-}})^n$ wave-function by giving $e^{\mp i(\gamma_1-\gamma_R)}$
to each factor of $x_{\mp\dot\pm}$. Summing over non-negative integers $m,n$, one obtains
\begin{equation}\label{bosonic-zero-1}
  \frac{1}{1-e^{i(\gamma_1-\gamma_R)}}\cdot\frac{1}{1-e^{-i(\gamma_1-\gamma_R)}}\ ,
\end{equation}
where each factor comes from monomials of $x_{+\dot{-}}$, $x_{-\dot{+}}$.
Of course, one obtains a divergent contribution as one expands the geometric series,
for a clear reason that there exist infinitely many states with given angular momentum.
If one wished, one could have given a factor $e^{-\epsilon\mp i(\gamma_1-\gamma_R)}$ before
summing over the states to get a regularized version of (\ref{bosonic-zero-1}), and then
send $\epsilon\rightarrow 0^+$.
The final expression (\ref{bosonic-zero-1}) is finite even after removing the regulator,
as is our index (\ref{com-index}). A similar partition function can be obtained for the
other $\mathbb{R}^2$ with zero modes $x_{\pm\dot\pm}\equiv x_3\mp ix_4$, having charges
$(\pm\frac{1}{2},\pm\frac{1}{2})$ under the two Cartans. The partition function for the
wavefunction coming from all four zero modes is given by
\begin{equation}\label{bosonic-zero}
  \frac{1}{(1-e^{i(\gamma_1-\gamma_R)})(1-e^{-i(\gamma_1-\gamma_R)})
  (1-e^{i(\gamma_1+\gamma_R)})(1-e^{-i(\gamma_1+\gamma_R)})}=
  \frac{1}{(2i)^4\sin^2\frac{\gamma_1+\gamma_R}{2}\sin^2\frac{\gamma_1-\gamma_R}{2}}
\end{equation}
Combining (\ref{fermion-zero}) and (\ref{bosonic-zero}), one obtains (\ref{com-index}),
proving our assertion that $I_{com}$ is indeed the index coming from the center-of-mass
super-multiplet.

For higher instanton numbers, one obtains the index after a similar but much more
tedious analysis of the path integral. Certainly one could have obtained it more
systematically by fully using techniques of \cite{Moore:1997dj}, as done in
\cite{Nekrasov:2002qd}. We did rather brutally at $k=2,3$ to make the structure of
Gaussian localization clear, heavily relying on mathematica for numerical calculations
of the determinants. To keep the notation simple, let us denote Young diagrams by
specifying the lengths of the rows. For instance, $(3,1)$ will mean $\Yboxdim7pt\yng(3,1)$\ .
Such a Young diagram with a subscript $(3,1)_i$ is for the instantons localized on the
$i$'th D4-brane.

At $k=2$, indices from various saddle points are
\begin{align}\label{two-index}
  I_{(1)_i(1)_j}& =I_{com}^2
  \frac{\sinh\frac{\mu_{ij}+i\gamma_1+i\gamma_2}{2}\sinh\frac{\mu_{ij}-i\gamma_1+i\gamma_2}{2}
  \sinh\frac{\mu_{ij}+i\gamma_1-i\gamma_2}{2}\sinh\frac{\mu_{ij}-i\gamma_1-i\gamma_2}{2}}
  {\sinh\frac{\mu_{ij}+i\gamma_1+i\gamma_R}{2}\sinh\frac{\mu_{ij}-i\gamma_1+i\gamma_R}{2}
  \sinh\frac{\mu_{ij}+i\gamma_1-i\gamma_R}{2}\sinh\frac{\mu_{ij}-i\gamma_1-i\gamma_R}{2}}
  \prod_{k(\neq i,j)}I(\mu_{ik})I(\mu_{jk})\nonumber\\
  \hspace*{-0.5cm} & =I_{com}(\gamma_1)^2I(\mu_{ij}+i\gamma_1+i\gamma_R)
  I(\mu_{ij}-i\gamma_1+i\gamma_R)
  \prod_{k(\neq i,j)}I(\mu_{ik})I(\mu_{jk})\nonumber\\
  I_{(2)_i},I_{(1,1)_i}  & =I_{com}\frac{\sin\frac{\pm 2\gamma_1+\gamma_2+\gamma_R}{2}
  \sin\frac{\pm 2\gamma_1-\gamma_2+\gamma_R}{2}}{\sin(\pm\gamma_1)\sin(\pm\gamma_1+\gamma_R)}\\
  & \hspace*{0.5cm}\times\prod_{k(\neq i)}\frac{\sinh\frac{\mu_{ki}-i\gamma_2+i\gamma_R}{2}
  \sinh\frac{\mu_{ki}+i\gamma_2+i\gamma_R}{2}
  \sinh\frac{\mu_{ki}\pm i\gamma_1-i\gamma_2+2i\gamma_R}{2}
  \sinh\frac{\mu_{ki}\pm i\gamma_1+i\gamma_2+2i\gamma_R}{2}}
  {\sinh\frac{\mu_{ki}}{2}\sinh\frac{\mu_{ki}+2i\gamma_R}{2}
  \sinh\frac{\mu_{ki}\pm i\gamma_1+i\gamma_R}{2}
  \sin\frac{\mu_{ki}\pm i\gamma_1+3i\gamma_R}{2}}\nonumber\\
  \hspace*{-0.5cm} \equiv&I_{com}(\gamma_1)I_{com}(2\gamma_1\pm\gamma_R)
  \prod_{k(\neq i)}I(\mu_{ik})I(\mu_{ik}\mp i\gamma_1-i\gamma_R)\ , \nonumber
\end{align}
where we have only shown the first arguments in the expressions $I_{com}$, $I$,
as the other two arguments $\gamma_2,\gamma_R$ always remain the same.
At $k=3$, one obtains
\begin{eqnarray}\label{three-index}
  \hspace*{-1cm}I_{(3)_i}&=&
  I_{com}(\gamma_1)I_{com}(2\gamma_1-\gamma_R)I_{com}(3\gamma_1-2\gamma_R)
  \prod_{j(\neq i)}I(\mu_{ij})I(\mu_{ij}+i\gamma_1\!-\!i\gamma_R)
  I(\mu_{ij}+2i\gamma_1\!-\!2i\gamma_R)\\
  \hspace*{-1cm}I_{(2,1)_i}&=&
  (I_{com}(\gamma_1))^2I_{com}(3\gamma_1)\prod_{j(\neq i)}I(\mu_{ij})
  I(\mu_{ij}+i\gamma_1-i\gamma_R)I(\mu_{ij}-i\gamma_1-i\gamma_R)\nonumber\\
  \hspace*{-1cm}I_{(1)_i(1)_j(1)_k} &=&
  (I_{com}(\gamma_1))^3[I(\mu_{ij}+i\gamma_1+i\gamma_R)I(\mu_{ij}-i\gamma_1+i\gamma_R)]
  [ij\rightarrow jk][ij \rightarrow ki]\prod_{l(\neq i,j,k)}
  I(\mu_{il})I(\mu_{jl})I(\mu_{kl})\nonumber\\
  \hspace*{-1cm}I_{(2)_i(1)_j} & =&I_{com}(\gamma_1)^2I_{com}(2\gamma_1\!-\!\gamma_R)
  I(\mu_{ij}+2i\gamma_1)I(\mu_{ij}-i\gamma_1+i\gamma_R)I(\mu_{ij})\!\!
  \prod_{k(\neq i,j)}\!\!I(\mu_{ik})I(\mu_{ik}+i\gamma_1\!-\!i\gamma_R)
  I(\mu_{jk})\nonumber
\end{eqnarray}
where $[ij\rightarrow jk]$ on the third line denotes replacing the $ij$ indices in the
factor in $[\ ]$ by $jk$, etc.

The general form of the index, including all cases above, is as follows.
For a saddle point given by the colored Young diagram
$\{Y_1,Y_2,\cdots,Y_N\}$, the index is given by
\begin{equation}\label{index-formula}
  I_{\{Y_1,Y_2,\cdots,Y_N\}}=\prod_{i,j=1}^N\prod_{s\in Y_i}
  \frac{\sinh\frac{E_{ij}-i(\gamma_2+\gamma_R)}{2}\sinh\frac{E_{ij}+i(\gamma_2-\gamma_R)}{2}}
  {\sinh\frac{E_{ij}}{2}\sinh\frac{E_{ij}-2i\gamma_R}{2}}\ ,
\end{equation}
where we should explain various quantities in the expression. $s$ denotes a box in
the Young diagram $Y_i$ in the above expression, and is labeled by a pair of positive integers
$(m,n)$ which count the position of the box from the upper-left corner of the Young diagram,
as we label matrix elements. For instance, the three boxes in the first row of $\Yboxdim7pt\yng(3,1)$ are labeled as $(1,1)$, $(1,2)$, $(1,3)$ from the left, and the box
in the second row is labeled as $(2,1)$. $E_{ij}$ is defined as
\begin{equation}
  E_{ij}=\mu_i-\mu_j+i(\gamma_1-\gamma_R)h_i(s)+i(\gamma_1+\gamma_R)(v_j(s)+1)\ ,
\end{equation}
where $h_i(s)$ and $v_j(s)$ denotes the distance from the box $s$ ($\in Y_i$) to
the right and bottom end of the $i$'th and $j$'th Young diagram, respectively. For
instance, if we take the pair of Young diagrams to be $Y_i=\Yboxdim7pt\yng(3,1)$ and $Y_j=\Yboxdim7pt\yng(1,1,1)$ , $s$ in the product $\prod_{s\in Y_i}$ of (\ref{index-formula})
can run over $(1,1)$, $(1,2)$, $(1,3)$, $(2,1)$. The values of $h$, $v$ are given by
$h_i(1,1)=2$, $h_i(1,2)=1$, $h_i(1,3)=0$, $h_i(2,1)=0$ and $v_j(1,1)=2$, $v_j(1,2)=-1$,
$v_j(1,3)=-1$, $v_i(2,1)=1$. See \cite{Bruzzo:2002xf} for more detailed explanations of
this formula. One can easily show that this formula reproduces all the expressions in
(\ref{single-index}), (\ref{two-index}), (\ref{three-index}) above.

From (\ref{index-formula}), one can see that the expression can be understood as the
instanton partition function of the 5d $\mathcal{N}\!=\!2^\ast$ theory compactified on a
circle \cite{Nekrasov:2003rj}. The last theory has $8$ real supercharges, which we call
$\mathcal{N}\!=\!2$ in 4 dimensional convenction. It has a massless vector multiplet and
a hypermultiplet with mass $m$. One considers this theory in the Coulomb phase with
VEV's $a_1,a_2,\cdots,a_N$ of the scalar in the vector multiplet, which break the $U(N)$
gauge symmetry to $U(1)^N$. To compute the instanton partition function, the system is put
in the Omega background with parameters $\epsilon_1,\epsilon_2$,
associated to the rotations on $12$ and $34$ planes, respectively. The two combinations
\begin{equation}
  \epsilon_L=\frac{\epsilon_1-\epsilon_2}{2}\ ,\ \
  \epsilon_R=\frac{\epsilon_1+\epsilon_2}{2}
\end{equation}
take values in the Cartan of $SU(2)_{1L}\times SU(2)_R$. We take all parameters
to be dimensionless by suitably multiplying the radius $R_5$ of the 5d circle. In
\cite{Nekrasov:2003rj,Bruzzo:2002xf,Hollowood:2003cv}, the partition functions of the
4d and 5d $\mathcal{N}\!=\!2^\ast$ theories were first presented for the self-dual Omega
background with $\epsilon_R=0$, i.e. when $\hbar\equiv\epsilon_1=-\epsilon_2$.
The generalization to the case with nonzero $\epsilon_R$, which will be the expression
to be compared with our index, has been discussed rather recently, and demands a careful
consideration on the mass parameter as argued in \cite{Okuda:2010ke}.

We first consider the denominator $\prod_{i,j}^N\prod_{s\in Y_i}\sinh\frac{E_{ij}}{2}
\sinh\frac{E_{ij}-2i\gamma_R}{2}$ of (\ref{index-formula}). This can be identified as a
contribution from the fields in the vector multiplet of $\mathcal{N}=2^\ast$ theory. In
the `4d limit' obtained by scaling the dimensionless parameters to be small, we erase
the $\sinh$'s and take the resulting polynomials as the denominator. Identifying
\begin{equation}\label{parameter-map}
  a_i=\frac{\mu_i}{2}\ ,\ \ -\epsilon_1=i\frac{\gamma_1-\gamma_R}{2}\ ,\ \
  \epsilon_2=i\frac{\gamma_1+\gamma_R}{2}\ ,
\end{equation}
where the parameters on the left hand sides are all made dimensionless by suitably
multiplying the radius of the circle, we recover the expressions for instanton partition
function for the 5d $\mathcal{N}\!=\!2$ super Yang-Mills theory. For instance, one
immediately recovers the finite product form (\ref{index-formula}) from eqns.(3.16) and
(3.17) of \cite{Bruzzo:2002xf} (after uplifting each factor into $\sinh$).
Now let us consider the numerator of (\ref{index-formula}). We can identify it as
the determinant from hypermultiplet of $\mathcal{N}\!=\!2^\ast$ theory. This numerator takes
a form similar to the denominator,
with shifts on the arguments of $\sinh$ by subtracting $i\frac{\gamma_2+\gamma_R}{2}$ to the
first $\sinh$ in (\ref{index-formula}), and adding it to the second $\sinh$. Comparing
with eqn.(3.26) of \cite{Bruzzo:2002xf}, with a recent modification in the hypermultiplet mass
contribution \cite{Okuda:2010ke}, we find that the $\sinh$ arguments in the numerator of the $\mathcal{N}\!=\!2^\ast$ partition function are shifted by $m+\epsilon_R$.
Since we already mapped $\epsilon_R=\frac{i\gamma_R}{2}$ from (\ref{parameter-map}),
our numerator is exactly the hypermultiplet contribution of the $\mathcal{N}\!=\!2^\ast$
theory if we identify
\begin{equation}\label{mass}
  m= i\frac{\gamma_2}{2}\ .
\end{equation}

\section{Uniqueness of $U(1)$ Kaluza-Klein modes}

In this section, we study the D0-brane index on a single D4-brane, or the $U(1)$
instanton index. Although $U(1)$ instantons are singular in ordinary field
theory, they play important roles in string theory. Also, under the non-commutative
deformation that we introduced, $U(1)$
instantons become regular solitons of classical field theory \cite{Nekrasov:1998ss}.
As the Kaluza-Klein states of M5-branes on a circle,
these instanton bound states are expected to be unique in each topological sector
given by the instanton number $k$. In other words, we expect only one
supermultiplet to exist in the single particle Hilbert space for each $k$.

The bound states of non-commutative $U(1)$ instantons have been studied
in \cite{Lee:2000hp} up to $k=2$, by studying the instanton
moduli space dynamics and constructing the wave-function for threshold bounds.
Such an approach would be very difficult for general multi-instantons, as one
should understand the metric and the normalizable harmonic forms on the
moduli space. The index in this paper is much easier to study. In particular,
the relation between our index and the instanton part of Nekrasov's
$\mathcal{N}\!=\!2^\ast$ partition function allows us to study these bound
states in great detail, relying on recent developments in topological string theory.

Before considering the general index, let us illustrate the structure of this index
for the cases with low instanton numbers, $k=1,2,3$. At single instanton sector,
one naturally obtains the index for one supermultiplet $I_{k=1}=I_{com}$ from
(\ref{single-index}), implying unique bound state with one unit of KK monentum.
This index at $k\!=\!1$ was also obtained in \cite{Dorey:2000zq}.

At $k\geq 2$, one has to remember that our index includes multi-particle contribution.
At $k=2$, collecting the contributions from the saddle points $\Yboxdim7pt\yng(2)$ and
$\Yboxdim7pt\yng(1,1)$ of (\ref{two-index}), one obtains
\begin{equation}\label{U(1)-2-instanton}
  I_{k=2}=\frac{\sin\frac{\gamma_1+\gamma_2}{2}\sin\frac{\gamma_1-\gamma_2}{2}}
  {\sin\frac{\gamma_1+\gamma_R}{2}\sin\frac{\gamma_1-\gamma_R}{2}}\left(
  \frac{\sin\frac{2\gamma_1+\gamma_2+\gamma_R}{2}
  \sin\frac{2\gamma_1-\gamma_2+\gamma_R}{2}}
  {\sin(\gamma_1)\sin(\gamma_1+\gamma_R)}+
  \frac{\sin\frac{2\gamma_1+\gamma_2-\gamma_R}{2}
  \sin\frac{2\gamma_1-\gamma_2-\gamma_R}{2}}
  {\sin(\gamma_1)\sin(\gamma_1-\gamma_R)}\right)\ .
\end{equation}
After some algebra, one can check that this expression can be written as
\begin{equation}\label{neutral-k=2}
  I_{k=2}=\frac{I_{com}(\gamma_1,\gamma_2,\gamma_R)^2+I_{com}(2\gamma_1,2\gamma_2,2\gamma_R)}{2}
  +I_{com}(\gamma_1,\gamma_2,\gamma_R)\ .
\end{equation}
The first term comes from two non-interacting identical particles,
each of them having instanton charge $1$. This is an expected contribution once we
have identified a single particle state at $k=1$ in the previous paragraph. The last
term of (\ref{neutral-k=2}) implies the existence of another single particle supermultiplet
at $k=2$, which shows the uniqueness of threshold bound state at $k=2$. This fact was
also shown in \cite{Lee:2000hp} by an explicit construction of the wave-function for
the threshold bound state on the Eguchi-Hanson moduli space.

At $k=3$, one obtains the following index from (\ref{three-index}):
\begin{eqnarray}
  \Yboxdim5pt I_{\yng(3)}&=&I_{com}
  \left(\frac{\sin\frac{2\gamma_1-\gamma_2-\gamma_R}{2}\sin\frac{2\gamma_1+\gamma_2-\gamma_R}{2}}
  {\sin\gamma_1\sin(\gamma_1-\gamma_R)}\right)
  \left(\frac{\sin\frac{3\gamma_1+\gamma_2-2\gamma_R}{2}\sin\frac{3\gamma_1-\gamma_2-2\gamma_R}{2}}
  {\sin(\frac{3\gamma_1-\gamma_R}{2})\sin(\frac{3\gamma_1-3\gamma_R}{2})}\right)\nonumber\\
  \Yboxdim5pt I_{\yng(2,1)}&=&\left(I_{com}\right)^2
  \left(\frac{\sin\frac{3\gamma_1+\gamma_2}{2}\sin\frac{3\gamma_1-\gamma_2}{2}}
  {\sin(\frac{3\gamma_1+\gamma_R}{2})\sin(\frac{3\gamma_1-\gamma_R}{2})}\right)\\
  \Yboxdim5pt I_{\yng(1,1,1)}&=&I_{com}
  \left(\frac{\sin\frac{2\gamma_1+\gamma_2+\gamma_R}{2}\sin\frac{2\gamma_1-\gamma_2+\gamma_R}{2}}
  {\sin\gamma_1\sin(\gamma_1+\gamma_R)}\right)
  \left(\frac{\sin\frac{3\gamma_1-\gamma_2+2\gamma_R}{2}\sin\frac{3\gamma_1+\gamma_2+2\gamma_R}{2}}
  {\sin(\frac{3\gamma_1+\gamma_R}{2})\sin(\frac{3\gamma_1+3\gamma_R}{2})}\right)\ .\nonumber
\end{eqnarray}
From these expressions, one can show after some algebra that
\begin{equation}
  \Yboxdim5pt I_{k=3}=I_{\yng(3)}+I_{\yng(2,1)}+I_{\yng(1,1,1)}=
  \frac{I_{com}(\gamma)^3+3I_{com}(\gamma)I_{com}(2\gamma)+2I_{com}(3\gamma)}{6}
  +I_{com}(\gamma)^2+I_{com}(\gamma)\ ,
\end{equation}
where $\gamma=(\gamma_1,\gamma_2,\gamma_R)$ is used as a collective symbol for the
three chemical potentials. The first term on the right hand side comes from three
identical particles, each particle with instanton number $1$. The second term
proportional to $I_{com}(\gamma)^2$ comes from $2$ particle states, one with instanton number
$1$ and another with $2$ (which we identified in the previous paragraph).
The last term confirms that there is a unique supermultiplet for the threshold bound state
of three instantons.

One can work more systematically by using the relation of our index
to the 5d $\mathcal{N}=2^\ast$ partition function and some recent development from
the topological string calculations. Namely, the $U(1)$ $\mathcal{N}\!=\!2^\ast$ theory in
5 dimension can be engineered by putting M-theory on a suitable Calabi-You 3-fold. The
instanton partition function as a function of $\epsilon_1,\epsilon_2,m$ was computed from
topological string theory, using the refined topological vertex technique
\cite{Iqbal:2008ra,Awata:2008ed,Poghossian:2008ge}. A nice feature of their result is that the
summation over the instanton saddle points was explicitly done. Following the notation of
\cite{Iqbal:2008ra}, one finds that the instanton part of the partition function
$Z_{\rm inst}$ (i.e. without the perturbative part) is given by
\begin{eqnarray}\label{top-vertex}
  Z&=&Z_{\rm pert}Z_{\rm inst}\nonumber\\
  Z&=&\prod_{k=1}^\infty\left[(1-Q_\bullet^k)^{-1}\prod_{i,j=1}^\infty
  \frac{(1-Q_\bullet^kQ_m^{-1}q^{i-\frac{1}{2}}t^{j-\frac{1}{2}})
  (1-Q_\bullet^kQ^{-1}q^{i-\frac{1}{2}}t^{j-\frac{1}{2}})}{(1-Q_\bullet^kq^{i-1}t^j)
  (1-Q_\bullet^kq^it^{j-1})}\right]\nonumber\\
  Z_{\rm pert}&=&\prod_{i,j=1}^\infty(1-Q_mt^{i-\frac{1}{2}}q^{j-\frac{1}{2}})\ ,
\end{eqnarray}
from eqn.(3.1) and the expressions below eqn.(3.5) in \cite{Iqbal:2008ra}.
Here, the three parameters $Q_\bullet,Q,Q_m$ are related by $Q_\bullet=QQ_m$, and $Q=e^{-T}$,
$Q_m=e^{-T_m}$ are related to the two K\"{a}hler parameters $T,T_m$ of the CY$_3$ which
yields the $\mathcal{N}\!=\!2^\ast$ theory. It will turn out that $T_m$ and $Q_\bullet$ are
the mass parameter and the instanton number chemical potential (or the coupling constant of
the gauge theory), respectively. $t,q$ are their Omega background parameters.
Their parameters are related to ours $q$ (fugacity for instanton number), $m$,
$\epsilon_1$, $\epsilon_2$ as
\begin{equation}
  [Q_\bullet]_{\rm theirs}=q\ ,\ \ [T_m]_{\rm theirs}=2m=i\gamma_2\ ,\ \
  [t]_{\rm theirs}=e^{2\epsilon_1}=e^{i(\gamma_R-\gamma_1)}\ ,\ \
  [q]_{\rm theirs}=e^{-2\epsilon_2}=e^{-i(\gamma_1+\gamma_R)}\ .
\end{equation}
As $Z_{\rm inst}$ is the multi-particle index, we should consider
the single particle index $z_{\rm sp}$ given by
\begin{equation}
  Z_{\rm inst}(Q_\bullet,Q_m,t,q)=
  \exp\left[\sum_{n=1}^\infty\frac{1}{n}z_{\rm sp}(Q_\bullet^n,Q_m^n,t^n,q^n)\right]
\end{equation}
to study how many bound states exist. As all the expressions in (\ref{top-vertex}) are given
by infinite products, it is easy to extract the closed form of $z_{\rm sp}$. One obtains
\begin{eqnarray}
  z_{\rm sp}&=&\sum_{k=1}^\infty Q_\bullet^k\left[1+\sum_{i,j=1}^\infty\left(
  q^{i-1}t^j+q^it^{j-1}-(Q_m^{-1}+Q^{-1})q^{i-\frac{1}{2}}t^{j-\frac{1}{2}}\right)\right]
  +\sum_{i,j=1}^\infty Q_mt^{i-\frac{1}{2}}q^{j-\frac{1}{2}}\nonumber\\
  &=&\frac{Q_\bullet}{1-Q_\bullet}\frac{1+qt-(qt)^{\frac{1}{2}}(Q_m^{-1}+Q_mQ_\bullet^{-1})}
  {(1-q)(1-t)}+\frac{Q_m(qt)^{\frac{1}{2}}}{(1-q)(1-t)}\\
  &=&\frac{Q_\bullet}{1-Q_\bullet}\frac{(1-(qt)^{\frac{1}{2}}Q_m)
  (1-(qt)^{\frac{1}{2}}Q_m^{-1})}{(1-q)(1-t)}\ \ \longrightarrow\ \
  \frac{q}{1-q}I_{com}(\gamma_1,\gamma_2,\gamma_R)\ ,\nonumber
\end{eqnarray}
where the first and second terms on the first line come from $Z$ and $Z_{\rm pert}^{-1}$,
respectively. We used the relation $Q_\bullet=QQ_m$ on the second line, and in the last
expression we changed the parameters to our $q,\gamma_1,\gamma_2,\gamma_R$.
Expanding the last expression in $q$, one finds that the coefficient of $q^k$ is
$I_{com}(\gamma_1,\gamma_2,\gamma_R)$ for all $k=1,2,3,\cdots$, proving that there indeed
exists unique bound state at each instanton number $k$.

One might wonder if one can do similar studies for the $U(N)$ instantons. Firstly,
it is unclear how many bound states we should expect in the symmetric phase based on
kinematics only. One normalizable harmonic form was constructed for $U(N)$ single instantons
\cite{Kim:2001kp}, which was interpreted as the first KK mode of the decoupled center-of-mass
tensor multiplet. In the Coulomb phase with unbroken $U(1)^N$ symmetry, our index gives $N$
neutral instanton bound states at all $k$. This is because there are $N$
non-interacting 6d tensor multiplets at low energy if we separate $N$ M5-branes. This result
at $k=1$ was also computed in \cite{Dorey:2000zq}.

\section{Degeneracy of self-dual strings from instantons}

In this section, we study a class of charged instanton bound states in the Coulomb
phase. As charged instantons are all mutually BPS, the long-range interactions vanish.
This implies that $I(q,\mu_I,\gamma_1,\gamma_2,\gamma_R)$ is given in terms of
the single particle index $z_{\rm sp}(q,\mu_I,\gamma)$ as
\begin{equation}
  I(q,\mu_I,\gamma)=\exp\left[\sum_{n=1}^\infty\frac{1}{n}
  z_{\rm sp}(q^n,n\mu_I,n\gamma)\right]\ ,
\end{equation}
As $z_{\rm sp}$ is an index for the single particle states, it will contain a factor
which comes from one set of position zero modes, taking the form of
\begin{equation}
  I_{com}=\frac{({\rm fermion\ zero\ modes})}
  {\sin\frac{\gamma_1+\gamma_R}{2}\sin\frac{\gamma_1-\gamma_R}{2}}\ .
\end{equation}
Among other things, the appearance of a factor $I_{com}$ in $z_{\rm sp}$ dictates
the small $\gamma_1$ and $\gamma_R$ behavior of the function $z_{\rm sp}$,
namely it diverges as $\frac{1}{(\gamma_1+\gamma_R)(\gamma_1-\gamma_R)}\sim
\frac{1}{\epsilon_1\epsilon_2}$. This pattern of divergence is indeed well-known.
In the context of Nekrasov's partition function, it is known that $\epsilon_1,\epsilon_2\rightarrow 0$
limit of the partition function takes the form \cite{Nekrasov:2002qd}
\begin{equation}
  \lim_{\epsilon_1,\epsilon_2\rightarrow 0}Z_{\rm inst}(q,a_I,\epsilon_1,\epsilon_2,m)=
  \exp\left[\frac{1}{\epsilon_1\epsilon_2}\mathcal{F}_{\rm inst}(q,a_I,m)\right]\ ,
\end{equation}
where $\mathcal{F}_{\rm inst}$ is the instanton part of the prepotential.
The general form of $z_{\rm sp}$ is quite complicated. For $SU(2)$ pure $\mathcal{N}\!=\!2$
Yang-Mills theory in 5 dimensions, \cite{Awata:2008ed} used their `trace identities'
in topological vertex formulation to rewrite the instanton partition function into
an infinite product form which clearly shows $z_{\rm sp}$. As far as
we can see, their technique is not applicable even to the $SU(2)$ $\mathcal{N}\!=\!2^\ast$
theory, due to the different topology of the toric diagrams for the two theories. We therefore
rely on numerical series expansions in powers of $q$, and also apply various tricks to simplify
the expressions. Some of these series expansions can be faithfully replaced by exact
expressions of $q$ and compared to the physics of self-dual strings.

Firstly, we are not interested in the single particle index $I_{com}$ of freely
moving particles, carrying no information on dynamics. In many cases,
we take $\epsilon_1=-\epsilon_2\equiv\hbar$ and the $\hbar\rightarrow 0$ limit.
Factoring out the $I_{com}\approx-\frac{\sinh^2m}{\hbar^2}$ factor from $z_{\rm sp}$,
the limit $\hbar\rightarrow 0$ will erase the spin information of the states in the
remaining pieces, leading to simplifications of many expressions below. On the other hand,
nonzero chemical potential $\gamma_2$ prevents complete cancelation between bosonic/fermionic
contributions. Taking $\gamma_2\!=\!0$ (at the point $\gamma_R=0$, or equivalently
$\epsilon_1=-\epsilon_2$) makes both $I_{com}$ and the remainder trivial. However, one finds
from the general expression of $I$ that taking $2m=i\gamma_2=i\pi$ changes all the $-1$ signs
in the fermionic degeneracies to $+1$ via $e^{i\gamma_2}=-1$, providing an expression
which looks like a partition function. To see why this happens, recall from
the instanton mechanics that all bosonic and fermionic degrees carry integral and
half-integral $J_{2L}$, respectively, apart from $\varphi_m$ which carries $\pm\frac{1}{2}$.
Had the last mode contributed nontrivially, one could not have the property
$(-1)^Fe^{2i\gamma_2J_{2L}}=+1$ which makes our index look like a partition function. However,
one can easily see that the determinant over $\varphi_m$ modes in all saddle points should
cancel with other fermionic determinants. This is because $\varphi_m$ expectation value is
zero at all saddle points, making the quadratic fluctuation term of this field to behave
like $\bar{x}x(\delta\varphi)^2\sim\zeta(\delta\varphi)^2$. Therefore, the determinant
for $\varphi_m$ would carry a dependence on the FI parameter $\zeta$, which
should cancel out in the final expression of the index.

From now on, in most cases we shall study the single particle index after taking
$\gamma_1=0$ and $\gamma_2=\pi$, factoring out the divergent $I_{com}$ and concentrating
on the `internal' contributions.

\subsection{$SU(2)$ self-dual strings}

For the $U(2)$ theory, one obtains (after eliminating $I_{com}$)
\begin{eqnarray}\label{SU(2)-single}
  \hspace*{-0.7cm}\left.z_{\rm sp}\frac{}{}\!\right|_{\gamma_1\!=\!0,\gamma_2\!=\!\pi}\!\!&\!=\!&
  2q\frac{(1+x)^2}{(1-x)^2}
  +2q^2\frac{(1+x)^2\left(1+12x+14x^2+12x^3+x^4\right)}{(1-x)^6}\nonumber\\
  \hspace*{-0.7cm}&&\!+2q^3\frac{1+72x+828x^2+4138x^3+12758x^4+27056x^5+41709x^6+48060x^7
  +\cdots+x^{14}}{(1-x)^8(1-x^3)^2}\nonumber\\
  \hspace*{-0.7cm}&&+2q^4\frac{1\!+\!262x\!+\!6755x^2\!+\!57708x^3\!+\!254801x^4\!+\!694298x^5
  \!+\!1242699x^6\!+\!1503976x^7\!+\!\cdots\!+\!x^{14}}{(1-x)^{14}}\nonumber\\
  \hspace*{-0.7cm}&&\!+\!\frac{2q^5}{(1\!-\!x)^{16}(1\!-\!x^5)^{2}}\!\left(1\!+\!840x\!+\!
  49064x^2\!+\!902680x^3\!+\!8303100x^4\!+\!47355570x^5\!+\!187537864x^6\right.\nonumber\\
  \hspace*{-0.7cm}&&\hspace{2cm}\left.\!\!+\!553053672x^7\!+\!1278050838x^8\!+\!
  2411818864x^9\!+\!3843375177x^{10}\!+\!5298097024x^{11}\right.\nonumber\\
  \hspace*{-0.7cm}&&\hspace{2cm}\left.\!\!+\!6403142196x^{12}\!+\!6818459180x^{13}\!+\!
  \cdots\!+\!x^{26}\right)+\cdots\nonumber
\end{eqnarray}
where $\mu\equiv\mu_1\!-\!\mu_2>0$ and $x\equiv e^{-\mu}<1$. The omitted terms in the
numerators can be restored from the fact that the coefficients are symmetric around the
middle terms, as manifestly shown up to $\mathcal{O}(q^2)$ on the first line.

At each order in $q$, the terms at $\mathcal{O}(x^0)$ have coefficient $2$, as explained
in the previous section. At $\mathcal{O}(x^1)$, one will obtain the degeneracy for an
M2 self-dual string stretched between two M5-branes.
Collecting the coefficient of
$x^1$, one obtains
\begin{equation}
  z_{\rm sp}\rightarrow 8q+40q^2+160q^3+552q^4+1712q^5+4896q^6+13120q^7+33320
  q^8+80872q^9+188784q^{10}+\cdots\ .
\end{equation}
In our analysis from instanton quantum mechanics, we can only probe BPS states
with nonzero instanton numbers. However, since we know that there should be a single
W-boson supermultiplet for $SU(2)$ at $q^0$, we add it by hand and obtain
\begin{equation}\label{complete-W-boson}
  z_{\rm sp}=1+8q+40q^2+160q^3+552q^4+1712q^5+4896q^6+13120q^7+33320
  q^8+80872q^9+188784q^{10}+\cdots\ .
\end{equation}
One finds that this series can be written as
\begin{equation}\label{SU(2)-string}
  I_{com}\cdot z_{\rm sp}=I_{com}\prod_{n=1}^\infty\frac{(1+q^n)^4}{(1-q^n)^4}\ .
\end{equation}
Restoring all chemical potentials, we find a more refined expression:
\begin{equation}\label{SU(2)-refined}
  \left(\frac{\sin\frac{\gamma_R+\gamma_2}{2}\sin\frac{\gamma_R-\gamma_2}{2}}
  {\sin\frac{\gamma_1+\gamma_R}{2}\sin\frac{\gamma_1-\gamma_R}{2}}\right)
  \prod_{n=1}^\infty\frac{(1-q^ne^{i(\gamma_2+\gamma_R)})(1-q^ne^{i(\gamma_2-\gamma_R)})
  (1-q^ne^{i(-\gamma_2+\gamma_R)})(1-q^ne^{i(-\gamma_2-\gamma_R)})}
  {(1-q^ne^{i(\gamma_1+\gamma_R)})(1-q^ne^{i(\gamma_1-\gamma_R)})
  (1-q^ne^{i(-\gamma_1+\gamma_R)})(1-q^ne^{i(-\gamma_1-\gamma_R)})}\ .
\end{equation}
One can understand this partition function from S-dual monopole strings after
compactifying the 5d theory on an extra circle
\cite{Douglas:2010iu,Lambert:2010iw,Tachikawa:2011ch}. The S-dual $SU(2)$ monopole string
is described by a free 1+1 dimensional QFT, as its moduli space $\mathbb{R}^3\times S^1$
is flat, with four bosonic and fermionic degrees of freedom. With zero momentum and winding
along the above circle in the target space, the compact boson can
be regarded as being non-compact so that we effectively get $\mathbb{R}^4$ as the moduli space.
This also coincides with the transverse space of a self-dual string along the M5-branes.
$SO(4)_1$ symmetry emerges in this case. The partition function for the four bosons and
four fermions yields (\ref{SU(2)-string}). To understand the spin contents in
(\ref{SU(2)-refined}), it suffices to understand the new center-of-mass factor
\begin{equation}\label{monopole-com}
  I_{com}=\frac{\sin\frac{\gamma_R+\gamma_2}{2}\sin\frac{\gamma_R-\gamma_2}{2}}
  {\sin\frac{\gamma_1+\gamma_R}{2}\sin\frac{\gamma_1-\gamma_R}{2}}
\end{equation}
as the remaining infinite product is obtained by giving nonzero momenta to these
zero modes. The bosonic zero modes simply yield
$\sin^{-2} \frac{\gamma_1+\gamma_R}{2}\sin^{-2}\frac{\gamma_1-\gamma_R}{2}$ as before.
To understand the fermion zero modes, we consider the broken supersymmetry of a
magnetic monopole, or more precisely its S-dual W-boson, as that should be what we
add as `1' in (\ref{complete-W-boson}). Using 10 dimensional spinors for the $16$
supersymmetry, the $\frac{1}{2}$-BPS condition for the W-boson stretched in $\varphi_5$
direction is given by a $\Gamma^{05}$ projector, where $5$ denotes the internal direction
along the scalar. In the 5d symplectic-Majorana spinor notation that we have been using,
$\Gamma^{05}$ acting on a 10d chiral spinor turns out to be
$\gamma^0\otimes\gamma^5$. So the W-boson preserves left-left or right-right spinors
$Q^a_\alpha$, $\bar{Q}^{\dot{a}}_{\dot\alpha}$ in the two $SO(4)_1\times SO(4)_2$ factors.
The broken supercharges $\bar{Q}^a_{\dot\alpha}$, $Q^{\dot{a}}_{\alpha}$ generate the
following factors of the index in $I_{com}$:
\begin{equation}
  \sin\frac{\gamma_R+\gamma_2}{2}\sin\frac{\gamma_R-\gamma_2}{2}
  \sin\frac{\gamma_1+\gamma_R}{2}\sin\frac{\gamma_1-\gamma_R}{2}\ .
\end{equation}
The first two $\sin$'s come from $\bar{Q}^a_{\dot\alpha}$, and the last two factors come from
$Q^{\dot{a}}_\alpha$. Combining this with the above bosonic contribution, one obtains
(\ref{monopole-com}), which further explains (\ref{SU(2)-refined}). This is also another concrete
example in which instantons provide the required KK tower of states along the M-theory circle.

At $x^n$ order, one obtains the degeneracy for $n$ identical $SU(2)$ strings
with nonzero momenta. From the above formula, one finds
\begin{eqnarray}\label{identical-monopole}
  x^2&:&0+16q+288q^2+2880q^3+21056q^4+125280q^5+\cdots=q\frac{d}{dq}\left[
  \prod_{n=1}^\infty\frac{(1+q^n)^8}{(1-q^n)^8}\right]\nonumber\\
  x^3&:&0+24q+1272q^2+26952q^3+360696q^4+3605520q^5+\cdots\nonumber\\
  x^4&:&0+32q+4160q^2+169600q^3+3842176q^4+60216000q^5+\cdots\nonumber\\
  x^5&:&0+40q+11080q^2+809760q^3+29471560q^4+692554440q^5+\cdots
\end{eqnarray}
and so on. We have added $0$'s at $\mathcal{O}(q^0)$ orders as we know
that $SU(2)$ magnetic monopole strings with many units of charges do not form any threshold
bound states. This is well-known from the dyon spectrum of 4d $\mathcal{N}\!=\!4$ Yang-Mills
theory \cite{Sen:1994yi}. Curiously, our formula predicts that there are threshold bound states
once we turn on nonzero momenta on the worldsheet.
It would be interesting to understand this phenomenon. One may start from
the 1+1 dimensional sigma model with $(4,4)$ supersymmetry, with the target space being
the moduli space of $SU(2)$ multi-monopoles. For instance, the relative moduli space
for two monopoles is the Atiyah-Hitchin space. One can calculate the index of this
2d theory. One would expect a contribution from
2-particle states. Subtracting this 2-particle contribution, it should be possible to
see if the above $\mathcal{O}(x^2)$ expression of (\ref{identical-monopole}) is obtained.
We leave it as a future work.

\subsection{$SU(N)$ self-dual strings}

One can also consider the charged bound states for larger gauge group, $U(N)$.
There appear many kinds of bound states, among which we only study a special kind
of states for simplicity.

$N^2$ microscopic degrees of freedom in Yang-Mills theory leave their remnant
in the Coulomb phase as $\frac{N(N-1)}{2}\sim N^2$ massive W-bosons (plus super-partners).
These degrees are all visible
perturbatively. In 4d $\mathcal{N}\!=\!4$ theory, which is S-duality invariant, it will
also be helpful to remind ourselves how the corresponding degrees for monopoles emerge.
From the classical magnetic monopole solutions, only $N-1$ `fundamental monopoles' are
visible, whose charges are labeled by chemical potentials $e^{-(\mu_1-\mu_2)}$,
$e^{-(\mu_2-\mu_3)}$, $\cdots, e^{-(\mu_{N\!-\!1}-\mu_N)}$ in the dual gauge group $U(N)$.
These may be viewed as D1-branes stretched between adjacent D3-branes in the Coulomb phase.
The way $\frac{N(N-1)}{2}$ monopole states emerge is by having unique threshold bound
states of the distinct fundamental monopoles, admitting states weighted by $e^{-(\mu_i-\mu_j)}$
with general $\mu_i>\mu_j$. This was shown for $SU(3)$ \cite{Lee:1996if}, and the general form
of the conjectured bound state wave-function for $SU(N)$ was studied in \cite{Yi:1996bb}.

It may also be interesting to consider self-dual strings compactified on a circle,
or the related magnetic monopole strings in 5d Yang-Mills theory on a circle which are
S-dual to our F1-D0 system. With zero momentum, one again expects there
to be $\frac{N(N-1)}{2}$ states from low dimensional physics. It would be interesting to
see what happens to the degeneracy of these objects with nonzero momentum, and most
interestingly with large enough momentum with which some remnants of 6d physics could be
visible. So in the remaining part of this section, we restrict our interest to the bound
states formed by one of $\frac{N(N-1)}{2}$ possible self-dual strings with
many units of momenta.

Without losing generality, let us only consider the string or W-boson connecting
the first and $N$'th D4-brane in the $U(N)$ theory. To generalize to the W-boson stretched
between $i$'th and $j$'th D4-branes, it just suffices to replace $N$ in the results below
by $j\!-\!i\!+\!1$, as the D4-branes outside the stretch of the string do not play any role.
For $U(3)$, the W-boson connecting the first and third D4-brane comes with the chemical
potential factor $e^{-(\mu_1-\mu_3)}$. We first obtain the single particle partition function
from the 5d $\mathcal{N}\!=\!2^\ast$ partition function, and then for simplicity set $\gamma_1=0$, $\gamma_2=i\pi$, factoring out the divergent $I_{com}$ part. Finally reading
off the coefficient of $e^{-(\mu_1-\mu_3)}$, one obtains
\begin{eqnarray}
  \hspace*{-0.5cm}z_{\rm sp}^{U(3)}&=&1+24q+264q^2+2016q^3+12264q^4+63504q^5
  +290976q^6\cdots\\
  \hspace*{-0.5cm}&=&\prod_{n=1}^\infty\left(\frac{1+q^n}{1-q^n}\right)^4 \times
  \Big(1+16q+96q^2+448q^3+1728q^4+5856q^5+18048q^6+\cdots\Big)\ .\nonumber
\end{eqnarray}
Doing a similar procedure for $U(4)$ single W-boson at $e^{-(\mu_1-\mu_4)}$, one obtains
\begin{align}
  z_{\rm sp}^{U(4)} & =1+40q+744q^2+8992q^3+82344q^4+
  \cdots\\
  & = \prod_{n=1}^\infty\left(\frac{1+q^n}{1-q^n}\right)^4 \times
  \Big(1+32q+448q^2+3968q^3+27008q^4+
  \cdots \Big)\ .\nonumber
\end{align}
The index for $U(5)$ single W-boson at $e^{-(\mu_1-\mu_5)}$ is
\begin{align}
  z_{\rm sp}^{U(5)} & = 1+56q+1480q^2+25184q^3+317288q^4+
  + \cdots \\
  & = \prod_{n=1}^\infty\!\!\left(\frac{1+q^n}{1-q^n}\right)^{\!4} \times
  \Big(1+48q+1056q^2+14656q^3+149568q^4+
  \cdots \Big)\ . \nonumber
\end{align}
In these expressions, we have added $1$ by hand at the beginning of the series
on the right hand sides. This is because there exists unique supermultiplet of these
W-bosons without instantons (or momentum), as explained above. We factored out the
center-of-mass fluctuation $\prod_{n=1}^\infty\frac{(1+q^n)^4}{(1-q^n)^4}$ as this
should exist for all self-dual strings fluctuating in the transverse space
$\mathbb{R}^4$. Then one finds that the remaining $U(4)$ and $U(5)$ contributions satisfy
\begin{eqnarray}
  1+32q+448q^2+3968q^3+27008q^4+\cdots&=&(1+16q+96q^2+448q^3+1728q^4+\cdots)^2\nonumber\\
  1+48q+1056q^2+14656q^3+149568q^4+\cdots&=&(1+16q+96q^2+448q^3+1728q^4+\cdots)^3
  \ .\nonumber
\end{eqnarray}
Namely, the remaining internal factor for $U(N)$ is given by the $N\!-\!2$'th power
of the universal factor, which is the $U(3)$ internal factor. One may view this as the
index having a single universal factor whenever the fundamental string crosses a D4-brane.

Now let us turn to the universal factor
\begin{equation}\label{universal}
  z_0=1+16q+96q^2+448q^3+1728q^4+5856q^5+18048q^6+\cdots\ .
\end{equation}
Quite remarkably, one can show that this series can be written as
\begin{equation}
  z_0=\oint\frac{dz}{2\pi i z}\exp\left[\sum_{n=1}^\infty\frac{1}{n}
  \left(f_B(q^n)+(-1)^{n-1}f_F(q^n)\right)\left(z^n+\frac{1}{z^n}\right)\right]\ ,
\end{equation}
where the bosonic and fermionic `letter partition functions' $f_B(q)$, $f_F(q)$ are given by
\begin{equation}\label{letter-index}
  f_B(q)=f_F(q)=\frac{2q^{1/2}}{1-q}=2q^{1/2}+2q^{3/2}+2q^{5/2}+\cdots\  .
\end{equation}
This expression implies that the series (\ref{universal}) can be regarded as coming
from 2 bosonic and 2 fermionic 2d degrees carrying instanton charge $\frac{1}{2}$ and
extra degeneracy labeled by $z^{\pm}$, with
$\frac{1}{1-q}$ coming from the standard infinite tower of modes on a circle.
$z$ is a phase, which is the chemical potential for an `emergent' $U(1)$ gauge symmetry.
The integral over $z$ is to project to the gauge singlets.
The factor $z$ and $\frac{1}{z}$ are for the fundamental and anti-fundamental
modes of $U(1)$, respectively.

So one finds that the partition function for the `longest' $SU(N)$ self-dual string
has the following closed form
\begin{equation}\label{distinct-monopole}
  z_{\rm sp}^{U(N)}=
  \prod_{n=1}^\infty \left(\frac{1+q^n}{1-q^n}\right)^4 \times
  \left[\oint\frac{dz}{2\pi i z}\
  \prod_{n=1}^\infty \left(\frac{(1+q^{\frac{2n\!-\!1}{2}}z)
  (1+q^{\frac{2n\!-\!1}{2}}z^{-1})}{(1-q^{\frac{2n\!-\!1}{2}}z)
  (1-q^{\frac{2n\!-\!1}{2}}z^{-1})}\right)^2\right]^{N-2}\ .
\end{equation}
We think this expression is interesting in the following sense. Firstly,
in the sector with zero momentum, we know that there are $\frac{N(N-1)}{2}$ BPS
W-boson states, which can be regarded as a remnant of the fact that Yang-Mills
theory in the unbroken phase has $N^2$ degrees of freedom. Now once we start to
put the momentum on the worldsheet, there turn out to be more `worldsheet degrees'
which can carry it. Namely, consider a self-dual string connecting $i$'th and $j$'th
D4-branes (with $i<j$). Its partition function is obtained from (\ref{distinct-monopole})
by replacing $N$ by $j\!-\!i\!+\!1$. The $4$ bosonic/fermionic degrees in the first factor
of (\ref{distinct-monopole}) simply comes from the natural fluctuation of the $4$
target space coordinates and their superpartners on the worldsheet. The second factor
implies a contribution from $4(j\!-\!i\!-\!1)$ extra bosonic/fermionic
degrees of freedom on the worldsheet. These degrees are not themselves `physical' in that
they carry positive or negative `charges' (with chemical potential $z$) with respect to
an emergent $U(1)^{j\!-\!i\!-\!1}$ gauge symmetry. In the regime with large momentum,
or when $q\rightarrow 1^-$, the integral over $z$ can be done using saddle point
approximation, which has a saddle point at $z=1$. This implies that
in the small wavelength limit, one finds that $4$ external plus $4(j\!-\!i\!-\!1)$
internal degrees of freedom are essentially unconstrained, somewhat similar to what
happens in the deconfined phase of gauge theories at high temperature.

It would also be interesting to collect all such worldsheet degrees of freedom on
$\frac{N(N-1)}{2}$ different W-bosons. Firstly there would be high temperature degrees of
freedom with momentum coming from internal modes. These are obtained first by specifying the
two end points of the W-boson, and then choosing one of the points at which the open string is
intersecting with other D4-branes. At each intersection, $4$ bosonic and fermionic degrees
can carry momentum. The number of possible intersections of ${}_NC_2$ different W-bosons and
D4-branes is ${}_NC_3=\frac{N(N-1)(N-2)}{6}$. As a 2d fermion behaves like half a bosonic
degree, one finds
\begin{eqnarray}
  &&n^{int}_B=n^{int}_F=4{}_NC_3=\frac{2}{3}N(N-1)(N-2)\nonumber\\
  &&n^{int}=n^{int}_B+n^{int}_F/2=N(N-1)(N-2)\ .
\end{eqnarray}
As for the `external' degrees on the $\frac{N(N-1)}{2}$ strings, coming
from the circle dependent fluctuations of the zero modes, one obtains
\begin{equation}
  n^{ext}_B=n^{ext}_F=4{}_NC_2=2N(N-1)\ \ \rightarrow\ \
  n^{ext}=n^{ext}_B+n^{ext}_F/2=3N(N-1)\ .
\end{equation}
Adding the two contributions, one obtains
\begin{equation}\label{N3}
  n=n^{int}+n^{ext}=N(N^2-1)\ ,
\end{equation}
which happens to be the coefficient of the anomaly of $A_{N\!-\!1}$ type $(2,0)$ theory
\cite{Harvey:1998bx}. Note that, at an algebraic level, the contributions $n^{int}$ and
$n^{ext}$ take the same forms as the two types of contributions in the counting of
$\frac{1}{4}$-BPS configurations of \cite{Bolognesi:2011rq}. In the limit where only one of the
 $5$ scalar fields takes nonzero expectation value, the 1/4 BPS junctions get degenerated to 1/2 BPS monopole strings while the junction point could move with the speed of light. The precise relation between the picture we find here and \cite{Bolognesi:2011rq} remains to be clarified.

It might be worthwhile to emphasize a role of the `emergent' $U(1)$ singlet conditions
in (\ref{distinct-monopole}). The `letters,' or the worldsheet degrees implied by
(\ref{letter-index}) all come with half-integral units of momenta, which are physically
forbidden. These letters also come with nonzero charges under the emergent $U(1)$'s.
After imposing the singlet conditions, one only acquires contributions from even numbers of
excitations of these letters, having integral momenta. In this way, one may feel inclined
to call these letters as `partons' of momentum on the self-dual strings. In the sense that
hidden gauge symmetries demand the partons to combine, they are somewhat similar to the
partons of 2+1 dimensional $\mathbb{CP}^N$ instantons discussed in \cite{Collie:2009iz}.

It is natural, although a bit speculative, to interpret these $U(1)$'s as gauge symmetries
of the M5-brane (or D4-brane) with which the self-dual strings intersect. This viewpoint is
natural if we view the self-dual strings as marginal bound states of `fundamental' self-dual
strings connecting adjacent M5-branes: this viewpoint is in particular relevant if we consider
magnetic monopole strings. The $U(1)$ singlet condition appears simply because the corresponding
M5-brane is not an endpoint of the M2-brane self-dual string, so that a nonzerero $U(1)$ charge
is forbidden.

It will be interesting to see if these letter indices indeed originate from physical
degrees of freedom in certain 1+1 dimensional model, derivable from string theory
or a theory of magnetic monopoles. There are many brane realizations of such
self-dual string systems. One can reduce the M2-M5 brane system to the intersecting
D2-NS5 brane system or D2-D4 system. The latter is a conventional D-brane realization of
magnetic monopole strings. The former would yield $U(1)^{N\!-\!1}$ theory on $N\!-\!1$
segments of D2-branes with bi-fundamental matters, similar to the Hanany-Witten
system \cite{Hanany:1996ie}. The latter would yield fundamental matters from the D2-D4 strings.
These models can flow in the IR to nontrivial 2d CFT's. In the literature, there have been
discussions on the possible fixed points \cite{Witten:1997yu}. When the classical
QFT has Coulomb and Higgs branches of moduli space (although their meanings become subtle
in 2d \cite{Witten:1997yu}), it has been argued that there are two CFT's
described by sigma models which have Coulomb or Higgs branch as the target space. When there
are no classical Higgs branch, there could be a `quantum Higgs branch' \cite{Witten:1997yu}
which could be understood as a theory on the threshold bound state of branes under
consideration.

One can also ask if the above $\frac{2N(N^2-1)}{3}$ bosonic/fermionic
degrees would still be the relevant basic constituents for other types of
charged instanton bound states, with appropriate singlet conditions. There are
many types of bound states having various electric charges, in which many W-bosons
bind together by turning on nonzero momentum. The simplest examples of
this sort were presented in the previous subsection in the $SU(2)$ theory. We have not fully
classified these bound states and studied them yet, which we hope to do in the near future.
From the viewpoint of the D2-D4 monopole strings, one can study the index of 2d QFT
for $SU(3)$ distinct monopoles, whose relative moduli space is a Taub-NUT space. Similar
to what we suggested for identical $SU(2)$ monopoles, one can subtract the known 2-particle
index from this index and see if the structures explored in this subsection emerges.

Finally, we point out that it will be interesting to seek for the connection between
the new worldsheet degrees that we found and the self-dual string anomaly, which was
indirectly calculated from the anomaly inflow method \cite{Berman:2004ew} based on earlier
works \cite{Intriligator:2000eq,Ganor:1998ve}. More concretely, \cite{Berman:2004ew}
considered various anomalies of self-dual strings when $G=SU(N)$ is broken to $H\times U(1)$
subgroup, namely, when one or more M5-branes are separated. The anomaly contributions come from
the M2-brane self-dual strings which have one ends on the M5-brane whose gauge symmetry is
the above $U(1)$. The coefficient of this anomaly is given by
\cite{Intriligator:2000eq,Ganor:1998ve}
\begin{equation}\label{self-dual anomaly}
  n_W\equiv|G|-|H|-1\in 2\mathbb{Z}\ ,
\end{equation}
where $|\ |$ is the dimension of a group.
The case with $H=SU(N-1)$ is having only one M5-brane separated.
For the maximally broken phase with $H=U(1)^{N-2}$, (\ref{self-dual anomaly}) is simply
$N^2-N$. This should come from (fermionic) 2d degrees of freedom living on the self-dual
strings which carry nonzero $U(1)$ charge of the separated out M5-brane. As we only find
$N-1$ self-dual strings connecting this M5-brane and other M5-branes, one might wonder
how to have $N^2-N$ worldsheet degrees to account for this anomaly. As we have found new
momentum-carrying degrees whose number grows as the intersections of M2-M5 increase, we
find that our degrees could naturally yield the desired $N^2$ degrees of freedom.
Further studies on self-dual or monopole strings could provide
a more concrete support of this observation.

\section{The instanton index in the symmetric phase}

Reviewing the derivation of our index in section 2 and appendices A, B, one finds
that setting the $U(N)$ VEV $v$ to zero does not change the calculation at all.
Note that the $U(N)$ symmetry is unbroken for $v=0$. We can still introduce
nonzero chemical potentials $\mu_1,\mu_2,\cdots,\mu_N$ for $U(1)^N\subset U(N)$
Cartans of this unbroken symmetry, and further take all of them to assume different values.
The path integral for the index is still perfectly localized, without having any
dangerous non-compact zero modes. This is actually the Omega deformation for the unbroken
$U(N)$ symmetry, similar to $(\epsilon_1,\epsilon_2)$ for the spatial $SO(4)$ symmetry.
So one can ask if our result can be used to learn something about the symmetric phase
of the $(2,0)$ theory on a circle.

An important aspect of the chemical potentials $\mu_i$ in the Coulomb phase
was that they were ordered in the same order as the nonzero VEV $v_i$:
namely $\mu_1>\mu_2>\cdots>\mu_N$ comes from $v_1>v_2>\cdots>v_N$ by requiring that
the index acquires damping factors from states with allowed electric charges.
So we expand all the $\sinh\left(\frac{\mu_i-\mu_j+\cdots}{2}\right)$ factors in the
denominator of our index in positive power series of $e^{-(\mu_i-\mu_j)}<1$
with $i<j$. Since the non-Abelian electric charges can come with arbitrary signs
as they do not appear in the BPS mass with zero VEV, we should not expand the index
this way. A good analogy comes from how we understood the center of mass index
$I_{com}=\frac{\sin\frac{\gamma_1+\gamma_2}{2}
\sin\frac{\gamma_1-\gamma_2}{2}}{\sin\frac{\gamma_1+\gamma_R}{2}
\sin\frac{\gamma_1-\gamma_R}{2}}$ in section 2 in a way symmetric in the sign flips
of $\gamma_1,\gamma_R$, as the spectrum is $SO(4)$ symmetric. Trying to expand the
`$\sin$' factors in the denominator in this democratic way, we have seen that the
resulting series diverges. This exactly reflects the infinitely many
wave-functions depending on center of mass coordinates unsuppressed by the spin
chemical potentials. We could however separate out these $I_{com}$ factors and classify
various terms in the index by particle numbers, extracting out the essential information
on threshold bound states of various sorts in sections 3 and 4. Now to `democratically'
expand the expression in the chemical potentials for the non-Abelian electric charge,
we define $\mu_i=i\alpha_i$. First of all, it is not obvious in general how to expand
various contributions from different saddle points in a way the spectrum is invariant
under various sign flips of all charges. To simplify the story,
if we turn off $\gamma_R=0$ and expand the resulting expression, we encounter
a similar divergence. For instance, let us consider the $SU(2)$ single instanton index
\begin{equation}
  I_{N=2,k=1}=2I_{com}\frac{\sin\frac{\alpha_1-\alpha_2+\gamma_2}{2}
  \sin\frac{\alpha_1-\alpha_2-\gamma_2}{2}}{\sin^2\frac{\alpha_1-\alpha_2}{2}}\ ,
\end{equation}
where the factor $2$ comes from two saddle points. An attempt to expand this in
$e^{i(\alpha_1-\alpha_2)}$ with $\alpha_1\leftrightarrow\alpha_2$ invariance yields
a divergence like $I_{com}$.

Unlike the case with Omega background $\gamma_1,\gamma_R$, we do not have a physical
understanding of these divergences. Perhaps a parton-like interpretation of the instantons
could tell us how to correctly treat this quantity and extract out useful information.
This is because, as suggested in \cite{Collie:2009iz}, the non-compactness of the internal
moduli space from instanton sizes (causing our divergence) could be implying some
multi-particle nature of instantons from partonic constituents. From our viewpoint,
the divergence apparently comes from instantons having many possible states with same
non-Abelian electric charges. Carefully defining an observable free of possible infrared
divergences could help cure this problem.

In the remaining part of this section, we turn to another interpretation of
our index in the symmetric phase. The D0-D4 quantum mechanics discussed in section 2
has variables ($\phi,\varphi^m$) which probe the Coulomb branch. At low energy, they can be
integrated out to yield a sigma model on the instanton moduli space. This model was studied
in \cite{Aharony:1997th,Aharony:1997an} to understand the $(2,0)$ theory compactified on
a circle, or more precisely the DLCQ $(2,0)$ theory compactified on a null circle. This sigma
model has a non-relativistic superconformal symmetry. From the 6 dimensional perspective,
this is the subgroup of the $OSp(6,2|4)$ superconformal symmetry of the (2,0) theory which
commutes with the momentum $P_-$ on a null circle \cite{Aharony:1997th,Aharony:1997an}.

Let us first consider the conformal symmetry. The relativistic conformal algebra
$SO(6,2)$ has generators $M_{AB}$, $A,B=0,1,2,\cdots,6,7$, with timelike directions
$0,7$. Apart from the $SO(5,1)$ Lorentz generators $M_{\mu\nu}$ with
$\mu,\nu=0,1,\cdots, 5$, the generators
\begin{equation}
  P_\mu=M_{6\mu}+M_{7\mu}\ ,\ \ K_\mu=-M_{6\mu}+M_{7\mu}\ ,\ \ \Delta=M_{67}
\end{equation}
are translation, special conformal transformation, dilatation.
Introducing the light-cone coordinates $x^\pm=x^0\pm x^5$, the non-relativistic
conformal algebra is given by a subgroup which commutes with $P_-=M_{6-}+M_{7-}$:
\begin{equation}
  H\sim P_+\ ,\ \ P_i\ ,\ \ M_{ij}\ ,\ \ G_i\sim
  M_{-i}\ ,\ \ K\sim K_-\ ,\ \ D=\Delta-M_{05}\ .
\end{equation}
$D$ is the non-relativistic dilatation generator.
In particular, from an $SL(2,\mathbb{R})$ subgroup
\begin{equation}
  [D,H]=-2iH\ ,\ \ [D,K]=2iK\ ,\ \ [K,H]=-iD\ ,
\end{equation}
we can form another combination
\begin{equation}\label{SL(2,R)}
  L_0=aH+a^{-1}K\ ,\ \ L_{\pm 1}=\frac{1}{2}(aH-a^{-1}K\mp iD)
\end{equation}
which satisfy
\begin{equation}
  [L_0,L_{\pm 1}]=\pm 2L_{\pm 1}\ ,\ \ [L_{+1}, L_{-1}]=-L_0\ .
\end{equation}
The spectrum of $H$ in conformal quantum mechanics is continuous,
while that of $L_0$ is discrete due to a harmonic potential coming from $K$
on the target space. We take $a=1$ from now on. In a conformal
theory with $SL(2,\mathbb{R})$ subgroup, one can use $L_0$ as the Hamiltonian to study
its discrete spectrum.

As explained in \cite{Nishida:2007pj}, local eigen-operators of the
dilatation operator $D$ can be mapped to the eigenstates of $L_0$. The arguments there
apply mainly to field theories, in which the vacuum is annihilated by $K$. For a mechanical
system, the arguments there can be slightly refined as follows. Under a similarity
transformation given by $M=e^{H/2}e^{-K}$, one can show that
\begin{equation}
  M^{-1}(iD)M=H+K\ .
\end{equation}
So the eigen-operators of $iD$ maps via $M$ to eigenstates of $L_0$. The last
operator with positive eigenvalue can be used to create states in our mechanical model.
As a simple example to cross-check, one can consider a free particle with
$H=\frac{p^2}{2}$, $K=\frac{x^2}{2}$, $D=-\frac{xp+px}{2}$. The variable $x$ has dimension
$-1$ under dilatation: $[iD,x]=-x$. By conjugating $x$ with $M$, one obtains
\begin{equation}
  M^{-1}xM=e^{K}\left(e^{-p^2/4}xe^{p^2/4}\right)e^{-K}=
  e^{x^2/2}\left(x+ip/2\right)e^{-x^2/2}=\frac{x+ip}{2}=a/\sqrt{2}\ ,
\end{equation}
where $a=\frac{x+ip}{\sqrt{2}}$ is the annihilation operator which has charge $-1$
under the harmonic oscillator Hamiltonian $L_0=\frac{p^2+x^2}{2}$.
One can also show $M^{-1}pM=\sqrt{2}ia^\dag$. Acting $a^\dag$ with positive eigenvalue $+1$
on the ground state creates the eigenstates of $L_0$.

The supersymmetric extension of this conformal symmetry is obtained by reducing
$OSp(6,2|4)$ to a subgroup which commutes with $P_-$. The $32$ supercharges are
grouped by their eigenvalues of $\Gamma^{67}$ (dilatation), whose sign determines
whether the supercharges are $Q$ or $S$. Both $Q$ and $S$ are again classified by
their eigenvalues of $\Gamma^{05}$, as $[P_\mu, S]\sim(\Gamma_\mu)Q$ should vanish
for $\mu=-$ which depends on the eigenvalue of $\Gamma^{05}$. Picking $Q$ to have $+,+$
and $S$ to have $-,-$ eigenvalue \cite{Aharony:1997an}, we obtain $8$ pairs of $Q,S$
type supercharges commuting with $P_-$, apart from $8$ more $Q$'s which also commute
with $P_-$ . Some of their algebra is given by
\begin{equation}
  2i\{\bar{Q}^{\dot{a}}_{\dot\alpha},\bar{S}_{\dot{b}}^{\dot\beta}\}=
  iD-4\delta_{\dot\alpha}^{\dot\beta}(J_{2R})^{\dot{a}}_{\ \dot{b}}
  -2\delta^{\dot{a}}_{\dot{b}}(J_{1R})_{\dot\alpha}^{\ \dot\beta}\ .
\end{equation}
We used the fact that a chiral $SO(6,2)$ spinor with a $\Gamma^{05}\Gamma^{67}$
projection reduces to a chiral $SO(4)$ spinor on $1234$, which we choose to be anti-chiral
(doublet in $SU(2)_{1R}$). We pay attention only to the supercharges charged under
$SU(2)_{2R}$, which contain the supercharges preserved by our path integral.
The above coefficients of R-charges can be easily fixed by, say, demanding it reproduce
the known BPS bound for relativistic $OSp(6,2|4)$ \cite{Bhattacharya:2008zy}.
Picking either of $Q=\bar{Q}^{\dot\mp\dot\pm}$, as we did in our index, we find that
the BPS bound for operators with positive dimensions is given by
\begin{equation}
  2i\{Q,S\}=iD\mp(4J_{2R}+2J_{1R})\ \rightarrow\ \ iD\geq \pm(4J_{2R}+2J_{1R})\ .
\end{equation}
The supercharge itself saturates this bound by having $D=1$,
$J_{2R}=\pm\frac{1}{2}$, $J_{1R}=\mp\frac{1}{2}$. The charge $J_R=J_{1R}+J_{2R}$
commutes with both $\bar{Q}^{\dot\mp\dot\pm}$. From
\begin{equation}
  [K,Q]=-iS\ ,\ \ [H,S]=iQ\ ,
\end{equation}
one finds that the supercharges under $M$ conjugation become
\begin{equation}
  M^{-1}QM=Q-iS\equiv\hat{Q}\ ,\ \ M^{-1}SM=-i/2(Q+iS)=-\frac{i}{2}\hat{S}\ .
\end{equation}
The superalgebra becomes
\begin{equation}\label{BPS-bound}
  \{\hat{Q},\hat{S}\}=L_0\mp(4J_{2R}+2J_{1R})\ .
\end{equation}
Thus, operators which diagonalize $iD$ and preserve $Q,S$ map to
eigenstates of $L_0$ which preserve $\hat{Q},\hat{S}$.

Now consider the following `superconformal index'
\begin{equation}\label{superconformal}
  I_{SC}={\rm Tr}\left[(-1)^Fe^{-\beta\{\hat{Q},\hat{S}\}}e^{-2i\gamma_RJ_R}
  e^{-2i\gamma_1J_{1L}-2i\gamma_2J_{2L}}e^{-i\alpha_i\Pi_i}\right]\ .
\end{equation}
The charges $J_R,J_{1L},J_{2L},\Pi_i$ commute with $\hat{Q},\hat{S}$. The imaginary time
evolution with period $\beta$ is provided by the new Hamiltonian $H+K$, where $K$ simply adds
a harmonic potential on the instanton moduli space. Integrating out the momentum variables
in the path integral representation, like what we did in section 2.2, one obtains a Euclidean
Lagrangian with extra harmonic potential with order $1$ coefficient, and time derivatives
twisted by $J_R,J_{1L},J_{2L},\Pi_i$ with coefficients $\frac{\gamma_R}{\beta}$,
$\frac{\gamma_1}{\beta}$, $\frac{\gamma_2}{\beta}$, $\frac{\alpha_i}{\beta}$ and also by
$2J_{1R}+4J_{2R}$ with an order $1$ coefficient. In the limit the regulator $\beta$ is taken
to zero, one finds that the extra harmonic potential and the $2J_{1R}+4J_{2R}$ twisting become
subleading compared to the terms proportional to other chemical potentials or those having
time derivatives with $\frac{d}{dt}\sim\frac{1}{\beta}$. \footnote{This argument works since
the path integral with nonzero chemical potentials does not have zero modes even before
adding $K$ to the Hamiltonian. Dimension of $\beta$ may look unclear at this point, but
this is simply because we have set a dimensionful  constant $a$ in (\ref{SL(2,R)}) to $1$.}
The path integral in this limit simply reduces to
our previous path integral in section 2.2. So our index admits another interpretation
in the symmetric phase, as counting operators saturating the superconformal BPS bound.

The fact that many terms in the previous paragraph become subleading in the
$\beta\rightarrow 0$ limit requires a careful interpretation of the resulting index.
Depending on whether we demand $L_0=\pm(2J_{1R}+4J_{2R})$ as our
superconformal BPS bound, the resulting $J_R=J_{1R}+J_{2R}$ is either non-negative or
non-positive. However, our index in section 2.2 can be expanded in two ways. It can
either be expanded in a Taylor series of $e^{-i\gamma_R}$ or $e^{i\gamma_R}$. These two
possible expansions naturally incorporate the two possibile BPS bounds, with positive
$J_R$ for BPS operators or negative $J_R$ for anti-BPS operators.

In this superconformal index interpretation, the nonzero chemical potential
$\gamma_R\sim\frac{\epsilon_1+\epsilon_2}{2}$ plays the most important role.
This is in curious contrast with the fact that in many cases, instanton calculus
has been most conveniently discussed in the `self-dual' Omega background with
$\epsilon_1=-\epsilon_2$. In particular, with nonzero $\gamma_R$, one finds that
the singularities that one encounters at $\mu_i=\mu_j$ for $SU(N)$ all disappears.
Namely, considering all examples (\ref{single-index}), (\ref{two-index}), (\ref{three-index}),
the singularities exist for each saddle point but completely cancel when we sum
over various contributions from different saddle points. This is consistent with
the fact that $e^{-i\gamma_R}$ is sufficient to guarantee convergence in the trace
over infinitely many states in (\ref{superconformal}) .

The index (\ref{index-formula}) actually has a contour integral representation,
as first presented for 4d $\mathcal{N}=2^\ast$ theory in \cite{Nekrasov:2002qd}.
The 5 dimensional version of this formula is given by
\begin{eqnarray}\label{index-contour}
  I_k&\sim&\frac{1}{k!}\oint\prod_{I=1}^k\left(d\phi_I\prod_{i=1}^N
  \frac{\sinh(\phi_I-a_i+m)\sinh(\phi_I-a_i-m)}
  {\sinh(\phi_I-a_i-\frac{\epsilon}{2})\sinh(\phi_I-a_i+\frac{\epsilon}{2})}
  \right)\prod_{I\neq J}\sinh\phi_{IJ}\\
  &&\times\prod_{I,J}\frac{\sinh(\phi_{IJ}-\epsilon)}
  {\sinh(\phi_{IJ}-\epsilon_1)\sinh(\phi_{IJ}-\epsilon_2)}\cdot
  \frac{\sinh(\phi_{IJ}+m+\frac{\epsilon_1-\epsilon_2}{2})
  \sinh(\phi_{IJ}+m-\frac{\epsilon_1-\epsilon_2}{2})}
  {\sinh(\phi_{IJ}+m-\frac{\epsilon}{2})\sinh(\phi_{IJ}+m+\frac{\epsilon}{2})}\ .
  \nonumber
\end{eqnarray}
where $\epsilon=2\epsilon_R=\epsilon_1+\epsilon_2$. It is convenient to define
$z_I=e^{2\phi_I}$, and consider the prescription for the poles to keep.
There are many poles from the denominator, and also from $d\phi_I\sim\frac{dz_I}{z_I}$
at the origins. To present the relevant poles, we take $\epsilon$ to be large and
positive, which makes a good sense in the context of superconformal index as
$e^{-i\gamma_R}=e^{-\epsilon}$ is the main convergence parameter.
If one only keeps the residues coming from the poles of
$\sinh(\phi_I-a_i-\frac{\epsilon}{2})$, $\sinh(\phi_I-a_i+\frac{\epsilon}{2})$
on the first line and $\sinh(\phi_{IJ}-\epsilon_1)$, $\sinh(\phi_{IJ}-\epsilon_2)$
on the second line, and also restrict to the poles which appear inside the unit
circle on the $z_I$ planes (satisfying $|z_I|<1$) with $\epsilon\gg 0$, then one
obtains (\ref{index-formula}). Note that there are many poles inside the unit
circle $|z_I|=1$ apart from the above ones, so the integral above cannot be regarded
as an integral over $-i\phi_I/2$ angle variables on the unit circles of $z_I$. Although
this prescription about poles is a well-developed fact, we checked that it reproduces
(\ref{index-formula}) for ($k=1,N\leq 4$), ($k=2,N=1,2$), ($k=3,N=1,2$).

One may try to understand the above formula by the following attempt to directly
count the BPS states in the instanton sigma model, generalizing \cite{Benvenuti:2010pq}.
In this sigma model, we only consider operators made of the fields
$a_m,q_{\dot\alpha},\lambda^i_\alpha,\psi^i$, while the fields
$\phi,\varphi^m,\bar\lambda^i_{\dot\alpha}$ are auxiliary. From the supersymmetry
transformations (\ref{SUSY-vector}), (\ref{SUSY-adjoint}), (\ref{SUSY-fundamental}),
we construct operators which are in the cohomology of, say,
$Q=\bar{Q}^{\dot{+}\dot{-}}$. The cohomologies made only of bosonic variables are
easy to understand, and have been studied in \cite{Benvenuti:2010pq}.
$Q$-closed variables saturating the BPS bound $D=-(2J_{1R}+4J_{2R})$ are $a_{\alpha\dot{+}}$,
$q_{\dot{+}}$ and $\bar{q}^{\dot{-}}$, 
having dimension $-1$. One may use $B_1,B_2\sim a_1+ia_2,a_3-ia_4$ defined in appendix A
to represent $a_{\alpha\dot{+}}$. Any $U(k)$ gauge invariant operators made of these
`BPS letters' are $Q$-closed. Among them, we should mod out $Q$-exact operators to count
the elements of $Q$-cohomology. The only bosonic $Q$-exact expression comes from the
second line of (\ref{SUSY-adjoint}), which is
\begin{equation}
  Q\bar\lambda^{\dot{-}}_{\dot\alpha}\sim D^{\dot{-}}_{\ \ \dot\alpha}\sim
  D_{\dot{+}\dot\alpha}\ .
\end{equation}
The real ADHM expression $D_{\dot{+}\dot{-}}\sim[B_1,B_1^\dag]+[B_2,B_2^\dag]+\cdots$
contains both BPS and non-BPS letters and are thus irrelevant. The complex ADHM expression
$D_{\dot{+}\dot{+}}\sim[B_1,B_2]+\bar{q}^{\dot{-}}q_{\dot{+}}$ contains BPS letters only
and should be modded out. The partition function for the two bosonic oscillators $B_1,B_2$
in $U(k)$ adjoint is given by the denominator of the first factor on the second line of
(\ref{index-contour}). The partition function for the BPS letters in $U(k)$ fundamental
is the denominator of the first line. The numerator of the first factor on the second line
is for the ADHM constraint, while the integral of $z_I$ over unit circles with the Haar
measure (given by the last factor on the first line) projects to $U(k)$ singlets.
This leads to the integrand
\begin{equation}
  \frac{1}{\sinh(\phi_I-a_i-\frac{\epsilon}{2})\sinh(\phi_I-a_i+\frac{\epsilon}{2})}
  \cdot\prod_{I,J}\frac{\sinh\phi_{IJ}\sinh(\phi_{IJ}-\epsilon)}
  {\sinh(\phi_{IJ}-\epsilon_1)\sinh(\phi_{IJ}-\epsilon_2)}\ ,
\end{equation}
which is that for the bosonic cohomology formula in \cite{Benvenuti:2010pq}. Although the
remaining factors of (\ref{index-contour}) seem to quite naturally map to partition functions
from fermionic
BPS letters as well as fermionic constraints which are superpartners of the ADHM constraint,
a detailed combinatoric understanding of (\ref{index-contour}) seems to be more challenging.
Most importantly, the complicated contour prescription explained after (\ref{index-contour})
is hard to understand from an explicit counting at the moment. Perhaps a subtlety in imposing
constraints \cite{Benvenuti:2006qr} should be properly understood. It will be nice to have
an elementary understanding of this pole prescription from a combinatoric viewpoint.

As $U(N)$ is also a gauge symmetry of the 5d and 6d theories, one would also
have to integrate over $a_i$ with an $SU(N)$ Haar measure to extract the spectrum
of gauge-invariant operators.

In the remaining part of this section, we make some consistency checks and a
preliminary study of this index. A more detailed analysis will be reported elsewhere.

Firstly, as consistency checks, one can compare our index with the counting of
a class of cohomologies in \cite{Aharony:1997an}. Also, one can compare the large $N$
index (at low energies) with the DLCQ supergraviton spectrum obtained from supergravity
on $AdS_7\times S^4$. For the latter, of course the DLCQ is a small radius limit so that
supergravity approximation is not reliable in general. One may however hope that the
spectrum is more robust in the BPS sector so that a naive supergravity calculation could
yield the correct result. In fact, we will explain that our index agrees with the BPS
spectrum of DLCQ gravity.

We start by considering the simplest case with $N=1$. There we expect that the spectrum can
be all understood as the KK modes of the free 6d tensor multiplet. In particular, at $k=1$,
\cite{Aharony:1997an} worked out a class of cohomology and found states in the vector
representation ${\bf 5}$ of $SO(5)$ which is in a singlet of $SU(2)_{1L}\times SU(2)_{1R}$.
They come with non-relativistic dimension $D=2$.
Acting the broken $8$ supercharges $Q^i_\alpha$, one generates fermions in
$({\bf 4}, {\bf 2}, {\bf 1})$ of $SO(5)\times SU(2)_{1L}\times SU(2)_{1R}$. Acting it
once more, one obtains a tensor in $({\bf 1}, {\bf 3}, {\bf 1})$.
Our index counts states preserving a specific supercharge $Q$ saturating the bound
$L_0\geq 2J_{1R}+4J_{2R}$. Decomposing states into representations of
$SU(2)_{2L}\times SU(2)_{2R}$ and only keeping those states saturating our bound, one obtains
\begin{equation}
  {\rm scalar}\ \rightarrow\ ({\bf 1},{\bf 2})_{\frac{1}{2}}\ ,\ \
  {\rm fermion}\ \rightarrow\ ({\bf 2}, {\bf 1})_{\frac{1}{2}}\ ,\ \
  {\rm tensor}\ \rightarrow\ \ {\rm none}\ ,
\end{equation}
where the entries denote $(SU(2)_{1L}, SU(2)_{2L})_{J_{2R}}$ representations and charges.
Collecting their contributions, and also multiplying the factors coming from derivatives
on $\mathbb{R}^4$ for descendants, one obtains the following contribution
\begin{equation}
  \frac{(e^{i\gamma_2}+e^{-i\gamma_2})e^{-i\gamma_R}-(e^{i\gamma_1}+e^{-i\gamma_1})e^{-i\gamma_R}}
  {(1-e^{-i\gamma_R+i\gamma_1})(1-e^{-i\gamma_R-i\gamma_1})}=I_{com}(\gamma_1,\gamma_2,\gamma_R)
\end{equation}
to our index. Furthermore, at $k>1$, all cohomologies found in \cite{Aharony:1997an} can be
understood as `multi-particle' excitations of those at $k=1$. So from \cite{Aharony:1997an},
one obtains the `single-particle' index
\begin{equation}
  I_{com}\frac{q}{1-q}=e^{-i\gamma_R}\frac{(e^{i\gamma_2}+e^{-i\gamma_2}-e^{i\gamma_1}
  -e^{-i\gamma_1})}{(1-e^{-i\gamma_R+i\gamma_1})(1-e^{-i\gamma_R-i\gamma_1})}\frac{q}{1-q}\ ,
\end{equation}
which completely agrees with our $U(1)$ instanton index.

We also study our index at general $N$ at $k=1$. After projecting to $SU(N)$ singlets
only, one obtains ($t\equiv e^{-i\gamma_R}$)
\begin{equation}\label{k=1 index 1}
  I_{k=1}=\frac{e^{i\gamma_2}+e^{-i\gamma_2}-e^{i\gamma_1}-e^{-i\gamma_1}}
  {(1-te^{i\gamma_1})(1-te^{-i\gamma_1})}\left[t+\sum_{n=1}^{N-1}(e^{in\gamma_2}
  +e^{-in\gamma_2})t^{n+1}-\chi_{\frac{N-2}{2}}(\gamma_2)t^{N+1}\right]\ ,
\end{equation}
which we checked till $N\leq 6$.
\begin{equation}
  \chi_j(\gamma_2)=e^{2ji\gamma_2}+e^{2(j-2)i\gamma_2}+\cdots+e^{-2ji\gamma_2}
  =\frac{e^{(2j+1)i\gamma_2}-e^{-(2j+1)i\gamma_2}}{e^{i\gamma_2}-e^{-i\gamma_2}}
\end{equation}
is the $SU(2)_{2L}$ character for the spin $j$ representation. This result contains
and extends the states counted in \cite{Aharony:1997an}, as we explain now. The above
result can be written as
\begin{equation}\label{k=1 index}
  I_{k=1}=\frac{e^{i\gamma_2}+e^{-i\gamma_2}-e^{i\gamma_1}-e^{-i\gamma_1}}
  {(1-te^{i\gamma_1})(1-te^{-i\gamma_1})}\left[\sum_{n=0}^{N-1}\chi_{\frac{n}{2}}(\gamma_2)
  t^{n+1}-\sum_{n=1}^{N-1}\chi_{\frac{n-1}{2}}(\gamma_2)t^{n+2}\right]\ .
\end{equation}
At general $N$ and $k=1$, \cite{Aharony:1997an} obtained cohomologies
which are in rank $n$ symmetric representations of  $SO(5)$ for $n=1,2,\cdots, N$, with
dimension $D=2n$. By restricting to states preserving our $Q$ and acting the
broken supersymmetry $Q_{a\alpha}$ which commute with our $Q$, in a similar manner as our
analysis for $N=1$ above, one obtains an index which accounts for the first summation of
(\ref{k=1 index}). The states contributing to the second summation stay beyond the class
of states considered in \cite{Aharony:1997an}, as they restricted to a particular subset of
primaries (in particular with $J_{1R}=0$).

However, one can easily see that the second contribution to (\ref{k=1 index}) should also
exist, by studying the large $N$ gravity dual index. The index (\ref{k=1 index 1}) or
(\ref{k=1 index}) at $N\rightarrow\infty$ becomes
\begin{equation}\label{large N}
  I_{N\rightarrow\infty,k=1}=\frac{e^{i\gamma_2}+e^{-i\gamma_2}-e^{i\gamma_1}-e^{-i\gamma_1}}
  {(1-te^{i\gamma_1})(1-te^{-i\gamma_1})}\ \frac{t-t^3}{(1-te^{i\gamma_2})(1-te^{-i\gamma_2})}
  \ .
\end{equation}
On the gravity side, one can start from the supergravity KK spectrum on $AdS_7\times S^4$
and restrict to states saturating our non-relativistic BPS bound after DLCQ. One may start
from, say, table 3 of \cite{Bhattacharya:2008zy} which decomposes the supergravity spectrum
on $AdS_7\times S^4$. The energy $\epsilon_0$ there may be understood as the generator
$H=M_{07}=-\frac{P_0+K_0}{2}$, and the compact generators $M_{mn}$ for $m,n=1,2,\cdots,6$ of
$SO(6,2)$ can be understood as $SO(6)$ in the table of \cite{Bhattacharya:2008zy}.
By a standard similarity transformation,
$H$ and $SO(6)$ generators map to $i\Delta=iM_{67}$ $\in SO(1,1)\subset SO(6,2)$ and $SO(5,1)$
generators. In particular, decomposing the $SO(6)$ generators into $M_{ab}$ for $a,b=1,2,\cdots,5$
and $M_{6a}$, one can show that $M_{6a}$ maps to $iM_{0a}$ boost generators. See, for instance,
eqn.(2.11) of \cite{Dolan:2002zh}. So we take one of the $SO(6)$ Cartans in \cite{Bhattacharya:2008zy}
and interpret it as $M_{05}$ boost eigenvalue, and subtract it to $\epsilon_0$ there to be
identified with our non-relativistic dimension $D$. By collecting the fields in their table
which saturate our BPS bound, one obtains the Kaluza-Klein fields of table \ref{sugra}.
\begin{table}[t!]
$$
\begin{array}{c|cccc|c}
  \hline &D&J_{1L}&J_{2L}&2(J_{1R}+J_{2R})&{\rm boson/fermion}\\
  \hline p\geq 1&2p&0&\frac{p}{2}&p&{\rm b}\\
  \hline p\geq 1&2p+1&0&\frac{p-1}{2}&p+1&{\rm f}\\
  p\geq 1&2p&\frac{1}{2}&\frac{p-1}{2}&p&{\rm f}\\
  \hline p\geq 2&2p+1&\frac{1}{2}&\frac{p-2}{2}&p+1&{\rm b}\\
  p\geq 2&2p&0&\frac{p-2}{2}&p&{\rm b}\\
  \hline p\geq 3&2p+1&0&\frac{p-3}{2}&p+1&{\rm f}\\
  \hline \cdot&3&0&0&2&{\rm b\ (fermionic\ constraint)}\\
  \hline
\end{array}
$$
\caption{BPS fields of supergravity}\label{sugra}
\end{table}
Collecting all, one obtains the following single particle index:
\begin{eqnarray}
  &&\hspace{-1.7cm}\sum_{p=1}^\infty t^p\chi_{\frac{p}{2}}(\gamma_2)-\sum_{p=1}^\infty
  \left(t^{p+1}+t^p\chi_{\frac{1}{2}}(\gamma_1)\right)\chi_{\frac{p-1}{2}}(\gamma_2)
  +\sum_{p=2}^\infty\left(t^{p+1}\chi_{\frac{1}{2}}(\gamma_1)+t^p\right)\chi_{\frac{p-2}{2}}
  (\gamma_2)-\sum_{p=3}^\infty t^{p+1}\chi_{\frac{p-3}{2}}(\gamma_2)+t^2\nonumber\\
  &=&(e^{i\gamma_2}+e^{-i\gamma_2}-e^{i\gamma_1}-e^{-i\gamma_1})
  \frac{t-t^3}{(1-te^{i\gamma_2})(1-te^{-i\gamma_2})}\ .
\end{eqnarray}
After multiplying the derivative (or wavefunction) factor in $\mathbb{R}^4$,
one obtains
\begin{equation}\label{gravity-single}
  I_{\rm sp}=\frac{e^{i\gamma_2}+e^{-i\gamma_2}-e^{i\gamma_1}-e^{-i\gamma_1}}
  {(1-te^{i\gamma_1})(1-te^{-i\gamma_1})}\ \frac{t-t^3}{(1-te^{i\gamma_2})(1-te^{-i\gamma_2})}\ .
\end{equation}
This is the single particle index for each instanton number (or DLCQ momentum) $k$.
Thus the full multi-particle index is obtained by multiplying $\frac{q}{1-q}$ to $I_{\rm sp}$
and then taking the Plethystic exponential. At $\mathcal{O}(q^1)$, one obtains $I_{\rm sp}$
which perfectly agrees with the instanton index (\ref{large N}). At larger $k$, we should
start from our instanton index in section 2, project to $SU(N)$ singlets, and then take
Plethystic logarithm to be compared with (\ref{gravity-single}) at each $\mathcal{O}(q^k)$.
We numerically find that this works well at $k=2$ till $\mathcal{O}(t^{N})$,
which we checked for $N=2,3,4$. This is all one can expect when
comparing with large $N$ gravity.

Our finite $N$ index (\ref{k=1 index}) is a simple generalization of the large $N$ index
by truncating the supergravity spectrum at $\mathcal{O}(t^N)$.

Finally, we study our index in the pure bosonic sector. One can obtain this subsector by
either restricting to bosonic variables for constructing states, or more systematically by
taking the limit $m\rightarrow\infty$, $q\rightarrow 0$ keeping $e^{Nm}q$
finite. This limit keeps states with largest $J_{2L}$ spin for given $k$.\footnote{From the
viewpoint of $\mathcal{N}\!=\!2$ partition function, this is simply the pure $\mathcal{N}\!=\!2$
SYM limit.} This sector seems to be discarding many states in the full theory: for instance,
at $k=1$, all states that we obtained in (\ref{k=1 index}) disappear except a single term in
the square parenthesis:
\begin{equation}
  I_{k=1}\rightarrow\frac{e^{Ni\gamma_2}t^{N}}{(1-te^{i\gamma_1})(1-te^{-i\gamma_1})}\ .
\end{equation}
This is also consistent with \cite{Benvenuti:2010pq}. There, all states except one came
in non-trivial representations of $SU(N)$ at $k=1$, which we project out. We shall
illustrate howeverer that even in this simplified sector there appears a curious large $N$
phase transition in the `6d limit' $k\rightarrow\infty$.

In the bosonic sector, the contour prescription becomes very simple as we explained above:
one simply keeps all the poles inside the unit circles for the variables $z_I=e^{2\phi_I}$.
The index can thus be written as ($t\equiv e^{-\epsilon}$)
\begin{eqnarray}
  I_{N,k}&=&\frac{e^{Nkm}t^{Nk}}{N!}\oint\prod_{i=1}^N\frac{d\alpha_i}{2\pi}\prod_{i<j}
  \left(2\sin\frac{\alpha_i-\alpha_j}{2}\right)^2\frac{1}{k!}\oint\prod_{I=1}^k
  \frac{d\beta_I}{2\pi}\prod_{I<J}\left(2\sin\frac{\beta_I-\beta_J}{2}\right)^2\\
  &&\times\prod_{i,I}\frac{1}{(1-te^{i(\alpha_i-\beta_I)})(1-te^{i(\beta_I-\alpha_i)})}
  \prod_{I,J}\frac{1-t^2e^{i(\beta_I-\beta_J)}}{(1-te^{i\gamma_1}e^{i(\beta_I-\beta_J)})
  (1-te^{-i\gamma_1}e^{i(\beta_I-\beta_J)})}\ .\nonumber
\end{eqnarray}
where $a_i=i\frac{\alpha_i}{2}$, $\phi_I=i\frac{\beta_I}{2}$ with $2\pi$ periodic angles
$\alpha_i$, $\beta_I$. The factor $(e^mt)^{Nk}$ is kept only in the second viewpoint of
this index explained in the previous paragraph.
Apart from the two Haar measures, the integrand on the second line
can be written as the Plethystic exponential
\begin{equation}
  \exp\left[\sum_{n=1}^\infty\frac{1}{n}f(t^n,n\gamma_1,n\alpha_i,n\beta_I)\right]
\end{equation}
of a letter index $f$ given by
\begin{equation}
  f=t\sum_{i,I}\left(e^{i(\alpha_i-\beta_I)}+e^{i(\beta_I-\alpha_i)}\right)
  +\left(t(e^{i\gamma_1}+e^{-i\gamma_1})-t^2\right)\sum_{i,j}e^{i(\beta_I-\beta_J)}\ .
\end{equation}
We firstly consider the large $k$ limit of this integral. A motivation for this could
be that this limit allows one to study the light-cone description of the uncompactified
$(2,0)$ theory \cite{Aharony:1997an}. Introducing the $\beta_I$ eigenvalue density
$\rho(\theta)\geq 0$ with $\theta\sim\theta+2\pi$, and Fourier expanding, one can replace
the integration over $\beta_I$ by that for the Fourier coefficients $\rho_n$ of $\rho(\theta)$
given by $\rho_n=\frac{1}{k}\sum_{I=1}^ke^{in\beta_I}$. The index becomes
\begin{eqnarray}
  I_{N,\infty}&=&\frac{1}{N!}\oint\prod_{i=1}^N\frac{d\alpha_i}{2\pi}\prod_{i<j}
  \left(2\sin\frac{\alpha_i-\alpha_j}{2}\right)^2\int\prod_{n=1}^\infty d\rho_nd\rho_{-n}\\
  &&\times\exp\left[-\sum_{n=1}^\infty\frac{1}{n}\left(k^2\rho_n\rho_{-n}(1-t^ne^{in\gamma_1})
  (1-t^ne^{-in\gamma_1})-kt^n\rho_n\sum_ie^{-in\alpha_i}-kt^n\rho_{-n}\sum_ie^{in\alpha_i}
  \right)\right]\ .\nonumber
\end{eqnarray}
Since the coefficients of $|\rho_n|^2$ are all positive, $\rho_n$ can be Gaussian-integrated
around $\rho_n=0$ at large $k$ to yield
\begin{eqnarray}
  I_{N,\infty}&=&\prod_{n=1}^\infty\frac{1}{(1-t^ne^{in\gamma_1})(1-t^ne^{-in\gamma_1})}
  \cdot\frac{1}{N!}\oint\prod_{i=1}^N\frac{d\alpha_i}{2\pi}\prod_{i<j}
  \left(2\sin\frac{\alpha_i-\alpha_j}{2}\right)^2\nonumber\\
  &&\times\exp\left[\sum_{n=1}^\infty\frac{1}{n}\frac{N^2t^{2n}\chi_n\chi_{-n}}
  {(1-t^ne^{in\gamma_1})(1-t^ne^{-in\gamma_1})}\right]\ ,
\end{eqnarray}
where we defined $\chi_n\equiv\frac{1}{N}\sum_{i=1}^Ne^{in\alpha_i}$. Now taking large $N$
limit (after large $k$ limit), the $\alpha_i$ integral can again be approximated as $\chi_n$
integral. Including the Haar measure, one obtains the following index
\begin{equation}
  \hspace*{-0.8cm}\prod_{n=1}^\infty\frac{1}{(1-t^ne^{in\gamma_1})(1-t^ne^{-in\gamma_1})}
  \int\prod_{i=1}^Nd\chi_nd\chi_{-n}\exp\left[-N^2\sum_{n=1}^\infty\frac{1}{n}\chi_n\chi_{-n}
  \left(1-\frac{t^{2n}}{(1-t^ne^{in\gamma_1})(1-t^ne^{-in\gamma_1})}\right)\right]\ .
\end{equation}
Now the Gaussian integral for $\chi_n$ can either have positive or negative coefficient,
depending on how close $t$ is to $1$. When any of the coefficients for certain $n$ is negative,
this implies a large $N$ phase transition in which $\chi_n$ assumes a nonzero saddle point value.
At sufficiently low $t$, all coefficients are positive and one obtains a large $N$ index which
is independent of $N$. As we increase $t$, the first coefficient which approaches zero is
that for $n=1$. One finds the phase transition `temperature' $t_c$ to be
\begin{equation}
  1-\frac{t^2_c}{(1-t_ce^{i\gamma_1})(1-t_ce^{-i\gamma_1})}=0\ \rightarrow\ \
  t_c=\frac{1}{2\cos\gamma_1}\ .
\end{equation}
Beyond this point, the `index entropy' scales like $N^2$. Note that this large
$N$ transition happens only when we take the large $k$ limit first. Of course, this
is much smaller than what one would expect for the true entropy of the $(2,0)$ theory,
which should scale like $N^3$.

Like the indices for 4 dimensional SCFT \cite{Kinney:2005ej}, this could be implying
that the index cannot see the true degeneracy due to boson-fermion cancelation.
However, the situation is more nontrivial here as we still get some sort of phase
transition (even in a subsector which discards many states), while the indices of
\cite{Kinney:2005ej} do not undergo any. It will be interesting
to see if the inclusion of all the fermionic degrees makes the phase structure more
similar to what we expect for the $(2,0)$ theory partition function, and in particular
if we can see the $N^3$ scaling.

\section{Discussions}

In this paper, we calculated and studied an index for the BPS threshold bound states
of instantons and W-bosons. They can be regarded as BPS states of pure momentum or self-dual
strings with momentum on M5-branes. We explicitly showed that the instanton sum provides the
full Kaluza-Klein spectrum of the pure $U(1)$ instantons and $SU(2)$ single self-dual strings.
We also disclosed interesting structures of the degeneracies of various self-dual strings.
Finally, we showed that our index can be calculated in the symmetric phase and also provided
an interpretation as the superconformal index of the instanton sigma model.

There are immediate works that one can do to further clarify the physics of the
self-dual strings of various sorts that we discussed in this paper. Firstly,
the bound states of many $SU(2)$ self-dual strings are predicted to exist with nonzero
momentum. We can make an alternative study of them from the moduli space dynamics of
magnetic monopole strings. The simplest case with two identical $SU(2)$ monopole strings
can be studied from the 2d sigma model with $(4,4)$ supersymmetry with the target space
given by
\begin{equation}\label{target}
  \mathbb{R}^3\times\frac{S^1\times\mathcal{M}_4}{\mathbb{Z}_2}\ ,
\end{equation}
where $\mathcal{M}_4$ is the Atiyah-Hitchin space. Without momentum on the worldsheet,
there are no bound states of two monopoles unless provided with odd units of momentum
(i.e. the electric charge) on $S^1$ above \cite{Sen:1994yi}. Our findings suggest that
there would be (threshold) bound states without electric charge but with nonzero
momentum along the monopole string. As discussed in section 4.1, calculating the index
from this 2d QFT and subtracting the 2-particle index could give a result which we can
compare with our instanton calculation.

As outlined in section 4.2, one can also study the threshold bound states of two distinct
monopole strings in the $SU(3)$ theory by studying the index of a sigma model with the
target space of the form (\ref{target}), where $\mathcal{M}_4$ is now the Taub-NUT space.
It would be interesting to see if such a calculation can shed more lights on the nature of
the degrees appearing in (\ref{letter-index}). It is also the degenerate limit of a monopole string junction where the strings become parallel. These new degrees of freedom are neutral
excitations connecting two distinct D2 branes at the middle D4 branes.

Our study of the index, using its relation to the $\mathcal{N}\!=\!2^\ast$
partition function, was often based on numerical expansions in $q$.  It should be desirable
to obtain exact expressions for various self-dual strings from our index. $SL(2,\mathbb{Z})$
properties of this quantity could be a key aspect \cite{Hollowood:2003cv}, as this will
turn the instanton sum into a KK sum over the circle in 5d. In particular, systematic
analytic studies seem to be needed to obtain exact forms of indices for more complicated
bound states, from which one might be able to check if the $N^3$ some of degrees we
observed in this paper are indeed the building blocks of all BPS bound states
in the Coulomb phase.

Perhaps the most important and interesting direction is to further study the index
in the symmetric phase to learn more about the UV fixed point of the theory.
One can first continue studying the superconformal index for the instanton sigma model.
Although this is an old problem after \cite{Aharony:1997th,Aharony:1997an}, there was
some recent interest in studying this system \cite{Maldacena:2008wh}. For instance, it
will be interesting to see if one can study from our index the thermodynamics of black
holes asymptotic to plane waves, which could be a supersymmetric version of the plane wave
black holes discussed in \cite{Maldacena:2008wh}.

It will also be interesting to see if our index contains any clue for better understanding
the instanton parton proposals \cite{Collie:2009iz,Bolognesi:2011nh} in the symmetric phase. For this, perhaps a proper physical
understanding of our index (not as the superconformal index but as the index defined in
our section 2) would be needed.
The simplest place to consider  is the $SU(2)$ single instanton, whose moduli space is
$\mathbb{R}^4\times\frac{\mathbb{R}^4}{\mathbb{Z}_2}$. This is also the moduli space
of two $U(1)$ instantons, although the meaning of $\mathbb{R}^4/\mathbb{Z}_2$
is different. Due to the same geometric structure of the two moduli spaces, the index
(\ref{single-index}) with $N=2$ in the
former sector has similarity with the latter index, (\ref{U(1)-2-instanton}). In fact,
substituting $\mu_1-\mu_2=i(\alpha_1-\alpha_2)=2i\gamma_1$ in (\ref{single-index}) yields
(\ref{U(1)-2-instanton}) for $N=2$.


One can also study partition functions of 5d SYM on various Euclidean curved
manifolds $\mathcal{M}_5$, and see if one can relate them to observables of the
$(2,0)$ theory on $\mathcal{M}_5\times S^1$. For instance, it will be interesting
to see if a suitable partition function of maximal SYM on $S^5$ can be identified as
the superconformal index of $(2,0)$ theory on $S^5\times S^1$ \cite{Bhattacharya:2008zy}.

\vskip 0.5cm

\hspace*{-0.8cm} {\bf\large Acknowledgements}
\vskip 0.2cm

\hspace*{-0.75cm} We are grateful to Stefano Bolognesi, Nick Dorey, Kazuo Hosomichi,
Daniel Jafferis, Sangmin Lee, Noppadol Mekareeya, Hirosi Ooguri, Costis Papageorgakis,
Jaemo Park, Soo-Jong Rey, Jaewon Song, David Tong, Masahito Yamazaki and Piljin Yi for
helpful discussions. This work is supported by the BK21 program of the Ministry of
Education, Science and Technology (HK, SK), the National Research Foundation of Korea
(NRF) Grants No. 2010-0007512 (HK, SK), 2009-0072755, 2009-0084601 (HK), 2006-0093850,
2009-0084601 (KL) and 2005-0049409 through the Center for Quantum Spacetime(CQUeST)
of Sogang University (SK, KL).

\appendix

\section{Saddle points}

In this appendix, we study the supersymmetric saddle points invariant under $Q$,
around which the path integral will localize (after taking $\beta\rightarrow 0$,
$\zeta,v_i\rightarrow\infty$ limit). All fermions are naturally set to zero
at the saddle points, while the bosonic variables are constrained by
\begin{eqnarray}\label{saddle}
  &&Q\eta=[\phi,\bar\phi]=0\ ,\ \ Q\Psi_m=[\phi,a_m]-\frac{2i(\gamma_1J_{1L}+\gamma_RJ_R)}{\beta}a_m=0\ ,\nonumber\\
  &&Q\Psi_{m\!+\!4}=[\phi,\varphi_m]-\frac{2i(\gamma_2J_{2L}+\gamma_RJ_R)}{\beta}\varphi_m=0
  \ ,\ \ Q\vec\chi=i\vec{\mathcal{E}}=0\ ,\ \ Q\chi_a=i\mathcal{F}_a=0\nonumber\\
  &&Q\chi^{\dot{a}}=\epsilon^{\dot{a}\dot\alpha}\left(x_{\dot\alpha}\phi-\frac{\mu}{\beta}
  x_{\dot\alpha}+\frac{2i\gamma_RJ_R}{\beta}x_{\dot\alpha}\right)=0\ ,
\end{eqnarray}
where we integrated out $\vec{H}$ and $h_a$, and $\mu$ is to be regarded as a diagonal
$N\times N$ matrix. The condition $\vec{\mathcal{E}}=0$ requires solving algebraic equations
involving the $3k^2$ real ADHM constraints. The general solution for $\vec{\mathcal{E}}=0$
is unknown, but imposing other conditions will let us to restrict to special points of the
instanton moduli space, which can be explicitly obtained.
These saddle points are actually well-known and are classified by the $N$-colored Young
diagrams \cite{Nekrasov:2002qd,Nekrasov:2003rj}. We provide an elementary review of this construction and illustrate them
for the cases with instanton numbers $k=1,2,3$, to be used in the 1-loop calculations.
To see this structure, it is desirable to choose complex variables $B_1,B_2$ as
\begin{equation}
  a_{\alpha\dot\beta}=\frac{1}{\sqrt{2}}(\sigma^m)_{\alpha\dot\beta}a_m=\frac{1}{\sqrt{2}}
  \left(\begin{array}{cc}ia_3+a_4&ia_1+a_2\\ia_1-a_2&-ia_3+a_4\end{array}\right)_{\alpha\dot\beta}
  \equiv\left(\begin{array}{cc}iB_2&iB_1^\dag\\iB_1&-iB_2^\dag\end{array}\right)\ ,
\end{equation}
the eigenvalues of $J_{1L}$ and $J_R$ are $(-\frac{1}{2},+\frac{1}{2})$ for $B_1\equiv\frac{1}{\sqrt{2}}(a_1+ia_2)$ and $(-\frac{1}{2},-\frac{1}{2})$ for
$B_2^\dag\equiv\frac{1}{\sqrt{2}}(a_3+ia_4)$, respectively. The saddle point equations
involving $a_m$ on the first line of (\ref{saddle}) and the ADHM constraint are then given by
\begin{eqnarray}\label{phi-B-eqn}
  &&[\phi,B_1]=\frac{i(\gamma_R-\gamma_1)}{\beta}B_1\ ,\ \
  [\phi,B_2]=\frac{i(\gamma_R+\gamma_1)}{\beta}B_2\nonumber\\
  &&[B_1,B_2]+\bar{x}^{\dot{-}}x_{\dot{+}}=0\ ,\ \ [B_1^\dag,B_1]+[B_2^\dag,B_2]
  +\bar{x}^{\dot{+}}x_{\dot{+}}-\bar{x}^{\dot{-}}x_{\dot{-}}=\zeta
\end{eqnarray}
with $\zeta>0$. Another nontrivial equation is the third line of (\ref{saddle}), which is
\begin{equation}\label{x-eqn}
  x_{\dot\pm}\phi-\frac{\mu \mp i\gamma_R}{\beta}x_{\dot\pm} =0\ .
\end{equation}
All other equations apart from $[\phi,\bar\phi]=0$ are satisfied by taking $\varphi_m=0$.
The saddle point value of $\bar\phi$ will not be completely determined by supersymmetry only,
apart from a constraint coming from the leftover equation $[\phi,\bar\phi]=0$.
We shall later determine it from its equation of motion in subsection A.2, around
which the 1-loop fluctuations are suppressed.

The equation (\ref{x-eqn}) requires $2N$ row vectors of $x_{\dot\pm}$ with dimension $k$
to be eigenvectors of $\phi$ with eigenvalue $\frac{\mu_i\mp i\gamma_R}{\beta}$
for the $i$'th row $x_{i\dot\pm}$ (where $i=1,2,\cdots,N$), if the vector is nonzero.
We consider the saddle point solution with a diagonal $k\times k$ matrix
$\phi$. This can be attained by using the gauge transformation of $U(k)$ together with
$[\phi,\bar\phi]=0$.\footnote{This is true if $\phi$ and $\bar\phi$ saddle point values are
conjugate to each other, without complexifying the variables. Later, we shall see that the
eigenvalues of $\bar\phi$ which solve the equation of motion are not conjugate to the
eigenvalues of $\phi$, which is basically due to the fact that our action is complex after
redefining $q_{\dot\alpha}$ variables to $x_{\dot\alpha}$. However, the
common diagonal form does not have to be relaxed.} Then, the eigenvector $x_{i\dot\pm}$ can
be taken to have at most one nonzero vector element if the vector is nonzero. Since
the eigenvalues $\frac{\mu_i\mp i\gamma_R}{\beta}$ with different $\pm$ signs can never be
equal, one finds that the two vectors $x_{i\dot{+}}$ and $x_{j\dot{-}}$ are always
orthogonal, namely $x_{\dot\pm}\bar{x}^{\dot\mp}=0$. Also, vectors with different $U(N)$
indices are orthogonal, $x_{i\dot\pm}\bar{x}^{j\pm}=0$ for $i\neq j$, since the eigenvalues
are different.

To find the full solution for the $k\times k$ matrices $\phi,B_1,B_2$, we consider the $k$
dimensional vector space on which these matrices act. This vector space can be spanned
by the bra (row vector) $\langle\lambda|$ which are taken to be eigenvectors of $\phi$: $\langle\lambda|\phi=\lambda\langle\lambda|$. $x_{i\dot\pm}$ that we discussed above are
part of this complete set. From the first line of (\ref{phi-B-eqn}), one finds that
the actions of $B_1,B_2$ to this vector change its eigenvalue as
\begin{equation}
  \langle\lambda|B_1\propto\left\langle\lambda-i\frac{\gamma_R-\gamma_1}{\beta}\right|\ ,\ \
  \langle\lambda|B_2\propto\left\langle\lambda-i\frac{\gamma_R+\gamma_1}{\beta}\right|\ .
\end{equation}
Similarly, acting $B_1^\dag$ or $B_2^\dag$ on the bra shifts the eigenvalue in opposite ways.

We first show that $x_{\dot{-}}$ is identically zero. Suppose otherwise. Then we
can start from $\bar{x}_{i\dot{-}}\propto|\frac{\mu_i+i\gamma_R}{\beta}\rangle$ and act
$B_1,B_2$ many times. One obtains different vectors in the complete set as we do so,
as the imaginary part of the eigenvalue proportional to $\gamma_R$ is all positive and
increases as one acts more $B_1,B_2$. As the vector space is finite $k$ dimensional,
this process should stop after multiplying $B_1,B_2$ finitely many times. In particular,
there should be a state $|\lambda\rangle$ obtained this way which is annihilated by both
$B_1$, $B_2$. Sandwiching the last equation of (\ref{phi-B-eqn}) with this state, one
obtains
\begin{equation}
  -\langle\lambda|(B_1B_1^\dag+B_2B_2^\dag+\bar{x}^{\dot{-}}x_{\dot{-}})|\lambda\rangle
  =\zeta\langle\lambda|\lambda\rangle\ .
\end{equation}
We used the fact $x_{\dot{+}}|\lambda\rangle=0$, as the eigenvalues of
$|\lambda\rangle$ and $\bar{x}_{i\dot{+}}$ have different signs in the imaginary part proportional to $\gamma_R$. As the left hand side is non-positive while the right hand
side is positive with $\zeta>0$, one obtains a contradiction and proves $x_{\dot{-}}=0$.

One can similarly start from $x_{i\dot{+}}\propto\langle\frac{\mu-i\gamma_R}{\beta}|$
and act $B_1,B_2,B_1^\dag,B_2^\dag$ many times to generate more vectors in the complete
set. We first show that the bra $x_{i\dot{+}}$ is annihilated by $B_1^\dag,B_2^\dag$.
To see this, we again act $B_1^\dag,B_2^\dag$ on it till we obtain a bra
$\langle\lambda|$ annihilated by both (from finite dimension of the vector space).
Again contracting the last equation of (\ref{phi-B-eqn}) with this state, one obtains
\begin{equation}
  -\langle\lambda|(B_1B_1^\dag+B_2B_2^\dag-\bar{x}^{\dot{+}}x_{\dot{+}})
  |\lambda\rangle=\zeta\langle\lambda|\lambda\rangle\ ,
\end{equation}
where we again used the fact $\langle\lambda|\bar{x}^{\dot{-}}=0$. If the state
$\langle\lambda|$ is obtained by acting one or more $B_1^\dag,B_2^\dag$, then the
eigenvalue of this state is different from all eigenvalues of $x_{i\dot{+}}$ due to
different imaginary parts, yielding $\langle\lambda|\bar{x}^{\dot{+}}=0$. Then we
again have a contradiction. The only possibility of nonzero $x_{\dot{+}}$ is thus
having it annihilated by both $B_1^\dag,B_2^\dag$, allowing the second term of the
left hand side to be nonzero and positive. This proves our claim.

\begin{figure}[t]
  \begin{center}
    \includegraphics[width=13cm]{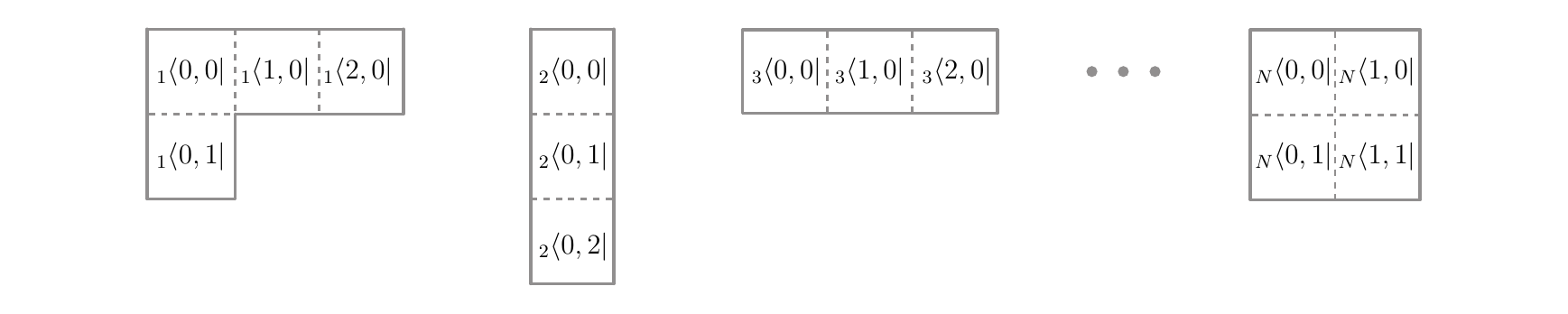}
    \caption{An $N$-colored Young diagram. Boxes map to
    the basis of $k$ dimensional vector space.}\label{colored-young}
  \end{center}
\end{figure}

Finally, we act $B_1,B_2$ on $x_{i\dot{+}}$ to obtain more vectors.
Since $x_{\dot{-}}=0$, we find from the third equation of (\ref{phi-B-eqn}) that
$[B_1,B_2]=0$. Therefore, we consider the normalized states
\begin{equation}
  {}_i\langle m,n|\propto x_{i\dot{+}}B_1^mB_2^n
\end{equation}
with $\phi$ eigenvalues $\frac{\mu_i-i(1\!+\!m\!+\!n)\gamma_R+i(m\!-\!n)\gamma_1}{\beta}$
for $m,n\geq 0$ and $i=1,2,\cdots,N$. This parametrization is non-redundant as the states
obtained by starting from different $x_{i\dot{+}}$ have different eigenvalues, from the
appearance of different $\mu_i$ in the eigenvalue. For certain values of $(m,n)$, the
state should be annihilated by both or one of $B_1,B_2$ to have finite dimensional vector
space. For given $i$, the possible set of vectors generated by acting $B_1,B_2$ are in
1-to-1 correspondence to the Young diagrams. See Fig. \ref{colored-young} for how each box
maps to a specific vector. The total number of boxes in the $N$ Young diagrams is the dimension
of the vector space, which should be $k$. Thus, the vector space maps to the
$N$-colored Young diagrams made of $k$ boxes \cite{Nekrasov:2003rj}.

For the actual construction of the solutions, one has to solve the last two equations of
(\ref{phi-B-eqn}), the ADHM conditions. It will be illustrated for small values of $k$
below.

\subsection{Examples}

At $k=1$, of course the ADHM constraint can be easily solved. Let us however construct
the solution following our logic above. Here, $x_{i\dot{+}}$ is simply a number for
each $i$. Only one  the $N$ numbers can be nonzero, which we take to be the $i$'th one.
Since the total vector space is $k=1$ dimensional, the vector $x_{i\dot{+}}$ itself is
annihilated by $B_1,B_2$, which are two complex numbers. For this to hold,
$B_1\!=\!B_2\!=\!0$. (This is also a simple consequence of the first two equations
of (\ref{phi-B-eqn}) at $k=1$.) One also finds $\phi=\frac{\mu_i-i\gamma_R}{\beta}$.
The last equation of (\ref{phi-B-eqn}) yields $x_{i\dot{+}}=\sqrt{\zeta}e^{i\theta}$,
where $\theta$ is the modulus for the broken $U(1)$ on the $i$'th D4-brane.
One thus finds $N$ different saddle points. We write the $i$'th saddle point as
$\Yboxdim7pt\yng(1)_{\ i}$ from the colored Young diagram notation. This can be regarded
as the saddle point for which the single instanton is bound to the $i$'th D4-brane.
It can also be eliminated by the $U(k)\rightarrow U(1)$ gauge symmetry.

At $k=2$, the two dimensional vectors $x_{i\dot{+}}$ can take following values.
Firstly, one may choose two nonzero vectors for different $i,j$ (which exists only
for $N\geq 2$). This corresponds to putting two instantons on different D4-branes,
and the resulting solution will turn out to be a simple `superposition' of the above
single instanton solutions. Using $U(2)$ gauge symmetry, we can take
\begin{equation}
  x_{i\dot{+}}=\lambda_1(1\ 0)\ ,\ \ x_{j\dot{+}}=\lambda_2(0\ 1)\ ,\ \
  \phi={\rm diag}\left(\frac{\mu_i-i\gamma_R}{\beta},\frac{\mu_j-i\gamma_R}{\beta}\right)\ .
\end{equation}
Since the two vectors in the complete set are already there, $B_1,B_2$ should annihilate
both $x_{i\dot{+}}$ and $x_{j\dot{+}}$, demanding $B_1\!=\!B_2\!=\!0$. Plugging the above
form of $x_{\dot{+}}$ into the real ADHM condition, one obtains
$|\lambda_1|=|\lambda_2|=\sqrt{\zeta}$. The remaining two phases of $\lambda_{1,2}$
are again from the $U(1)$ symmetries of the two D4-branes, and can also be eliminated
by the unbroken $U(1)^2\subset U(2)$ gauge symmetry for two instantons. In the Young
diagram notation, these ${}_NC_2$ saddle points are given by
$\Yboxdim7pt(\yng(1)_{\ i},\yng(1)_{\ j})$.

Secondly, one can choose only one of the $N$ vectors to be nonzero: among the $N$
possible saddle points, let us take $x_{i\dot{+}}$ to be nonzero and write
$x_{i\dot{+}}=\lambda\langle 1|$. As we need one more vector to form a complete set
for $k=2$, we allow either $B_1$ or $B_2$ to act on it nontrivially, corresponding
to the colored Young diagrams $\Yboxdim7pt\yng(2)_{\ i},\yng(1,1)_{\ i}$, respectively.
In the first case, let us take
\begin{equation}
  \langle 2|\propto\langle 1|B_1\ ,\ \ B_1=c|1\rangle\langle 2|\ ,
\end{equation}
where $|1\rangle,|2\rangle$ form an orthonormal complete set. From the ADHM equations,
one finds
\begin{equation}\label{saddle-2-1}
  x_{i\dot{+}}=\sqrt{2\zeta}(1\ 0)\ ,\ \ \phi={\rm diag}(\frac{\mu_i-i\gamma_R}{\beta},
  \frac{\mu_i-2i\gamma_R+i\gamma_1}{\beta})\ ,\ \
  B_1=\left(\begin{array}{cc}0&\sqrt{\zeta}\\0&0\end{array}\right)\ ,\ \ B_2=0\ .
\end{equation}
where we killed some variables which can be killed by an unbroken subgroup of $U(2)$.
Similarly, for the second saddle point, one obtains
\begin{equation}\label{saddle-2-2}
  x_{i\dot{+}}=\sqrt{2\zeta}(1\ 0)\ ,\ \ \phi={\rm diag}(\frac{\mu_i-i\gamma_R}{\beta},
  \frac{\mu_i-2i\gamma_R-i\gamma_1}{\beta})\ ,\ \ B_1=0\ ,\ \
  B_2=\left(\begin{array}{cc}0&\sqrt{\zeta}\\0&0\end{array}\right)\ .
\end{equation}
The above two saddle points have two instantons bound to the same $i$'th D4-brane,
and are essentially the $U(1)$ 2-instantons embedded to $U(N)$ in $N$ different ways.

It is interesting to compare our result with the general $U(1)$ two instantons studied in \cite{Lee:2000hp}. The convention in \cite{Lee:2000hp}  can be understood in our setting as
replacing our $\zeta>0$ by $-\zeta$ in ADHM condition. In our notation, the general ADHM
2-instanon solution is given by\footnote{The ADHM variables are related as
$(B_0,B_1)_{\rm theirs}=((B_2)^\dag,(B_1)^\dag)_{\rm ours}$, $J=x_{\dot{-}}$, $I^\dag=x_{\dot{+}}$, $\zeta_{\rm theirs}=\zeta_{\rm ours}$. Also, the solution of
\cite{Lee:2000hp} presented below is related to ours by a $U(2)$ gauge transformation
of exchanging $I=1,2$ rows/columns.}
\begin{equation}
  B_2^\dag=w_1{\bf 1}_2+\frac{z_1}{2}\left(\begin{array}{cc}1&\sqrt{\frac{2\beta}{\alpha}}\\
  0&-1\end{array}\right),\
  B_1^\dag=w_2{\bf 1}_2+\frac{z_2}{2}\left(\begin{array}{cc}1&\sqrt{\frac{2\beta}{\alpha}}\\
  0&-1\end{array}\right),\ x_+=\sqrt{\zeta}\left(\sqrt{1-\beta}, \sqrt{1+\beta}\right),\
  x_-=0
\end{equation}
where $\alpha\equiv\frac{|z_1|^2+|z_2|^2}{2\zeta}$ is the dimensionless parameter for the relative separation of two instantons, and $\beta\equiv\frac{1}{\alpha+\sqrt{1+\alpha^2}}$.
$x_{\dot\pm}$ are $N\times k=1\times 2$ matrices. Our solution can be viewed as a special
case of the general solution in which the center of mass position $w_1,w_2$ and the
relative separation $z_1,z_2$ are taken to be zero. Namely, in this limit the general
solution reduces to
\begin{equation}\label{2-instantons}
  B_2^\dag=\left(\begin{array}{cc}0&\sqrt{\zeta}\frac{z_1}{\sqrt{|z_1|^2+|z_2|^2}}\\
  0&0\end{array}\right)\ ,\ \
  B_1^\dag=\left(\begin{array}{cc}0&\sqrt{\zeta}\frac{z_2}{\sqrt{|z_1|^2+|z_2|^2}}\\
  0&0\end{array}\right)\ ,\ \ x_+=\left(0\ \ \sqrt{2\zeta}\right)\ ,\ \ x_-=0.
\end{equation}
The projective variables $\frac{z_1}{\sqrt{|z_1|^2+|z_2|^2}}$, $\frac{z_2}{\sqrt{|z_1|^2+|z_2|^2}}$ at $z_1,z_2=0$ parametrize the 2-sphere at the center
of the Eguchi-Hanson moduli space. Our chemical potentials further restrict the moduli on the
2-sphere, either at the north or south poles, $z_1/z_2=0$ or $\infty$. The two cases are
precisely our two solutions, (\ref{saddle-2-1}) and (\ref{saddle-2-2}), after a $U(2)$
gauge-transformation of exchanging the rows/columns.

Finally, let us explain the case with $k=3$. The colored Young diagrams of the form
\begin{equation}
  \Yboxdim8pt(\yng(1)_{\ i},\yng(1)_{\ j},\yng(1)_{\ k})\ ,\ \
  (\yng(2)_{\ i},\yng(1)_{\ j})\ ,\ \ (\yng(1,1)_{\ i},\yng(1)_{\ j})
\end{equation}
with different $i,j,k$ can all be understood as superpositions of $U(1)$ instantons
with $k\leq 2$ studied above. The remaining cases are $\Yboxdim7pt\yng(3)_{\ i},
\yng(2,1)_{\ i},\yng(1,1,1)_{\ i}$, which are also embeddings of $U(1)$ 3 instantons
to $U(N)$ in $N$ different ways.

For $\Yboxdim7pt\yng(3)$ , the vector space is spanned by $x_{1\dot{+}}\sim\langle 1|$,
$x_{1\dot{+}}B_1\sim\langle 2|$ and $x_{1\dot{+}}B_1^2\sim\langle 3|$.
The matrices take the following form:
\begin{equation}
  B_1=c_1|1\rangle\langle 2|+c_2|2\rangle\langle 3|\ ,\ \ B_2=0\ .
\end{equation}
For $\Yboxdim7pt\yng(2,1)$ , the vector space is spanned by $x_{1\dot{+}}\sim\langle 1|$,
$x_{1\dot{+}}B_1\sim\langle 2|$ and $x_{1\dot{+}}B_1B_2\sim\langle 3|$.
The matrices take the following form:
\begin{equation}
  B_1=c_1|1\rangle\langle 2|\ ,\ \ B_2=c_2|1\rangle\langle 3|\ .
\end{equation}
The case with $\Yboxdim7pt\yng(1,1,1)$ , has the role of $B_1,B_2$ changed
from the first case. The vector space is spanned by $x_{1\dot{+}}\sim\langle 1|$,
$x_{1\dot{+}}B_2\sim\langle 2|$ and $x_{1\dot{+}}B_2^2\sim\langle 3|$.
The matrices take the following form:
\begin{equation}
  B_1=0\ ,\ \ B_2=c_1|1\rangle\langle 2|+c_2|2\rangle\langle 3|\ .
\end{equation}
Plugging the above form into the ADHM equations, one obtains
\begin{eqnarray}
  &&\phi={\rm diag}\left(\frac{\mu_i-i\gamma_R}{\beta},\frac{\mu_i+i\gamma_1-2i\gamma_R}{\beta},
  \frac{\mu_i+2i\gamma_1-3i\gamma_R}{\beta}\right)\\
  &&B_1=\left(\begin{array}{ccc}0&\sqrt{2\zeta}&0\\0&0&\sqrt{\zeta}\\0&0&0\end{array}\right)
  \ ,\ \ B_2={\bf 0}_{3\times 3}\ ,\ \ x_{1\dot{+}}=\sqrt{3\zeta}\ (1\ 0\ 0)\nonumber
\end{eqnarray}
for $\Yboxdim7pt\yng(3)$ ,
$\phi_1=\frac{\mu_i-i\gamma_R}{\beta}$, $\phi_2=\frac{\mu_i-i\gamma_1-2i\gamma_R}{\beta}$, $\phi_3=\frac{\mu_i+i\gamma_1}{\beta}$ with
\begin{eqnarray}
  &&\phi={\rm diag}\left(\frac{\mu_i-i\gamma_R}{\beta},\frac{\mu_i+i\gamma_1-2i\gamma_R}{\beta},
  \frac{\mu_i-i\gamma_1-2i\gamma_R}{\beta}\right)\\
  &&B_1=\left(\begin{array}{ccc}0&\sqrt{\zeta}&0\\0&0&0\\0&0&0\end{array}\right)\ ,\ \
  B_2=\left(\begin{array}{ccc}0&0&\sqrt{\zeta}\\0&0&0\\0&0&0\end{array}\right)\ ,\ \
  x_{1+}=\sqrt{3\zeta}\ (1\ 0\ 0)\nonumber
\end{eqnarray}
for $\Yboxdim7pt\yng(2,1)$ , and
\begin{eqnarray}
  &&\phi={\rm diag}\left(\frac{\mu_i-i\gamma_R}{\beta},\frac{\mu_i-i\gamma_1-2i\gamma_R}{\beta},
  \frac{\mu_i-2i\gamma_1-3i\gamma_R}{\beta}\right)\\
  &&B_1={\bf 0}_{3\times 3}\ ,\ \
  B_2=\left(\begin{array}{ccc}0&\sqrt{2\zeta}&0\\0&0&\sqrt{\zeta}\\0&0&0\end{array}\right)
  \ ,\ \ x_{1\dot{+}}=\sqrt{3\zeta}\ (1\ 0\ 0)\nonumber
\end{eqnarray}
for $\Yboxdim7pt\yng(1,1,1)$ .

\subsection{The value of $\bar\phi$}

Let us consider the linear fluctuations of the action in $\delta\phi$ to determine the
saddle point value of $\bar\phi$. This field is not determined from supersymmetry only.
Physically, this is natural as $A_\tau$ has to be constrained by the $U(k)$ Gauss' law
constraint, which generally is an extra input even for supersymmetric configurations.
By varying $\delta\phi$, one obtains
\begin{equation}\label{gauss}
  0=-\frac{1}{2}\left[a_m,[\bar\phi,a_m]+\frac{2i(\gamma_1J_{1L}+\gamma_RJ_R)}{\beta}
  a_m\right]+\frac{1}{2}\left\{\bar\phi,\bar{x}^+x_+\right\}
  +\bar{x}^+\frac{\mu-i\gamma_R}{\beta}x_+-2\bar{x}^+vx_+\ .
\end{equation}
Rewriting $a_m$ with $B_1,B_2$, the first term on the right hand side can be rewritten as
\begin{eqnarray}\label{gauss-adjoint}
  \left[a_m,[\bar\phi,a_m]+\frac{2i(\gamma_1J_{1L}+\gamma_RJ_R)}{\beta}a_m\right]&=&
  [B_1,[\bar\phi,B_1^\dag]]+[B_1^\dag,[\bar\phi,B_1]]+[B_2,[\bar\phi,B_2^\dag]]
  +[B_2^\dag,[\bar\phi,B_2]]\nonumber\\
  &&-\frac{2i(\gamma_1-\gamma_R)}{\beta}[B_1^\dag,B_1]+\frac{2i(\gamma_1+\gamma_R)}{\beta}
  [B_2^\dag,B_2]\ .
\end{eqnarray}
We should solve this equation with diagonal $\bar\phi$, which is required
from one of the saddle point equation $[\phi,\bar\phi]=0$.

Here, note that all the other $U(k)$ adjoint variables $\phi,B_1,B_2$ take block diagonal forms
with $N$ possible blocks in their saddle point values, depending on the divisions of $k$ instantons
to $N$ possible D4-branes. It turns out that $\bar\phi$ equation can also be solved in this
block diagonal form. It suffices for us to consider the $i$'th block only, associated with
the $i$'th D4-brane and $i$'th VEV $v_i$.
In a direct study for all cases with $k\leq 3$, we found that the saddle point values
in the $i$'th block satisfy $\bar\phi=2v_i-\phi$. It is easy to generally show that
$\bar\phi=2v_i-\phi$ is the solution in this block. Firstly, inserting $\bar\phi=2v_i-\phi$,
one finds that (\ref{gauss-adjoint}) is exactly zero by using the first line of (\ref{phi-B-eqn}).
Then considering the remaining terms in (\ref{gauss}), and remembering that
$x_{\dot{+}}\phi=\frac{\mu-i\gamma_R}{\beta}\phi$ also implies $\phi\bar{x}^{\dot{+}}=
\bar{x}^{\dot{+}}\frac{\mu-i\gamma_R}{\beta}$ with our solutions, one finds that (\ref{gauss})
holds. The full solution is obtained by superposing these solutions.

\section{Determinants}

We study the 1-loop determinant around the saddle points found in the previous sections,
making it clear why Gaussian approximation suffices. The saddle points always satisfy
$\varphi_m=0$. Later, when we discuss the single instanton sector or the saddle points
in which all instantons are located on different D4-branes, further simplification would
arise since $a_m=0$.

We consider the quadratic fluctuations around a generic saddle point. We can separate
the problem into bosonic terms and fermionic terms. The bosonic fluctuation is given by
\begin{align}
  & \hspace*{-0.5cm} L^{(2)}_B  =
  \frac{1}{8}\left(2\delta\dot\varphi_5+[\phi,\delta\bar\phi]-[\bar\phi,\delta\phi]\right)^2
  -\frac{1}{2}[a_m,\delta\varphi_n]
  [a_m,\delta\varphi_n]+\bar{x}^+x_+\delta\varphi_m\delta\varphi_m
  \\ & \hspace*{-0.5cm}
  +\frac{1}{2}\left(\!\delta\dot{a}_m\!+\![\phi,\delta a_m]\!-\!
  [a_m,\delta\phi]\!-\!\frac{2i(\gamma_1J_{1L}\!+\!\gamma_RJ_R)}{\beta}\delta a_m\!\right)\!\!
  \left(\!\delta\dot{a}_m\!-\![\bar\phi,\delta a_m]\!+\![a_m,\delta\bar\phi]\!-\!
  \frac{2i(\gamma_1J_{1L}+\gamma_RJ_R)}{\beta}\delta a_m\right)\nonumber\\ & \hspace*{-0.5cm}
  +\frac{1}{2}\left(\delta\dot{\varphi}_m+[\phi,\delta\varphi_m]
  -\frac{2i(\gamma_2J_{2L}+\gamma_RJ_R)}{\beta}
  \delta\varphi_m\right)\left(\delta\dot{\varphi}_m-[\bar\phi,\delta\varphi_m]
  -\frac{2i(\gamma_2J_{2L}+\gamma_RJ_R)}{\beta}\delta\varphi_m\right)\nonumber\\ & \hspace*{-0.5cm}
  +\frac{1}{2}\left(\bar{x}^+\delta x_++\delta\bar{x}^+ x_+-[B_1,\delta B_1^\dag]-[\delta B_1,B_a^\dag]-(1\rightarrow 2)\right)^2+
  2\left|\delta\bar{x}^- x_++[B_1,\delta B_2]-[B_2,\delta B_1]\right|^2\nonumber\\ & \hspace*{-0.5cm}
  +\frac{1}{2}\left\{\phi+\partial_\tau,\bar\phi-\partial_\tau
  \right\}\delta\bar{x}^{\dot\alpha}\delta x_{\dot\alpha}+(\phi-\bar\phi+2\partial_\tau)
  \delta\bar{x}^{\dot\alpha}\frac{\mu-2i\gamma_RJ_R}{\beta}\delta x_{\dot\alpha}
  -\delta\bar{x}^{\dot\alpha}\frac{(\mu-2i\gamma_RJ_R)^2}{\beta^2}
  \delta x_{\dot\alpha}\nonumber\\ & \hspace*{-0.5cm}
  -2\left((\phi+\partial_\tau)\delta\bar{x}^{\dot\alpha}
  -\delta\bar{x}^{\dot\alpha}\frac{\mu-2i\gamma_RJ_R}{\beta}
  \right)v\delta x_{\dot\alpha}+\frac{1}{2}\{\delta\phi,\delta\bar\phi\}\bar{x}^+x_+
  -2\delta\phi\left(\delta\bar{x}^+vx_+
  +\bar{x}^+v\delta x_+\right) \nonumber\\ & \hspace*{-0.5cm}
  +\frac{1}{2}(\{\delta\phi,\bar\phi-\partial_\tau\}+
  \{\phi+\partial_\tau,\delta\bar\phi\})\left(\delta\bar{x}^+x_++\bar{x}^+\delta x_+\right)
  +(\delta\phi-\delta\bar\phi)\left(\delta\bar{x}^+\frac{\mu-i\gamma_R}{\beta}x_++\bar{x}^+
  \frac{\mu-i\gamma_R}{\beta}\delta x_+\right) \nonumber \ ,
\end{align}
where we used the facts $\varphi_m\!=\!0$, $x_-\!=\!0$ at the saddle points. All charge operators
are understood to act on the variables on their right, and the time derivatives in
$\{\phi+\partial_\tau,\bar\phi-\partial_\tau\}$ are acting on $\delta\bar{x}^{\dot\alpha}$
and all other objects in between.

To analyze the fermionic fluctuation, it is slightly inconvenient to work with the
cohomological variables. So we work directly with the original variables, while at the final
stage the background bosonic variables will be rewritten in cohomological formulation.
One obtains
\begin{align}
  & \hspace*{-0.5cm} L^{(2)}_F =
  \frac{1}{2}(\bar\lambda_a^{\ \dot\alpha})^\dag\!
  \left(\!\dot{\bar\lambda}_a^{\ \dot\alpha}\!-\![\bar\phi,\bar\lambda_a^{\ \dot\alpha}]
  \!-\!\frac{2i(\gamma_2J_{2L}+\gamma_RJ_R)}{\beta}\bar\lambda_a^{\ \dot\alpha}\!\right)\!+\!
  \frac{1}{2}(\lambda^{\dot{a}}_{\ \alpha})^\dag\!\left(\dot{\lambda}^{\dot{a}}_{\ \alpha}\!-\![\bar\phi,\lambda^{\dot{a}}_{\ \alpha}]
  \!-\!\frac{2i(\gamma_1J_{1L}+\gamma_RJ_R)}{\beta}\lambda^{\dot{a}}_{\ \alpha}
  \right)\nonumber\\ & \hspace*{-0.5cm}
  +\frac{1}{2}(\lambda_{a\alpha})^\dag\left(\dot\lambda_{a\alpha}
  +[\phi,\lambda_{a\alpha}]-\frac{2i(\gamma_1J_{1L}\!+\!\gamma_2J_{2L})}{\beta}
  \lambda_{a\alpha}\right)+\frac{1}{2}(\bar\lambda^{\dot{a}\dot\alpha})^\dag
  \left(\dot{\bar\lambda}^{\dot{a}\dot\alpha}+[\phi,\bar\lambda^{\dot{a}\dot\alpha}]
  -\frac{2i\gamma_RJ_R}{\beta}\bar\lambda^{\dot{a}\dot\alpha}\right)\nonumber\\ & \hspace*{-0.5cm}
  +(\xi_a)^\dag\left(\dot\xi_a-\xi_a\phi+\frac{\mu}{\beta}\xi_a
  -\frac{2i\gamma_2J_{2L}}{\beta}\xi_a\right)+(\xi^{\dot{a}})^\dag\left(\dot\xi^{\dot{a}}
  -2v\xi^{\dot{a}}+\xi^{\dot{a}}\bar\phi+\frac{\mu-2i\gamma_RJ_R}{\beta}\xi^{\dot{a}}\right) \nonumber \\ & \hspace*{-0.5cm}
  +\frac{i}{2}\left((\lambda^{\dot{a}}_{\ \alpha})^\dag
  [(\sigma^m)_{\alpha\dot\beta}a_m,\bar\lambda^{\dot{a}\dot\beta}]-(\bar\lambda_a^{\ \dot\alpha})^\dag[(\bar\sigma^m)^{\dot\alpha\beta}a_m,\lambda_{a\beta}]
  -(\bar\lambda^{\dot{a}\dot\alpha})^\dag
  [(\bar\sigma^m)^{\dot\alpha\beta}a_m,\lambda^{\dot{a}}_{\ \beta}]+ (\lambda_{a\alpha})^\dag[(\sigma^m)_{\alpha\dot\beta}a_m,\bar\lambda_a^{\ \dot\beta}]
  \right) \nonumber \\ & \hspace*{-0.5cm}
  -\sqrt{2}i\left((\bar\lambda_a^{\ \dot\alpha})^\dag\bar{x}^{\dot\alpha}\xi_a-(\xi_a)^\dag x_{\dot\alpha}\bar\lambda_a^{\ \dot\alpha}+(\bar\lambda^{\dot{a}\dot\alpha})^\dag\bar{x}^{\dot\alpha}\xi^{\dot{a}}-
  (\xi^{\dot{a}})^\dag x_{\dot\alpha}\bar\lambda^{\dot{a}\dot\alpha}\right)\ .
\end{align}
The fourth and fifth lines are conjugate to each other.

In the bosonic part of the quadratic action, note that all the coefficients are quadratures
of $\frac{\mu^i}{\beta}$, $\frac{\gamma_{1L}}{\beta}$, $\frac{\gamma_{2L}}{\beta}$,
$\frac{\gamma_R}{\beta}$, $\sqrt{\zeta}$, $v^i$, or $\partial_\tau\sim\frac{1}{\beta}$,
where the last expression holds as the time circle has circumference length $\beta$. Since
the action is $\int d\tau L^{(2)}_B$, there is an extra factor of $\beta$ multiplied to these
quadratures. It is guaranteed that the resulting Gaussian measures are steep once we set
$\beta^{-1}\sim v^i\sim\sqrt{\zeta}\rightarrow\infty$. Recall that we are allowed to take
these limits since index does not depend on the values of $\beta,v^i,\zeta$, being
parameters of the theory or a regulator. Thus, the path integral over bosonic variables are
localized around the saddle points. Once bosonic variables are localized, fermionic action
is exactly quadratic in our theory so that we can completely rely on Gaussian approximation
to calculate the index.

Below, we shall elaborate on the 1-loop calculation in the single instanton sector,
as this is relatively simple and sheds some light on some important structures. In the
single instanton sector, we also pay detailed attention to the regularization/cancelation
of divergent parts and the gauge fixing. We have treated two instantons and three instantons
cases in similar manner, being less rigorous on the gauge fixings. Since the analysis becomes
exceedingly messier for two and three instantons, we relied mostly on numerical evaluation
of the determinant to get the index for two and three instantons: we just present the results
for $k=2,3$ in the main text.

For single instantons, we can set $a_m=0$ in the background
and furthermore ignore all commutators of $k\times k$ matrices. The quadratic bosonic
fluctuations around the $i$'th saddle point consist of following parts.
\begin{enumerate}

\item $\delta a_m$: The action is given by
$$
  \beta\left(\delta a_{+\dot\pm}\right)^\ast\left(\frac{2\pi i n}{\beta}+\frac{i(\gamma_1\pm\gamma_R)}{\beta}\right)\left(-\frac{2\pi i n}{\beta}-\frac{i(\gamma_1\pm\gamma_R)}{\beta}\right)\delta a_{+\dot\pm}\ ,
$$
for the mode $\delta a_{+\dot\pm}$ coming with time dependence
$e^{-\frac{2\pi i\tau}{\beta}}$. The determinant is given by
\begin{equation}
  \left[\mathcal{N}^4\sin^2\frac{\gamma_1+\gamma_R}{2}\sin^2\frac{\gamma_1-\gamma_R}{2}
  \right]^{-1}\ ,
\end{equation}
where $\mathcal{N}\equiv-\frac{2i}{\beta^{1/2}}\prod_{n\neq 0}
\left(\frac{-2\pi i n}{\beta^{1/2}}\right)$.

\item $\delta\varphi_m$: The action for the $n$'th Fourier mode is
$$
  \left(\delta \varphi_{+\dot\pm}\right)^\ast\left[-\left(-\frac{2\pi i n}{\beta}
  -\frac{i(\gamma_2\pm\gamma_R)}{\beta}\right)^2+\zeta\right]\delta \varphi_{+\dot\pm}\ ,
$$
whose determinant is given by
\begin{equation}
  \left[\mathcal{N}^4\prod_{\pm}\sin\left(\frac{\gamma_2\pm\gamma_R}{2}
  +i\sqrt{\frac{\zeta\beta^2}{2}}\right)\sin\left(\frac{\gamma_2\pm\gamma_R}{2}
  -i\sqrt{\frac{\zeta\beta^2}{2}}\right)\right]^{-1}\ .
\end{equation}

\item $\delta x_{\dot\alpha j}$ with $j\neq i$: The action for $n$'th Fourier mode is
$$
  \delta\bar{x}^{\dot\pm}_j\delta x_{\dot\pm j}\left(\frac{(\mu_i-i\gamma_R)\!-\!
  (\mu_j\mp i\gamma_R)}{\beta}-\frac{2\pi i n}{\beta}\right)\left(2(v_i\!-\!v_j)-\frac{(\mu_i-i\gamma_R)\!-\!
  (\mu_j\mp i\gamma_R)}{\beta}+\frac{2\pi i n}{\beta}\right)\ ,
$$
and the determinant is given by the inverse of
\begin{align}
  \prod_{j\neq i}\mathcal{N}^4\sinh\left(\frac{\mu_j\!-\!\mu_i}{2}\right)
  \sinh\left(\frac{\mu_j\!-\!\mu_i\!+\!2i\gamma_R}{2}\right) & \nonumber \\ \times
  \sinh\left(\frac{\mu_j\!-\!\mu_i}{2}-2\beta(v_j\!-\!v_i)\right) &
  \sinh\left(\frac{\mu_j\!-\!\mu_i\!+\!2i\gamma_R}{2}-2\beta(v_j\!-\!v_i)\right)\ .
\end{align}

\item $\delta\phi,\delta\bar\phi,\delta x_{\pm i}$: The $\delta x_{-i}$ part of the action is
$$
  \delta\bar{x}^-_j\delta x_{-j}\left(-\frac{2i\gamma_R}{\beta}-\frac{2\pi i n}
  {\beta}\right)\left(\frac{2\pi i\gamma_R}{\beta}+\frac{2\pi in}{\beta}\right)
  +2\zeta\left|\delta x_{-i}\right|^2
$$
for $n$'th Fourier mode, leading to the determinant
\begin{equation}
  \left[\mathcal{N}^2\sin\left(\gamma_R+i\sqrt{\frac{\zeta\beta^2}{2}}\right)
  \sin\left(\gamma_R-i\sqrt{\frac{\zeta\beta^2}{2}}\right)\right]^{-1}\ .
\end{equation}
The fluctuation of $x_{+i}$ is taken to be
$$
  x_{+i}=e^{i\theta}\left(\sqrt{\zeta}+\frac{\delta r}{\sqrt{2}}\right)\ ,
$$
since $\theta$ is an exactly flat direction.
The remaining part of the Lagrangian is
$$
  \frac{1}{2}\left(\delta\dot{r}\right)^2+\zeta\left(\delta r\right)^2+\zeta\left(\dot\theta+\delta A_\tau\right)^2
  +\frac{1}{2}\left(\delta\dot\varphi_5\right)^2+\zeta(\delta\varphi_5)^2\ .
$$
This part requires gauge fixing. We choose the gauge $\theta=0$. The
Faddeev-Popov determinant is simply $1$. The integration measure is given by
$$
  \int[\sqrt{2\zeta} dr][d(\delta A_\tau)d(\delta\varphi_5)]\exp
  \left[-\int d\tau\left(\frac{1}{2}\left(\delta\dot{r}\right)^2+\zeta(\delta r)^2
  +\zeta\left(\delta A_\tau\right)^2+\frac{1}{2}(\delta\dot\varphi_5)^2
  +\zeta\left(\delta\varphi_5\right)^2\right)\right]\ .
$$
Contribution from  $r,A_\tau,\varphi_5$ is given by
\begin{equation}
  \left[\mathcal{N}^2\sinh^2\sqrt{\frac{\zeta\beta^2}{2}}\right]^{-1}\ .
\end{equation}
\end{enumerate}
For the fermions, one obtains the following contributions.
\begin{enumerate}

\item $\lambda_{a\alpha}$: The action consists purely of kinetic term. Taking care of
 the realith condition for fermions, the determinant is given by
\begin{equation}
  \mathcal{N}^2\sin\frac{\gamma_1\!+\!\gamma_2}{2}\sin\frac{\gamma_1\!-\!\gamma_2}{2}\ .
\end{equation}

\item $\lambda^{\dot{a}}_{\ \alpha}$: The determinant is
\begin{equation}
  \mathcal{N}^2\sin\frac{\gamma_1+\gamma_R}{2}\sin\frac{\gamma_1-\gamma_R}{2}\ .
\end{equation}

\item $\xi^{\dot{a}}_j$ with $j\neq i$: The determinant is
\begin{equation}
  \prod_{i\neq i}\mathcal{N}^2\sinh\left(\frac{\mu_j\!-\!\mu_i}{2}-2(v_j\!-\!v_i)\beta\right)
  \sinh\left(\frac{\mu_j\!-\!\mu_i\!+\!2i\gamma_R}{2}-2(v_j\!-\!v_i)\beta\right)\ .
\end{equation}

\item $\bar\lambda^{\dot{a}\dot\alpha},\xi^{\dot{a}}_i$: The action is given by
$$
  \left(\bar\lambda^{\dot\pm\dot{+}}\ \xi^{\dot\pm}_i\right)^\ast
  \left(\begin{array}{cc}-\frac{2\pi i n}{\beta}+\frac{i\gamma_R}{\beta}
  \pm\frac{i\gamma_R}{\beta}&-\sqrt{2\zeta}ie^{-i\theta}\\
  \sqrt{2\zeta}ie^{i\theta}&-\frac{2\pi i n}{\beta}+\frac{i\gamma_R}{\beta}\pm\frac{i\gamma_R}{\beta}\end{array}\right)
  \left(\begin{array}{c}\bar\lambda^{\dot\pm\dot{+}}\\ \xi^{\dot\pm}_i\end{array}\right)\ .
$$
The determinant is given by
\begin{equation}
  \mathcal{N}^4\sinh^2\sqrt{\frac{\zeta\beta^2}{2}}
  \sin\left(\gamma_R+i\sqrt{\frac{\zeta\beta^2}{2}}\right)
  \sin\left(\gamma_R-i\sqrt{\frac{\zeta\beta^2}{2}}\right)\ .
\end{equation}

\item $\xi_{aj}$ with $j\neq i$: The determinant is
\begin{equation}
  \prod_{j\neq i}\mathcal{N}^2\sinh\frac{\mu_j\!-\!\mu_i\!-\!i\gamma_2\!+\!i\gamma_R}{2}
  \sinh\frac{\mu_j\!-\!\mu_i+i\gamma_2\!+\!i\gamma_R}{2}\ .
\end{equation}

\item $\bar\lambda_{a}^{\ \dot\alpha},\xi_{ai}$: Action is given by
$$
  \left(\bar\lambda_{\pm}^{\ \dot{+}}\ \xi_{\pm i}\right)^\ast
  \left(\begin{array}{cc}-\frac{2\pi i n}{\beta}\mp\frac{i\gamma_2}{\beta}
  +\frac{i\gamma_R}{\beta}&-\sqrt{2\zeta}ie^{-i\theta}\\
  \sqrt{2\zeta}ie^{i\theta}&-\frac{2\pi i n}{\beta}\mp\frac{i\gamma_2}{\beta}
  +\frac{i\gamma_R}{\beta}\end{array}\right)
  \left(\begin{array}{c}\bar\lambda_{\pm}^{\ \dot{+}}\\ \xi_{\pm i}\end{array}\right)\ .
$$
The determinant is given by
\begin{equation}
  \mathcal{N}^4\prod_{\pm}\sin\left(\frac{\gamma_2\pm\gamma_R}{2}+i\sqrt{\frac{\zeta\beta^2}{2}}
  \right)\sin\left(\frac{\gamma_2\pm\gamma_R}{2}-i\sqrt{\frac{\zeta\beta^2}{2}}\right)\ .
\end{equation}

\end{enumerate}
Combining bosonic and fermionic contributions, one finds that many terms depending on
$\beta,v^i,\zeta$ all cancel out, as it should. After all cancelation, one obtains the
following index associated with the $i$'th saddle point:
\begin{equation}
  I_i=\left(\frac{\sin\frac{\gamma_1\!+\!\gamma_2}{2}\sin\frac{\gamma_1\!-\!\gamma_2}{2}}
  {\sin\frac{\gamma_1+\gamma_R}{2}\sin\frac{\gamma_1-\gamma_R}{2}}\right)\prod_{j(\neq i)=1}^N
  \left(\frac{\sinh\frac{\mu_j\!-\!\mu_i\!-\!i\gamma_2\!+\!i\gamma_R}{2}
  \sinh\frac{\mu_j\!-\!\mu_i\!+\!i\gamma_2\!+\!i\gamma_R}{2}}{\sinh\frac{\mu_j\!-\!\mu_i}{2}
  \sinh\frac{\mu_j\!-\!\mu_i\!+\!2i\gamma_R}{2}}\right)\ .
\end{equation}
The first part comes from the contribution of center of mass supermultiplet. The full
contribution at $k\!=\!1$ is simply the summation over the indices from $N$ different saddle
points, $I_{k=1}=\sum_{i=1}^NI_i$.

\end{document}